%% file: 00-main.tex
\documentclass[sigconf]{acmart}
\AtBeginDocument{%
  }

\copyrightyear{2026}
\acmYear{2026}
\setcopyright{cc}
\setcctype{by}
\acmConference[CHI '26]{Proceedings of the 2026 CHI Conference on Human Factors in Computing Systems}{April 13--17, 2026}{Barcelona, Spain}
\acmBooktitle{Proceedings of the 2026 CHI Conference on Human Factors in Computing Systems (CHI '26), April 13--17, 2026, Barcelona, Spain}
\acmDOI{10.1145/3772318.3791600}
\acmISBN{979-8-4007-2278-3/2026/04}

\usepackage{microtype}

\begin{document}
\title{Feedback by Design: Understanding and Overcoming User Feedback Barriers in Conversational Agents}

\graphicspath{{figures/}}

\author{Nikhil Sharma}
\authornote{This work was done during the author’s internship at Adobe}
\email{nsharm27@jhu.edu}
\orcid{0009-0004-9183-6811}
\affiliation{%
  \institution{Johns Hopkins University}
  \city{Baltimore}
  \state{MD}
  \country{USA}
}

\author{Zheng Zhang}
\email{zhengzhang@adobe.com}
\affiliation{%
  \institution{Adobe Inc.}
  \city{San Jose}
  \state{CA}
  \country{USA}
}

\author{Daniel Lee}
\email{dlee1@adobe.com}
\affiliation{%
  \institution{Adobe Inc.}
  \city{San Jose}
  \state{CA}
  \country{USA}
}

\author{Namita Krishnan}
\email{namitak@adobe.com}
\affiliation{%
  \institution{Adobe Inc.}
  \city{San Jose}
  \state{CA}
  \country{USA}
}

\author{Guang-Jie Ren}
\email{gren@adobe.com}
\affiliation{%
  \institution{Adobe Inc.}
  \city{San Jose}
  \state{CA}
  \country{USA}
}

\author{Ziang Xiao}
\email{ziang.xiao@jhu.edu}
\affiliation{%
  \institution{Johns Hopkins University}
  \city{Baltimore}
  \state{MD}
  \country{USA}
}

\author{Yunyao Li}
\email{yunyaol@adobe.com}
\affiliation{%
  \institution{Adobe Inc.}
  \city{San Jose}
  \state{CA}
  \country{USA}
}

\renewcommand{\shortauthors}{Sharma et al.}

\begin{abstract}
High-quality feedback is essential for effective human–AI interaction. It bridges knowledge gaps, corrects digressions, and shapes system behavior; both during interaction and throughout model development. Yet despite its importance, human feedback to AI is often infrequent and low quality. This gap motivates a critical examination of human feedback during interactions with AIs. To understand and overcome the challenges preventing users from giving high-quality feedback, we conducted two studies examining feedback dynamics between humans and conversational agents (CAs). Our formative study, through the lens of Grice’s maxims, identified four Feedback Barriers---Common Ground, Verifiability, Communication, and Informativeness---that prevent high-quality feedback by users. Building on these findings, we derive three design desiderata and show that systems incorporating scaffolds aligned with these desiderata enabled users to provide higher-quality feedback. Finally, we detail a call for action to the broader AI community for advances in Large Language Models capabilities to overcome Feedback Barriers.
\end{abstract}




 \maketitle
\input{sections/01-introduction}
\input{sections/02-related-work}

\input{sections/03-formative-study}

\input{sections/04-design-desidarata}

\input{sections/06-study-2}

\input{sections/07-results}

\input{sections/08-discussion}

\input{sections/09-conclusion}

\begin{acks}
This work is partially supported by Science of trustworthy AI award from Schmidt Sciences.
\end{acks}

\bibliographystyle{ACM-Reference-Format}
\bibliography{ref}
\newpage
\appendix

\input{sections/10-appendix}
\end{document}

%% file: sections/01-introduction.tex
\section{Introduction}
The rapid progress of foundation models~\cite{chatgpt} and agentic AI~\cite{durante2024agentaisurveyinghorizons} have elevated Conversational Agents (CAs) into one of the central paradigms for Human–AI interaction. These systems can now reason over complex knowledge bases and dynamic contexts to address multifaceted user goals, while leveraging human feedback to refine their responses in multi-turn conversations. In multi-turn goal-oriented tasks, where users and model have a shared goal, high-quality human feedback is pivotal to keep CAs aligned with the user’s evolving intent and requirements, enabling the successful pursuit of goals across extended conversations.

Human feedback also plays a critical role in today's machine learning (ML) research \cite{doi:10.1177/00222429241276529,zou2025collaborativeintelligencehumanagentsystems,10.1145/3610206,vats2025surveyhumanaicollaborationlarge}. Techniques like Direct Preference Optimization (DPO) \cite{rafailov2024directpreferenceoptimizationlanguage} or Reinforcement Learning from Human Feedback (RLHF) \cite{ouyang2022training} extract preference signals from human feedback to steer model behavior. With more powerful and embedded models in diverse high-stakes domains, human feedback quality will influence system behavior and outputs \cite{Min_2025,kirk-etal-2023-past, fernandes-etal-2023-bridging}. 

However, recent literature reveals a troubling disconnect: despite the critical importance of human feedback, people rarely provide high-quality feedback when interacting with CAs in real-world settings \cite{shi2025wildfeedbackaligningllmsinsitu,DonYehiya2024NaturallyOF,mysore-etal-2025-prototypical}.
For instance, \textit{WildFeedback}\cite{shi2025wildfeedbackaligningllmsinsitu}, a dataset of in-the-wild human feedback occurring over 1 million ChatGPT conversations, shows that out of 1.02 million conversations, only 38,992 (3.89\%) contain feedback. Furthermore, the feedback people gave was often ambiguous and incomplete, such as ``The flow is not natural'', ``wrong'' and``incorrect try again''. This pattern persists even during complex, multi-turn tasks where high-quality feedback would be most beneficial for achieving user goals. The result demonstrates a critical gap: while AI systems are designed to learn and adapt from human input, users consistently fail to provide clear, specific feedback to improve the interaction outcome. Although synthetic feedback has emerged as a scalable alternative\cite{lee2023rlaif}, synthetic feedback cannot fully capture the contextual nuances and diverse cultural perspectives that humans bring\cite{Tao_2024,faux-polyglot} when interacting with CAs --- making human feedback irreplaceable in today's AI training paradigm.

In contrast to human feedback towards CAs, in many everyday settings, humans provide frequent, high-quality feedback; In education and organizations, people routinely exchange timely, specific feedback. Research demonstrates that humans are adept at giving effective feedback \cite{Topping01012009, Henderson2025Comparing}. In collaborative work environments, humans regularly provide contextual, actionable feedback that helps others improve performance \cite{hattie2007power}. Even in informal settings, people engage in effective feedback behaviors --- from coaching sports teams to mentoring colleagues --- suggesting that the capacity for quality feedback is inherent to human social interaction~\cite{kluger1996effect}. This raises a critical question: if humans are capable feedback providers in other domains, what prevents them from providing high-quality feedback when interacting with Conversational Agents?

HCI research has studied how to elicit effective feedback in human–AI interaction~\cite{honeycutt2020soliciting,ou2022artemis,vaccaro2024meta,ahn2021trust,dietvorst2015algorithm}. Proposed scaffolds range from mixed-initiative systems that dynamically negotiate control with users~\cite{horvitz1999principles,allen1999mixedinitiative,horvitz1999continual} to interactive machine learning frameworks that lower the barrier for corrective input~\cite{amershi2014power,fails2003interactive,simard2017machine}. These approaches emphasize design strategies such as exposing system state, structuring user input, and sharing initiative in dialogue to improve both quality and frequency of user feedback. However, most techniques were developed for rule-based or hybrid systems that lack the capacity for multi-round complex tasks or incorporating nuanced feedback during interaction~\cite{chen2017survey,zhang2020recent}. 

The emergence of LLM-powered conversational agents with their capacity to have natural multi-turn conversations, prompts new questions: What barriers limit user feedback with LLM-powered Conversational Agents \textbf{(RQ1)}, and how should we design scaffolds to foster sustained, high-quality feedback in richer, more effective interactions \textbf{(RQ2)}?

To explore the feedback dynamics in Conversational Agents, we conducted two studies. To answer \textbf{RQ1}, we carried out a formative study with 16 participants, who regularly use CAs for diverse, goal-driven tasks such as creative writing, coding, image editing, research, and search. Through in-depth interviews, we identified common breakdowns in human-AI interaction, how users provide feedback to the models, and what prevents users from giving high-quality feedback. Through the lens of Grice's maxims\cite{grice1975logic}, which describe cooperative principles that human-human conversations follow, we identify and define four key \textbf{Feedback Barriers}.

In our second study, we designed and implemented \emph{FeedbackGPT}, which incorporates six model-agnostic scaffolds, each targeting a specific feedback barrier identified in Study 1. To answer \textbf{RQ2}, we recruited 20 participants for a within-subject study where they completed two goal-oriented tasks using the baseline interface (ChatGPT) and FeedbackGPT. Through a mixed-methods analysis, we found that providing scaffolds that minimize feedback barriers enabled users to provide higher-quality and frequent feedback.

In summary, our work makes three primary contributions:
\begin{itemize}
    \item \emph{First}, we provide empirical insights into the \textbf{Feedback Barriers} that deter users from providing high-quality feedback during interactions with Conversational Agents.
    \item \emph{Second}, we present the design and implementation of FeedbackGPT, a system incorporating a set of model and task-agnostic scaffolds that effectively encourages users to provide more frequent and higher-quality feedback.
    \item \emph{Third}, we contribute a set of practical design recommendations to elicit high-quality user feedback during interactions with Conversational Agents.
\end{itemize} 

%% file: sections/02-related-work.tex
\section{Related Works}

\subsection{Feedback in Human–AI Interaction}
Interactive systems have long required human feedback for task completion and alignment. In the design and evaluation of human–AI systems, feedback is typically defined as information provided \emph{after} the system has acted, indicating how well its behavior met a predefined goal~\cite{hattie2007power,shute2008formative}. Research distinguishes two main forms of feedback: \emph{explicit feedback}, such as ratings or corrections, and \emph{implicit feedback}, derived passively from users’ behaviors (e.g., clicks, skips, dwell time). Early human-in-the-loop (HITL) and interactive-machine-learning (IML) research treat feedback as critical for incremental improvement of models and interfaces~\cite{wang-etal-2021-putting,MosqueiraRey2023HITL}  

In HCI and ML literature, eliciting effective user feedback has long been central. Mixed-initiative systems emphasized shared control between human and agent, enabling users to continuously steer system behavior rather than issuing one-shot commands~\cite{horvitz1999principles,allen1999mixedinitiative}. Interactive ML extends this by treating users as collaborators who iteratively label, correct, and re-weight examples such that models can adapt in situ~\cite{amershi2014power,10.1145/3185517,Pfeuffer2023XILADR}. Complementing this, work on explanatory debugging shows that when systems expose their reasoning, people can understand and debug model behavior more effectively and provide more targeted corrective feedback~\cite{Stumpf2009InteractingMeaningfully,10.1145/2678025.2701399}.

Similarly, research on recommender systems treats feedback as the backbone of personalization. Recent work on interactive and conversational recommender create interfaces that encourages users to critique and refine suggestions over time rather than passive consumption~\cite{10.1145/1297231.1297263,He2016InteractiveRecommenders,10.1145/3453154}. For instance, ~\citet{xiao2021let} study how a music streaming service's voice assistant can elicit explicit feedback and show that interface-level choices strongly shape when and how users are willing to provide corrections. These lines of work demonstrate that thoughtfully designed feedback channels can substantially improve model alignment with user preferences.

LLM-powered CAs inherit this history as mixed-initiative interactive systems—but they differ in key respects. Unlike traditional IML or recommender systems, conversational agents support open-ended, multi-turn tasks in which goals, strategies, and evaluation criteria can evolve over time. Recent studies of human–LLM collaboration document challenges in maintaining shared goals, deciding when to trust or correct the model, and negotiating initiative during complex workflows~\cite{lin2023beyond,ju2025collaborating,zhang2024collaboration}. Our work builds on this tradition by focusing specifically on the \emph{feedback phase} of collaboration: we empirically characterize how and why feedback breaks down in CA interactions and derive interface-level scaffolds that make feedback easier to express, interpret, and incorporate.

\subsection{User interaction in LLM-based systems}
Users interact with LLM-powered systems by providing two types of inputs: user prompts and user feedback. Both play distinct roles in LLM-powered systems: prompts act as \emph{prospective} control signals that try to prevent errors before the model acts, while feedback functions as a \emph{retrospective} evaluation after a response is produced. In the ML and alignment literature, both are used as signals for steering and improving models. Prompt-engineering work focuses on crafting effective \emph{initial} instructions, conceptualizing prompts as ex-ante, task-specific templates that specify goals, constraints, and output formats without changing model parameters~\cite{sahoo2025systematicsurveypromptengineering,white2023promptpatterncatalogenhance}. Human-in-the-loop prompt optimization methods such as CoTAL treat prompt design itself as an iterative learning problem, combining chain-of-thought prompting with active learning so that users can refine grading prompts over time~\cite{cohn2025cotalhumaninthelooppromptengineering}. However, empirical work such as~\citet{10.1145/3544548.3581388} shows that even technically savvy non-experts struggle to translate their goals into effective first-turn prompts, making \emph{ex-post} information (feedback) essential to steer the model.

ML work typically treats feedback primarily as a training signal extracted from large-scale human--LLM chat logs. WildFeedback automatically identifies natural-language critiques and preferences in in-situ user--ChatGPT conversations and uses them to build preference datasets, but finds that most usable supervision comes from short retries and terse follow-ups rather than rich, structured critique~\cite{shi2025wildfeedbackaligningllmsinsitu}. Reinforcement Learning from Human Interaction (RLHI) similarly repurposes the WildChat-1M corpus as a source of interaction-level supervision, distinguishing follow-up messages that revise previous answers from those that start new tasks and emphasizing that only a minority of turns carry explicit corrective signals~\cite{jin2025erarealworldhumaninteraction}. Complementary work on Feedback shows that loosely structured comments (e.g., thumbs up/down, short replies) are noisy, sparse, and skewed toward coarse accept/reject judgments~\cite{DonYehiya2024NaturallyOF}.

Taken together, this ML-centric line of work demonstrates the promise of learning from real-world interactions but also reveals a core limitation: high-quality feedback data are scarce, fragmentary, and rarely capture the rich structure needed for nuanced alignment. Existing methods largely treat feedback as labels attached to turns, without asking \emph{why} users rarely provide detailed, well-formed feedback in the first place. Our work complements these efforts by approaching the problem from an interaction standpoint: we empirically investigate how feedback breaks down in everyday CA use and how interface scaffolds can elicit more structured, high-quality feedback that could eventually serve as better training signals.

\subsection{Human–AI Interaction with LLM-Powered Conversational Agents}
\label{sec:problems-in-feedback}
From an HCI perspective, LLM-powered conversational agents represent a shift from narrow, task-oriented, rule-based dialog systems to general-purpose, mixed-initiative interfaces that support increasingly complex goals. Users now turn to CAs for search, writing, coding, tutoring, emotional support, and everyday problem solving~\cite{Lee2022EvaluatingHM,yang2024human,Gao_2024,10.1145/3747588,kang-etal-2024-large,generative-echo-chamber,wang2024largelanguagemodelseducation,10.1145/3746058.3758376}. Unlike earlier systems, these agents are deployed as always-on collaborators that can flexibly shift roles (e.g., explainer, co-author, critic) within a single session. Recent alignment surveys argue that this context demands a \emph{bidirectional} view of alignment: models must adapt to human goals, but humans also cognitively and behaviorally adapt to powerful assistants over time~\cite{kirk2025socioaffective,jiang2025onewayinfluencebidirectionalopinion,10.1145/3706599.3716291,li2025we}. In such mixed-initiative collaborations, feedback is central to maintaining shared goals and coordinating complex tasks across multiple turns.

A growing body of HCI and NLP work examines how people actually interact with CAs in the wild. Frameworks like HALIE re-conceptualize evaluation around the full interaction trace, first-person experience, and criteria such as ownership and enjoyment, rather than only task accuracy on static prompts~\cite{Lee2022EvaluatingHM}. Large-scale analyses of WildChat and related corpora study information-access conversations “in the wild,” mapping what users ask for and which utterances contain check-worthy claims across over a million ChatGPT conversations~\cite{Joko2025WildClaimsIA}. Educational and sociocultural studies show that students and language learners appropriate CAs as tutors or cultural informants, often over-trusting fluent responses unless explicitly encouraged to adopt more critical interactional practices~\cite{zhao2024risk,dai2025critic}. Recent work on prototypical human--AI collaboration behaviors clusters multi-turn chat traces into patterns such as delegation, co-editing, and verification, and observes that users frequently treat CAs as collaborators while offering surprisingly little explicit evaluative feedback~\cite{mysore-etal-2025-prototypical}.

Multi-turn interaction further amplifies the importance of feedback. Human--LLM interaction work increasingly treats decomposition, planning, and refinement across turns as central design problems, arguing that “pure prompting” places unrealistic burdens on end users and makes it difficult to leverage ongoing feedback effectively~\cite{yang2024human}. Multi-turn evaluation studies such as ~\citet{laban2025llmslostmultiturnconversation} show that when instructions are underspecified and information is revealed gradually, models tend to over-commit to early guesses and struggle to course-correct as users add clarifications, leading to substantial performance degradation compared to single-turn baselines. Yet in these evaluations, the user side is typically simulated or abstracted away: multi-turn phenomena are analyzed to understand model robustness and task performance, not to characterize how real users decide when and how to provide feedback.

In summary, HCI work has begun to evaluate LLM-powered CAs as mixed-initiative, multi-turn collaborators and to analyze real-world usage traces, but has devoted relatively little attention to the micro-level interactional work of giving feedback: how people formulate critiques, when they abandon repair, and what barriers prevent them from offering more informative guidance. Existing studies typically focus on task outcomes, user satisfaction, or coarse preferences, treating feedback as an undifferentiated scalar or short text tag. Our work addresses this gap by shifting from a purely outcome- or data-centric view of feedback to an interaction-centric one: we systematically code turn-level feedback acts and conversational breakdowns, identify the barriers that prevent users from providing high-quality feedback, and design interface-level scaffolds to make such feedback easier to express, interpret, and incorporate, complementing recent calls for bidirectional, interaction-level alignment~\cite{DBLP:journals/corr/abs-2503-00858,10.1145/3706599.3716291,yang2024human}.
 
\subsection{The Science of Effective Feedback}
\label{sec:science-of-effective-feedback}
LLM-powered conversational agents are explicitly designed to emulate human conversational partners at the interaction layer. Prior work has shown that computers are social actors~\cite{nass1994computers} and characterizes CAs as natural-language user interfaces that “emulate human-to-human communication” and “mimic human conversation” through text, speech, and other modalities~\cite{Seeger2021TextingWH, Pinxteren2020HumanlikeCI}. Recent analyses of LLM-based conversational agents show that they convincingly reproduce human-like dialogue and prosocial behaviors, even without genuine social understanding~\cite{peter2025anthropomorphic,dai2025critic,Ashery2025Conventions,KirkebyHinrup_Stenseke_2025}. From a user’s perspective, interacting with a CA therefore resembles interacting with another person: they formulate requests, interpret responses, and decide whether and how to provide feedback through the same conversational resources they use with humans. While the underlying mechanisms differ radically, this emulation justifies importing insights from the science of human–human feedback and communication as a \emph{normative baseline} for evaluating interaction-level quality.

Decades of research in education and psychology establish feedback as a powerful driver of learning and performance. Meta-analyses by ~\citet{kluger1996effect} and by ~\citet{hattie2007power} show that well-designed feedback interventions can yield large positive effects on achievement across domains. Effective feedback is typically timely, specific, and goal-referenced: it helps learners understand the gap between their current performance and a desired standard and offers concrete information about how to close that gap~\cite{hattie2007power,shute2008formative,Wisniewski2020FeedbackMetaAnalysis}. Conversely, vague, overly complex, or poorly timed feedback can overload learners, be ignored, or even harm performance~\cite{kluger1996effect,shute2008formative}. These results highlight that feedback is not merely any response, but structured information that reduces uncertainty about how to improve.

More recent work emphasizes that feedback must also be \emph{usable}. ~\citet{carless2018developing} introduce the notion of ``feedback literacy''---the capacity of learners to interpret, evaluate, and act on feedback. ~\citet{Nicol01042006} similarly argue that students need opportunities to compare their work against explicit criteria, discuss feedback, and regulate their own revisions if feedback is to lead to improvement. These accounts treat feedback as a dialogic, iterative process rather than a one-shot transmission: givers and receivers jointly negotiate meaning, and both must invest effort for feedback to translate into better outcomes. For CAs, this suggests that simply providing a feedback channel (e.g., a thumbs-down button or a free-text box) is insufficient; the interaction must scaffold users in articulating, interpreting, and acting on feedback.

Beyond effectiveness and usability, human communication research explains \emph{how} such feedback is coordinated in conversation. Conversation analysis and pragmatics specify the cooperative norms that make feedback exchanges possible. Grice's maxims and subsequent work on common ground describe how interlocutors keep contributions truthful, relevant, sufficiently informative, and clear, while grounding mutual understanding through backchannels and repairs~\cite{grice1975logic,clark1991grounding}. In NLP and HCI, these maxims have been adopted as concrete design and evaluation heuristics for chatbots and analytical agents. For example, \citet{xiao2020tell} operationalize response quality in a chatbot-based open-ended survey by coding answers along Gricean dimensions such as informativeness, relevance, specificity, and clarity, and use these codes to compare chatbot-elicited responses with traditional web forms. Similarly, \citet{Panfili_2021} analyze human interactions with voice assistants like Alexa through a Gricean lens, showing that users experience off-topic or nonsensical replies as violations of relevance and clarity, and arguing that such violations are especially disruptive for human–AI collaboration. More broadly, recent work explicitly instantiates Gricean maxims as criteria for judging whether system responses are appropriately informative, honest, on-topic, and comprehensible, and for structuring evaluation tasks for conversational models~\cite{krause-vossen-2024-gricean-maxims,10.1145/3491102.3501972,10.5555/3709347.3743817,miehling2024languagemodelsdialogueconversational,park2024pragmatic,xiao2020if,ge2023should}.

In parallel, social and organizational psychology clarifies why people do not always give feedback, even when norms are understood. Feedback seeking and giving are modeled as cost–benefit trade-offs: people provide detailed feedback only when expected benefits outweigh cognitive, temporal, and social costs~\cite{Anseel2015FeedbackSeekingMetaAnalysis}. This perspective complements conversational norms by highlighting motivational barriers shaping cooperative feedback.

We adopt these human--human frameworks as a normative baseline for analyzing human--CA feedback. Given that conversational agents are built to participate in human-like dialogue~\cite{Seeger2021TextingWH,Pinxteren2020HumanlikeCI,peter2025anthropomorphic,Ashery2025Conventions,KirkebyHinrup_Stenseke_2025}, they will be judged against human conversational expectations, even if their internal processes are non-human. In our analysis, Gricean maxims and feedback-effectiveness criteria serve as a lens for characterizing where CA--user interactions fall short of cooperative, high-quality feedback exchanges. We use this lens to understand the feedback dynamics in human--CA communication and to motivate the Feedback Barriers and design implications developed in this paper.

%% file: sections/03-formative-study.tex
\section{Formative Study}
\label{sec: formative}
\citet{shi2025wildfeedbackaligningllmsinsitu} highlights divergence between in-the-wild human feedback for CAs and how humans naturally provide feedback to other humans. To understand why such divergences exist, we ran an in-person formative study with a total of 16 participants. Out of the 16, we had 11 experienced CA users of complex goal-oriented tasks and 5 inexperienced CA users for complex goal-oriented tasks.; Through analysis of the N=11 experienced user interviews we discover the Feedback Barriers (defined in the section below) that deter users from providing ``high-quality'' feedback during interactions with CAs \textbf{(RQ1)}. We also generalize our findings with N=5 inexperienced users of CA who don't have experience in task-oriented dialogues.

\subsection{Method and Participants}
We conducted 30-minute, one-on-one interviews with 16 participants(Experienced, N = 11; Inexperienced, N = 5). The 11 experienced participants were from large international tech companies and academic institutions, places actively exploring Conversational Agents for complex goal-oriented tasks. All experienced participants had current or recent experience with Conversational Agents across various tasks, including research, creative writing, coding, image editing, decision-making, and essay writing. The inexperienced participants were school teachers, retired, house-helpers and high school graduates. These participants had little experience using CAs for complex goal-oriented tasks. Instead, their primary use cases included medical help, recommendations for clothing, language learning, image touch-ups and general factual questions. All participant demographics and detailed usage scenarios are summarized in Table \ref{tab:participants}. 

The interviews followed a semi-structured format centered around the following themes:
\begin{itemize}
    \item \textbf{Conversational breakdowns}: What are the common challenges that participants experienced during conversations with CAs?
    \item \textbf{Breakdown Repair (Feedback)}: What strategies did the participants use resolve the conversational breakdowns? 
    \item \textbf{Feedback effectiveness}: What were the obstacles participants faced while providing feedback to CAs?
    \item \textbf{Feedback evaluations}: What strategies do the participants use to evaluate the success of their feedback?
\end{itemize}

All participants volunteered for the study. The study was approved by company's internal review committee. Interviews were transcribed and anonymized with consent from the participants.

\subsection{Analysis}
\label{sec:thematic-analysis}
We followed the reflexive thematic analysis approach by ~\citet{Braun25012023}. The semi-structured format allowed for flexible exploration of emerging themes and follow-up questions based on participants’ responses. The first author coded the interviews inductively. The first author then drafted a codebook, which was discussed and refined with the research team. The first author then coded the transcripts again after which further iterations of discussion and refinement took place, resulting in an updated codebook. The first author proceeded to derive thematic clusters from the coded transcripts to build up the wider findings and discussion. We report our findings below with the following format: \emph{Quote\_participant$\alpha$ (P$\alpha$; supported by P$i$,P$j$...)}.

\subsection{Positionality Statement}
The research team primarily consisted of authors with different positionalities. The authors had two distinct institutional backgrounds: academic and industrial institutions. All researchers except the third author have gone through postgraduate training at academic institutions in STEM backgrounds in the fields of Human-Computer Interaction(HCI) and Machine Learning (ML). The diverse institutional and professional backgrounds of the team members created a rich, interdisciplinary environment for this project, but also introduced distinct perspectives on research goals and practices. The scholars from academic institutions focused on delivering generalizable insights on understanding the feedback barriers users faced and the scholars from industry focused on actionable insights on overcoming feedback barriers. The collaboration required continuous, open discussion and reflexivity to bridge these different priorities. For instance, the early research questions were discussed to balance both the broader understanding of feedback barriers and actionable insights for the industry to overcome the feedback barriers. During data collection in user studies, the research team had regular discussions that made explicit how each researcher's individual positionality might impact the data collection and interpretation. By acknowledging these differences, we aimed to conduct robust studies leveraging the diverse backgrounds of the scholars. The final research was a synthesis of perspectives from both the academic and industrial institutions, which was strengthened by a deliberate process of reconciling diverse backgrounds and institutions.

\subsection{Findings - Descriptive}
\subsubsection{What are the common breakdowns while interacting with CAs?}
We first asked participants to recall instances when conversations with the CAs were unsatisfactory or when outputs diverged from what they had expected. Participants especially experienced participants most often pointed out \textit{context loss} (10/16 participants) as the dominant interaction failure, where the agent gradually drifted away from the original goal of the task. One participant explained that ``the biggest thing is \emph{context drift} … it really just forgets what the initial goal of this whole conversation was'' (P2;P4, P5, P7, P9). Others emphasized how long, interleaved sessions made it hard to keep the agent grounded: ``You may have to spend hours telling it the right context or start a new chat'' (P9; P2, P4, P8).

Beyond context drift, participants frequently reported quality issues. When using CAs for research and coding, they encountered fabricated or unreliable content: “It will make things up that I never mentioned” (P8; P1, P9, P10, P11). Many participants also described tensions around instruction-following and sycophancy~\cite{sharma2025understandingsycophancylanguagemodels, generative-echo-chamber}. While models attempted to comply, they often did so in ways that ignored nuance or failed to challenge users’ assumptions: ``Sometimes the AI agrees even when I’m wrong—whereas a person would say, ‘That doesn’t make sense’'' (P7; P2, P4, P10); ``Cursor just takes your command as law and doesn’t really ask any clarifications'' (P2; P4, P7, P9); ``[ChatGPT] just tells me what I want to hear, and that makes me uncomfortable'' (P12; P13, P14, P16). Together, these failures—loss of context, unreliable content, and over- or under-compliance—capture where current CAs fall short of maintaining shared understanding with users. 

\subsubsection{How do users give feedback?} When the conversation derailed, participants employed pragmatic repair tactics, ranging from low-effort repeated-inputs to abandoning the agent altogether. A common first move was to simply rephrase or repeat the request: ``Most of the time, out of sheer laziness, I just reprompt it'' (P2; P1, P3, P4, P7, P8, P10, P11). Others tried to incrementally refine inputs or decompose tasks: ``Other times I’ll break the task into smaller tasks and manually guide the AI step-by-step''(P4; P3, P5, P7, P9), or front-load more detail in order to steer the model: ``For coding tasks I front-load a very detailed prompt—the clearer my input, the better it performs''(P11; P2, P4, P6).

When the strategies highlighted above failed, participants often reset the interaction or switched tools: ``Sometimes it can be so bad that you just restart the whole conversation'' (P2; P3, P4, P6, P9) or ``If that fails, I try a different LLM—I hop around because one of them will probably solve it'' (P6; P1, P5, P7). Under time pressure, many resorted to accepting imperfect outputs and completing the task themselves: ``It’s faster to do it myself than to keep prompting, so I accept a sub-optimal reply and fix it manually'' (P5; P1, P7, P9, P10, P11). Abandonment was especially common with all inexperienced users because they were not aware of the affordances to provide feedback: ``I didn't know it [Feedback] was possible''(P12;P13,P14,P15) and ``It [ChatGPT] never asked me to''(P13;P12,P16).

Across the strategies, participants' feedback tended to be reactive, local to the most recent turn, and short-lived. They nudged outputs just enough to move the task forward but rarely invested in rich and structured feedback. This pattern raises a natural question: how do these everyday feedback practices compare to canonical notions of \emph{high-quality} feedback from education and psychology~\cite{hattie2007power,locke2013new,shute2008formative}, which emphasize goal-referenced, actionable, articulate, and progressive feedback? We examine this next.

\subsubsection{Does user feedback follow the properties of \textbf{high-quality} feedback?}
\label{sec: high-quality-feedback}
CAs require high-quality human feedback to complete complex tasks and bridge knowledge gaps. We define feedback quality along four dimensions derived from the literature in education and psychology--- goal-referenced, actionable, articulated, and progressive feedback~\cite{hattie2007power,locke2013new,shute2008formative}. Below, we examine how participants’ feedback practices towards CAs align and diverge from the properties of high-quality feedback.

\begin{itemize}
    \item \textbf{Goal-referenced}: High-quality feedback is coupled with the immediate goal and explicitly references the criteria for success. Educational literature emphasizes that feedback should be anchored in clear success criteria so that learners can see ``how [they] are doing in relation to the goal'' rather than receiving comments on tangential aspects~\cite{hattie2007power}. Goal-referenced feedback thus helps close the gap between current performance and the desired outcome. Although users actively strive to maintain context, for instance by re-emphasizing instructions: ``No, I want you to do what I said earlier'' (P1); participants highlight that when a conversation derails, users often fail to provide goal-referenced, corrective feedback to steer it back on track. Their most common strategy is to abandon relevance altogether by ending the interaction. A participant noted that when conversation goes poorly, ``it can be so bad that you just restart the whole conversation'' (P2; P4, P7, P9). This restart abandons formative feedback disrupting goal-referenced feedback loops central to improvement as highlighted by education literature.
    
    \item \textbf{Actionable}: Effective feedback is specific and actionable. In education and formative assessment, feedback is most powerful when it gives learners concrete information about what to change and how to change it, rather than global evaluations (e.g., ``good job'')~\cite{shute2008formative}. ~\citet{hattie2007power} highlight that feedback should not only describe current performance but also offer ``where to next'' guidance—strategies or revisions that help close the gap to the goal. In sync with the literature, participants who achieve better outcomes often do so by providing granular instructions, such as ``break[ing] the task into smaller tasks'' (P4) or front-loading a ``very detailed prompt'' for coding tasks (P11). However, user feedback frequently lacks the level of detail needed for strong actionability. Participants often default to high-level, vague instructions, such as asking the AI to ``Improve the flow'' or ``Fix the vocabulary'' (P1; P3, P4, P6, P9). This tendency towards low actionability is also evident in low-effort tactics. For example, few participants admitted: ``I just reprompt it, it's easier'' (P8; P2, P7, P10, P11) rather than providing granular corrections. From the perspective of feedback research, these behaviors resemble evaluative comments without process-level guidance: they signal dissatisfaction but offer little information about specific next steps or strategies, limiting the CAs ability to meaningfully adjust.

    \item \textbf{Articulated}: High-quality feedback is clearly articulated, with minimal ambiguity, to support common grounding. Formative feedback research stresses that comments should be easy to interpret; when feedback is vague or opaque, learners must expend extra effort just to decode what is being asked of them, which can make feedback feel ``useless'' or frustrating~\cite{shute2008formative}. In our study, users’ feedback was often under-articulated, with participants citing struggles to express the problem clearly: ``Sometimes I know what I want semantically, but I can’t phrase it so the model understands'' (P3; P7, P8, P11). This difficulty was a recurring theme, with one user confirming that ``explaining why [a response is wrong] is hard'' (P10; P3, P7, P8). This inability to verbalize the issue results in feedback that is inherently unclear and difficult for the CAs to act upon, mirroring classroom findings where underspecified comments leave learners uncertain about how to respond or improve~\cite{shute2008formative}.

    \item \textbf{Progressive}: High-quality feedback is delivered over time as part of a progressive formative cycle. Education research shows that the most impactful feedback is not a one-off judgment at task completion, but information that supports revision and reattempts~\cite{hattie2007power,shute2008formative}. However, feedback volume must be calibrated: too many comments at once can increase cognitive load and reduce effectiveness~\cite{shute2008formative}. One participant stated bluntly, ``I don't give AI feedback, not as frequently. And sometimes not at all'' (P3; P5, P6). Thresholds for abandonment were low; users will \emph{abstain} from giving more feedback if the AI ``gets basic things wrong more than twice'' (P8; P1, P5), or will ``abandon the task entirely if it isn't working within 15–20 minutes'' (P11; P6, P8). This pattern results in a sparse feedback landscape in stark contrast to the progressive, formative cycles envisioned in the literature.
\end{itemize}
Notably, prior work distinguishes between task-level feedback (about correctness), process-level feedback (about strategies), and self-regulation feedback (about monitoring one’s own learning)~\cite{hattie2007power}. In our interviews, participants’ comments to CAs were almost entirely task-level—nudging specific outputs—rather than process-oriented guidance that might help calibrate how they and the CA collaborate over time. Taken together, these accounts suggest that while users sometimes provide feedback that is goal-focused, actionable, clearly articulated, and progressive, their everyday practices more often fall short along one or more of these dimensions and rarely rise beyond local, task-level corrections. This education- and psychology-driven lens focuses on the \emph{content} and \emph{form} of feedback itself; next we complement these findings with a conversational lens based on Grice’s Maxims to understand the underlying barriers that keep users from reaching this ideal.

\subsection{Findings - What prevents users from giving ``high-quality'' feedback?} 
Our descriptive findings above, found that users rarely provide ``high-quality feedback'' in practice. To understand what prevents users from providing high-quality feedback, we distill participant accounts into recurring challenges users face when providing feedback. As described in related works (Sec~\ref{sec:science-of-effective-feedback}), CAs are explicitly designed for human communication and often perceived as social actors. Hence to uncover the barriers users faced while interactions with CAs, we use Grice's Maxims\cite{grice1975logic} which are cooperative principles for successful human communication ---\emph{Relation} (staying on topic), \emph{Quality} (being truthful and well-evidenced), \emph{Manner} (being clear and orderly), and \emph{Informativeness} (providing the right amount of information), as an analytical lens to unpack these challenges.

Critically, violations are \emph{reciprocal}: model failures (e.g., drift, hallucination, verbosity) trigger user behaviors (e.g., abandonment, under-specification, vague repair) that themselves breach cooperative principles. This reciprocal degradation explains why feedback cycles collapse and motivates \emph{scaffolds} that address both model limitations and human cognitive constraints that sustain cooperative feedback cycles. Using the four maxims, first author coded all interviews: each interview sentence was annotated once for user-side maxim violations and once for CA-side maxim violations. The detailed codebooks are provided in the supplementary material.

\subsubsection{Maxim of Relation: CAs shift the conversational goal and users abandon the interaction}
The maxim of Relation requires each conversational move to be relevant to the shared goal of the interaction. On the model side, participants described how even corrective feedback often failed to re-anchor the CA to the shared goal. One participant summarized this drift: ``It just goes even more down a hole you don’t want it to'' (P7; P2, P4, P8). Others noted that repeated attempts to restate context or constraints did not prevent the model from circling around an off-target interpretation: ``If I give it feedback more than two or three times, it starts iterating its own previous answers instead of finding newer directions'' (P8; P5, P6, P9). In longer interactions, users felt they had to keep re-establishing common ground: ``You may have to spend hours telling it the right context or start a new chat'' (P9; P2, P4). These patterns show the CA failing to use feedback to restore relevance to the user’s goal, violating the maxim of Relation.

On the user side, participants responded to this misalignment by withdrawing from the interaction rather than offering more targeted, goal-focused feedback. As one participant remarked, ``Sometimes it’s honestly easier to restart … or just code it myself.'' (P4; P2, P5, P6, P7, P8). Another was more blunt: ``It’s faster to do it myself than to keep prompting, so I accept a sub-optimal reply and fix it manually.''(P5; P3, P9, P11). Instead of feedback pulling the conversation back toward a shared goal, both sides drift: the CA continues off-topic, and the user abandons the attempt to repair, breaking the relational thread that effective feedback depends on~\cite{clark1991grounding}.

\subsubsection{Maxim of Quality: CAs hallucinate and users propagate unverified outputs}
The maxim of Quality requires speakers to provide contributions that are truthful and well-evidenced.
On the model side, participants consistently flagged fabricated or hallucinated outputs. One participant noted, ``Out of five papers, it makes up one or two''(P8; P1, P10, P11), while another explained, ``If it's research-related or fact-heavy, I need to check whether links or cited papers are real. So that takes more effort'' (P5; P6). Such failures to ground information in evidence erode trust. These hallucinations and loosely directly increase the cost of evaluating whether feedback has been correctly incorporated. In human conversation, such lapses would undermine trust; in CAs, they impose additional verification burdens on users discouraging them from offering more nuanced, quality-oriented feedback. 

From the user perspective, although participants didn't state false claims themselves, they described relaxing their evidential standards under time and effort constraints. As P6 put it ``I don’t check every line. I look at the high-level structure … it's more of a gut check'' (P6; P5, P7, P9). Another participant admitted, ``When I’m short on time … I just copy and paste it without thinking about the feedback'' (P9; P7, P10). Under deadline pressure or low stakes, users propagated model outputs with minimal verification rather than offering corrective, evidence-based feedback. Thus, model-side Quality violations (hallucinations) and user-side shortcuts (gut checks, skipped verification) reinforce each other, undermining the overall reliability of the feedback loop.

\subsubsection{Maxim of Manner: CAs rarely follow up and users struggle to articulate}
\label{sec:formative:maxim of manner}
The maxim of Manner requires contributions to be clear, concise, and easy to interpret.
On the model side, participants, described a pattern of surface politeness without genuine clarification or adjustment. As P2 noted, “Cursor … just takes your command as law and doesn’t really ask any clarifications,” (P2; P3, P4, P6, P7). Even repeated inputs often failed to change the underlying behavior: “Even if I re-prompt it, it apologized but repeats the same answer” (P6; P2, P4, P7). Others experienced this as stylistic clutter: ``It feels overly polite—almost fake. A person would just say 'Okay, let me try this,' not wrap it in unnecessary fluff.''(P1; P3, P6). Because CAs lack paralinguistic cues such as tone, facial expressions, and body language, users also struggled to infer whether the system had truly understood their feedback: ``human-to-human conversations, you give someone feedback, you can sometimes see it in their eyes. If someone’s clueless, you can see they’re confused.''(P2; P1, P7, P13, P12, P14, P16). Together, these issues violate Manner by making model responses harder to interpret and diagnose.

On the user side, participants admitted difficulty articulating their needs. One participant explained, ``Sometimes I know what I want semantically, but I can’t phrase it so the model understands''(P3; P7, P8, P11). Another reflected, ``In the beginning it’s hard—I know what I want, but I don’t have the right adjectives''(P7; P3, P10). P10 similarly noted that ``Explaining why [the model was wrong] is hard''(P10; P3, P8). In more visual tasks, articulation barriers were even sharper: ``It doesn’t get visualization positions right and I find it difficult to verbalize them.'' (P8). As a result, even when users recognize problems, the feedback they provide is often vague or underspecified, limiting its usefulness for the CA and constituting a user-side violation of Manner.

\subsubsection{Maxim of Quantity: CAs are either over or under-informative and users consistently are under-informative}
\label{sec:formative:quantity}
The maxim of Quantity calls for providing neither too much nor too little information relative to the task at hand.
On the model side, participants experienced both over- and under-informativeness from the AI. On one side, the CA omitted important elements: ``It picks and chooses two or three things … I have to keep adding the rest of my metrics,''(P7; P4, P5, P6, P9). On the other, it sometimes did more than asked for: ``It’s usually trying to do more things than I asked it to do, messing up other parts of my code'' (P4; P7, P9). In both cases, users needed to compensate—either by repeatedly adding missing details or undoing unwanted edits—before they could even begin to provide focused feedback on essential parts. Front-loading detailed context was seen as burdensome, particularly for novices or those learning from the AI’s responses: ``Sometimes I want to give feedback but I hold back because I’m also learning from its response'' (P9).

On the user side, Quantity violations showed up primarily as the tendency to provide too little or insufficient feedback. Many participants defaulted to minimal tweaks: ``Most of the time, out of sheer laziness, I just reprompt it''(P2; P1, P3, P6, P7, P8, P9, P10, P11). Others described escalating frustration rather than providing richer context: ``I usually yell at it three or four times first … if that doesn’t work, I start a new chat'' (P6; P2, P3, P7, P9). A few participants did add more detail when repeated retries failed---for example, ``If retry fails, I add more context or ask it to review the entire code-base before trying again'' (P4, P3, P5, P9)---but this was the exception rather than the norm. Overall, users rarely calibrated the amount of feedback they provided to what the CA would need to recover, instead oscillating between very short inputs and complete abandonment. This combination of model-side over/under-informativeness and user-side under-informativeness makes it difficult to sustain the progressive, appropriately scoped feedback cycles as articulated in formative assessment research~\cite{hattie2007power,shute2008formative,sadler1989formative,Black01031998}. 

\subsubsection{The Feedback Barriers}
\label{sec:barriers}
From the observed reciprocal violations of Grice's maxims, we derive four key \textbf{Feedback Barriers} that prevent users from giving high-quality feedback to CAs today.

\paragraph{\textbf{Barrier of Common Ground}} 
\begin{itemize}
    \item Model: The model fails to maintain the stated task goal. This manifests as \textit{context drift}, ignoring explicit constraints, or silently substituting user's objective for hallucinated ones.
    \item User: The user disengages from the conversational goal, thereby severing the collaborative relationship. Visible as \textbf{explicit abandonment} (``I'll just do it myself'') or \textbf{restarting the chat}, which voids all shared context.
    \item Impact: This barrier forces users to constantly re-align CAs, increasing the cognitive cost of interaction. It places the entire burden of \emph{disseminating} and maintaining shared goal on the user, making interaction feel unproductive, making users: ``spend hours telling it the right context or start a new chat'' (P9; P2, P5, P7, P8).
\end{itemize}

\paragraph{\textbf{Barrier of Verifiability}}
\begin{itemize}
    \item Model: The model provides information that is not grounded in verifiable evidence. This includes generating \textit{hallucinations}, citing non-existent sources, or presenting outdated information as current fact.
    \item User: The user fails to uphold evidential standards in the workflow. This is not about lying, but rather \textbf{skipping verification} i.e. accepting potentially flawed outputs, or propagating unverified information.
    \item Impact: This barrier forces the user into the role of a fact-checker, undermining trust. Without affordances to \emph{evaluate} whether feedback was incorporated faithfully, the verification effort becomes prohibitively costly, forcing users to adopt shortcuts: ``I don't check every line. I look at the high-level structure … it's more of a gut check'' (P6; P5, P7, P9). 
\end{itemize}

\paragraph{\textbf{Barrier of Communication}}
\begin{itemize}
    \item Model: The model communicates in an unclear or unhelpful way. This includes evasive politeness (``fluff''), apology loops, and critically, \textbf{failing to ask clarifying questions} when the user's request is ambiguous.
    \item User: The user fails to provide clear and unambiguous feedback. This stems from the inherent \textbf{difficulty of articulating} what is wrong, providing vague instructions (e.g., ``make it better''), or being unable to verbalize the desired correction.
    \item Impact: This barrier leads to frustrating cycles of guesswork. The high cognitive effort required to continue the conversation without useful probes from the AI makes \emph{sustained engagement} feel irrational, leading users to abandon the task when perceived costs outweigh benefits. As one user explained, ``If I’ve learned that it takes seven prompts to get a good result for a certain task, I probably won't even try anymore'' (P3; P1, P5, P8, P11).
\end{itemize}

\paragraph{\textbf{Barrier of Informativeness}}
\begin{itemize}
    \item Model: The model provides an inappropriate amount of information. This can be under-informativeness (omitting key steps; addressing only a subset of a input) or information overload (unrequested changes; providing tangential details).
    \item User: The user provides insufficient information for the model to make a meaningful correction. This includes low-effort responses within the conversation, such as \textbf{simple re-prompts} (``try again''), or feedback that lacks necessary context (``that's not right'').
    \item Impact: This barrier traps the user in an inefficient guessing game. Lacking low-friction ways to \emph{disseminate} structured feedback about what to change, where, and at the right granularity, users default to providing minimal information; this perpetuates a cycle of low-quality outputs: ``Unless I give very specific, file-level instructions I hesitate, because otherwise it messes everything up'' (P11; P2, P7, P8, P10).
\end{itemize}

\subsubsection{Synthesis and Implications.}
Our formative study reveals that the feedback dynamics in human-AI interactions are systematically fragile. These failures are not one-sided but reflect a reciprocal collapse of the cooperative principles from the model and the user. Model failures (drift, fabrication, ambiguity, over- and under-informativeness) elicit low-quality user feedback such as abandonment and under-informativeness. In turn, these behaviors starve the model of the goal-referenced, actionable, articulated, and ongoing feedback necessary for alignment and task success. 

The four \textbf{Feedback Barriers} we identified, the \emph{\textbf{Barrier of Common Ground, Barrier of Verifiability, Barrier of Communication and Barrier of Informativeness}}, provide a compact framework for diagnosing these dysfunctional cycles across core phases of CA interaction, including how users disseminate feedback, how models and users engage with it, and how both parties evaluate outputs. Together, they demonstrate that without proper support, the cognitive and interaction costs\cite{Kurniawan2022PathOfLeastResistance,Anderson2003PsychologyOfDoingNothing} of providing good feedback quickly outweigh the perceived benefits for the user.

To make human-AI collaboration truly effective, we must design systems that explicitly break down these Feedback Barriers. In the next section, we translate this analysis into design desiderata aimed at mitigating the feedback barriers and building scaffolds aimed at restoring the cooperative communication principles with CA.

%% file: sections/04-design-desidarata.tex
\section{Designing systems to resolve the Feedback barriers}
The preceding analysis positions the four \emph{Feedback Barriers} as concrete obstacles that prevent users from providing high-quality feedback for effective human-AI collaboration.

Overcoming the feedback barriers requires addressing disruptions in the three fundamental areas of human-AI interaction impacted by the barriers: First, they hinder \emph{how users disseminate feedback}—that is, how easily they can express goals, corrections, and constraints. Second, they undermine \emph{how users and models engage with feedback}, including how well the system interprets, clarifies, and responds to user-provided input. Third, they compromise the user's ability to \emph{evaluate feedback outcomes}, making it difficult to verify whether the system correctly incorporated their corrections. Together, the Feedback Barriers impede Dissemination, Engagement, and Evaluation, creating a cycle in which users cannot contribute the very feedback necessary to improve the interaction.

Building on these three interaction areas and the ways the Feedback Barriers distort them, we now translate our findings into three design desiderata that any conversational agent must satisfy to break down these barriers and support high-quality feedback.

\subsection{The Design Desiderata}
Each desideratum corresponds to one or more Feedback Barriers and outlines how systems can reduce the interaction costs that currently prevent users from giving high-quality feedback.

\subsubsection{Foster a Persistent and Legible Shared Frame of Reference.}
The Barrier of Common Ground arises when users and CAs fail to maintain a stable shared understanding of the task\cite{clark1991grounding}. Our study shows that the Barrier of Common Ground arises from the ephemeral, turn-by-turn nature of current Conversational Agents, which frequently leads to context drift. This failure, forces user into a constant high-effort loop of re-orienting the agent, a significant source of interaction cost\cite{Anderson2003PsychologyOfDoingNothing, Kurniawan2022PathOfLeastResistance}. The first desideratum, therefore, is a system that must \textbf{maintain and expose a persistent, shared understanding of the task}. This aligns with foundational principles of external cognition, where visible, persistent artifacts offload user's cognitive burden of remembering and tracking task state\cite{scaife1996external, kirsh1994distinguishing, zhang1994representations, norman1993things}. Instead of relying solely on the transient chat log, systems should externalize goals and constraints into a shared, editable artifact. When feedback can be anchored to this stable frame of reference, it supports the pillar of Dissemination by allowing users to articulate their intent with precision and confidence.

\subsubsection{Design for Proactive and Low-Friction Interaction.} 
The Barriers of Communication and Informativeness emerge when users must shoulder the full burden of expressing corrections, determining granularity, and repairing misunderstandings—leading to premature abandonment. The second desideratum aims to \textbf{design for reciprocal, mixed-initiative interaction that lower the cost of articulation and correction.} Instead of requiring users to craft increasingly specific inputs or diagnose model failures, systems should actively participate in maintaining communicative clarity. This includes prompting users for missing constraints, surfacing ambiguities, proposing candidate interpretations, and offering structured actions (e.g., “Apply change only to Section 2”). This aligns with interactive machine learning, which frames human–AI collaboration as a reciprocal, mixed-initiative process\cite{amershi2014power,allen1999mixedinitiative,horvitz1999principles}. By proactively seeking clarification when user input is ambiguous and offering structured, contextual actions, the system can bridge the ``gulf of execution'', reducing the cognitive load required for users to express their intent\cite{Westbrook2015Cognitive}.

\subsubsection{Provide Transparent and Verifiable Reasoning.}
The Barrier of Verifiability emerges when users cannot inspect how the CA interpreted their feedback or verify that it was faithfully incorporated. This opacity pushes users toward either blind acceptance or exhaustive manual checking—both of which hinder sustained feedback. To counter this, the third desideratum requires systems to \textbf{offer transparent and verifiable reasoning, making the feedback loop visible and auditable.} This is a core tenet of Explainable AI (XAI), where making a system's internal logic understandable helps users build accurate mental models and calibrate their trust\cite{gunning2019darpa,miller2019explanation, turner2022calibrating,lee2004trust}. Instrumenting the interaction to make outcomes legible---for example, by presenting changes with a visual diff---bridges the ``gulf of evaluation'' by making the effects of a user's feedback immediately apparent. By making the connection between feedback and outcome direct and ``scannable'', the system enables low-cost evaluation, enabling users to assess both the models output and the efficacy of their feedback.

\subsection{Operationalizing Design Desiderata to Breakdown the Feedback Barriers}
\label{sec:operationalizing}
\begin{figure*}[t]
    \centering
    \includegraphics[width=0.95\textwidth]{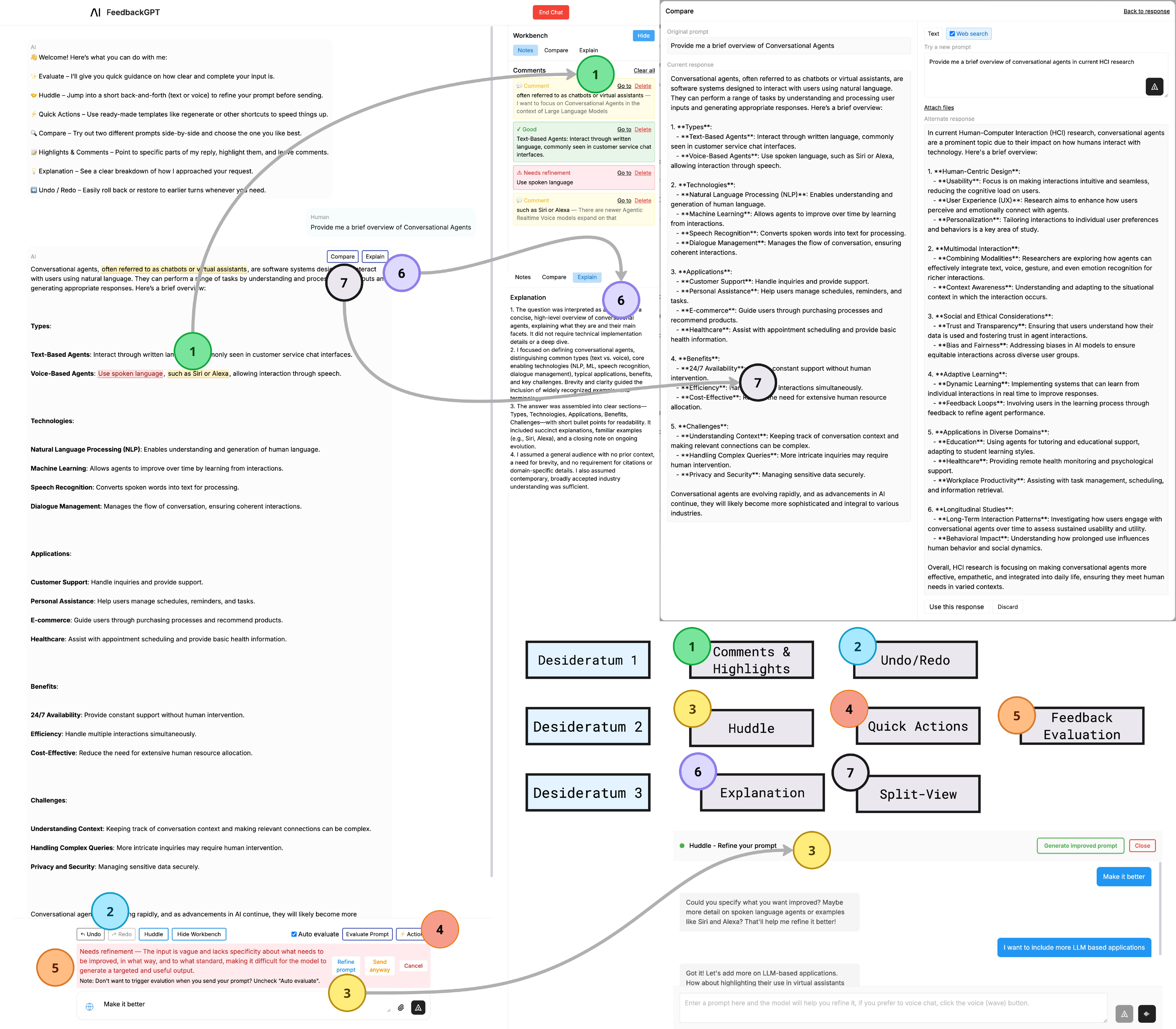}
    \caption{Overview of FeedbackGPT. The system consists of three new interface components, a sidebar, a panel and a popup. The interface give users seven different scaffolds for the three desideratum as described in Sec \ref{sec:operationalizing}}
    \Description[Overview of the scaffolded system]{The system consists of three new interface components, a sidebar, a panel and a popup. The interface give users seven different scaffolds for the three desideratum as described in Sec \ref{sec:operationalizing}}
    \label{fig:feedbackgpt}
\end{figure*}

To operationalize the design desiderata, we built a lightweight, web-based conversational agent system (FeedbackGPT) that uses a familiar ChatGPT-like interface (Figure \ref{fig:feedbackgpt}). We introduced these scaffolds without modifying model weights or core model behavior, ensuring that improvements arise from interaction design rather than model tuning. In what follows, we describe and motivate the scaffolds and the system, and detail implementation choices so the system is reproducible and interpretable. Because the formative study surfaced many potential scaffolds, we prioritized those that were (i) complementary to existing chat-based workflows, (ii) feasible within current CA affordances, (iii) model- and domain-agnostic, and (iv) directly addressed the identified Feedback Barriers. These scaffolds represent one possible instantiation of many; our goal is to demonstrate how scaffolds that operationalize the design desiderata bridge the feedback barriers leading to higher-quality feedback.

Below, we describe each scaffold by (1) identifying the design desideratum it satisfies, (2) explaining which Feedback Barrier it mitigates, and (3) detailing how the feature achieves this in practice.

\paragraph{Interface components}
FeedbackGPT preserves the familiar chat stream and composer while introducing inline controls, a persistent right-hand sidebar, and an auxiliary bottom panel.

\subsubsection{\textbf{Scaffolds for Desideratum 1: Foster a Persistent and Legible Shared Frame of Reference}}
These features are designed to overcome the \textbf{Barrier of Common Ground} by making task state visible and anchoring feedback to specific contexts, thereby stabilizing the shared frame of reference between user and CA.

\paragraph{\textbf{Inline Comments \& Highlights:}} Inspired by collaborative document tools, Inline Comments \& Highlights allow users to anchor feedback to specific spans in the models output via highlights and in-place comments\cite{bacchelli2013expectations}. By tying feedback to concrete text rather than to vague references (e.g., “the second paragraph”), this scaffold operationalizes the desideratum of a persistent shared frame: feedback lives on as an explicit, inspectable part of the conversation state. These anchors are collected in a sidebar and passed back as structured constraints in the next input, forcing the model to acknowledge and address each locus. In doing so, Inline Comments \& Highlights directly mitigate the Barrier of Common Ground by reducing context drift and ensuring that feedback remains grounded in a shared, stable representation of the task.

\paragraph{\textbf{Undo and Redo:}} Absence of an undo feature in today's CAs deters experimentation and restricts user control\cite{bacchelli2013expectations}. Undo and Redo extend the persistent frame of reference by letting users step back through conversation snapshots without discarding the entire interaction. By capturing conversation states and exposing them through simple buttons, the system lets users explore alternative feedback paths---for instance, trying different critiques or different constraints---without the risk of losing hard-won context. This reduces the need to abandon and restart chats when feedback goes wrong, thereby weakening the Barrier of Common Ground and making it safer to iteratively refine feedback.

\subsubsection{\textbf{Scaffolds for Desideratum 2: Design for Proactive and Low-Friction Interaction}}
These features are designed to overcome the \textbf{Barriers of Communication and Informativeness} by lowering the effort required to provide high-quality feedback and by turning feedback refinement into a collaborative, back-and-forth process rather than a one-sided monologue.

\paragraph{\textbf{Feedback Huddle:}} When a user's feedback is vague or hard to articulate (a key aspect of the Barrier of Communication), the Feedback Huddle opens a panel where the same model has new instructions to proactively ask targeted, clarifying questions; In this space, users can iteratively refine broad reactions (e.g., “this doesn’t sound right”) into concrete, actionable instructions following properties of high-quality feedback (\ref{sec: high-quality-feedback}). By sharing responsibility for clarification and prompting the user with specific follow-up questions, Feedback Huddle operationalizes the desideratum of reciprocal, low-friction interaction and helps users overcome articulation difficulties, thereby addressing both Communication and Informativeness barriers.

\paragraph{\textbf{Quick Actions:}} Quick Actions address the Barrier of Informativeness by turning common, high-value feedback patterns into one-tap operations. Users can invoke reusable and editable templates---for example, “regenerate using only the highlighted changes”---without repeatedly retyping long instructions. Because even small repetition costs deter users from providing detailed feedback~\cite{Westbrook2015Cognitive}, these shortcuts lower the marginal effort of adding rich constraints, making it more likely that users will provide sufficiently informative feedback to guide the CA.

\paragraph{\textbf{Feedback Evaluation:}} Feedback Evaluation helps users overcome ambiguity in their own feedback by providing real-time guidance before a message is sent. The system evaluates the user’s input against prompting heuristics and communicative principles, then offers brief suggestions to improve specificity, structure, and clarity. This directly targets user-side violations of the maxims of Manner and Quantity: instead of expecting users to instinctively produce high-quality feedback, the interface teaches them how to refine their messages with minimal extra effort, advancing the desideratum of low-friction, reciprocal feedback exchange.

\subsubsection{\textbf{Scaffolds for Desideratum 3: Provide Transparent and Verifiable Reasoning}}
These features are designed to overcome the \textbf{Barrier of Verifiability} by making the model's reasoning processes and the impact of user feedback transparent and auditable.

\paragraph{\textbf{Explanation:}} Opaque outputs force users into the role of a fact-checker. The Explanation scaffold offers a one-click control that prompts the model to detail the context it used, which instructions it prioritized, and how it interpreted the user’s feedback to generate a response. Inspired by explanatory debugging\cite{10.1145/2678025.2701399,6645235}, this feature operationalizes the desideratum of transparent, verifiable reasoning: it helps users build a more accurate mental model of the system, diagnose the root cause of errors, and determine whether poor outcomes stem from their feedback, the models reasoning, or both. This approach differs from the ``thinking'' or intermediate reasoning traces produced by some models, which are primarily internal-facing and often hidden from users. Explanations are user-directed and generated after the response, allowing them to remain grounded in the final output and, consequently, more faithful to what the model actually produced.

\paragraph{\textbf{Split-View Comparison:}} To prevent users from blindly accepting the first plausible output, the Split-View Comparison feature allows side-by-side evaluation of different response versions generated from different feedback\cite{tohidi2006getting}. By making comparative evaluation cheap and visuually salient, this scaffold supports the desideratum of verifiable reasoning: users can directly see how alternative feedback strategies change the models behavior, and are encouraged to critically assess the outcome of their feedback rather than propagating a potentially flawed response.

\subsubsection{\textbf{Implementation}}
To develop our system we used \textit{Next.js} framework for the frontend, \textit{Node.js with express} for the backend and finally \textit{MongoDB} for the database. We used typescript as our language for developing the system. 

We leveraged the responses API from OpenAI to interface with the models allowing tool use, file upload and web search. Our primary model was \textit{GPT-4o} for text based tasks and \textit{gpt-4o-realtime-preview-2024-12-17} for voice mode.

For the conversation in the main single-stream chat window, we did not modify the default system prompt, to ensure we did not change the underlying model behavior. When a user provided highlights and comments, we added them as (context, comment) pairs to the user input. Green highlights were translated as “good” comments and red highlights as “needs improvement” comments.

During huddle mode, we change the system prompt to enable the model to provide feedback based on the OpenAI prompting guidelines. We also instructed the model to reply in short 1-2 line responses to make the interaction more conversational. Similarly, for evaluation of a user input we used a custom system prompt where the model evaluated the input based on the OpenAI prompting guidelines (See Supplementary Material for prompt details).

%% file: sections/06-study-2.tex
\section{Study 2: How do affordances that resolve Feedback Barriers change user feedback?}

To evaluate how our scaffolds that provide affordances to minimize feedback barriers affect the user feedback-quality, we ran a one-hour within-subject in-person study (virtual for participants not available in-person) with 20 participants where participants interacted with ChatGPT and FeedbackGPT to complete two writing tasks. Below we describe our study design and measures.

\begin{figure*}[t]
    \centering
    \includegraphics[width=0.95\textwidth]{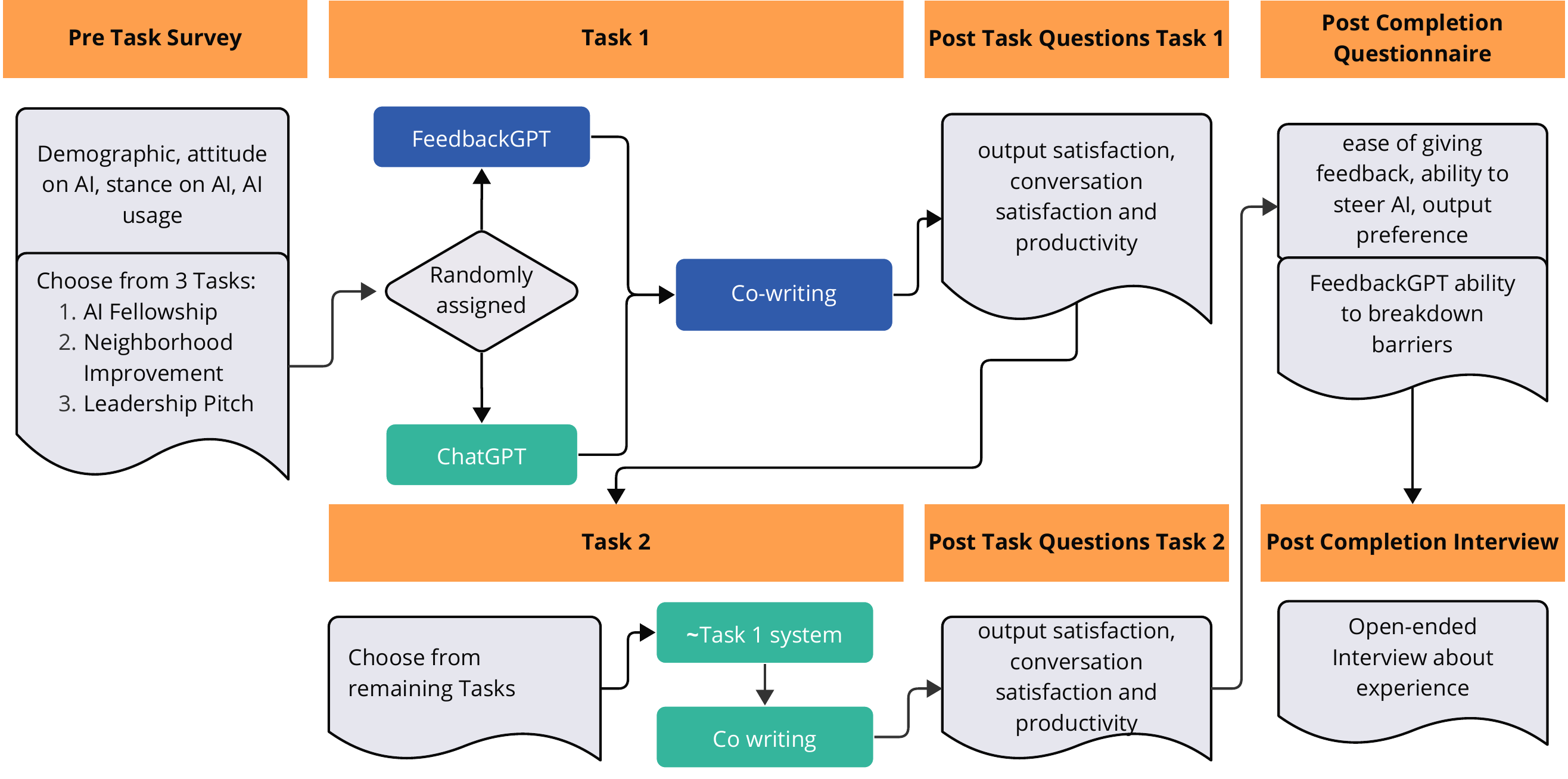}
    \caption{Overall study procedure for Study 2. In the pre-task survey, participants answered questions regarding their prior experience with conversational AI and their prior attitude and demographic questions. Participants then chose a task from three available topics. They were then randomly assigned a system with which they complete the task. Following which, they were asked three post-task questions about their results, conversation and productivity. They then choose a second task from the remaining two options and were assigned the system \textbf{not} assigned in task 1 and completed the task with the post task questions. After which, participants were asked three general questions about their experience with FeedbackGPT and then were asked eight questions about their experience with FeedbackGPT aimed to capture the effects of feedback barriers in FeedbackGPT.}
    \Description[Overall study procedure for second study]{In the pre-task survey, participants answered questions regarding their prior experience with conversational AI and their prior attitude and demographic questions. Participants then chose a task from three available topics. They were then randomly assigned a system with which they complete the task. Following which, they were asked three post-task questions about their results, conversation and productivity. They then choose a second task from the remaining two options and were assigned the system \textbf{not} assigned in task 1 and completed the task with the post task questions. After which, participants were asked three general questions about their experience with FeedbackGPT and then were asked eight questions about their experience with FeedbackGPT aimed to capture the effects of feedback barriers in FeedbackGPT.}
    \label{fig:study2 procedure}
\end{figure*}

\subsection{Study Procedure}
The study procedure as shown in Figure \ref{fig:study2 procedure}, includes five parts: a pre-task survey, first co-writing task,survey for first task, second co-writing task, survey for second task, post-task survey and finally an open-ended interview about their experience with the two systems they interacted with.

In the pre-task survey, participants were asked to rate their prior experiences with AI and their attitudes toward Conversational Agents. Following which participants completed two tasks each with a different topic and system. In each cycle, participants were randomly assigned a system which was either the baseline (ChatGPT) or FeedbackGPT. They had to spend a minimum of 20 mins interacting with the system and complete the task they chose. After completion of the task, they were asked to rate the output quality, conversation quality, and productivity on a 5-point Likert scale.

After completion of both the tasks, participants completed the post-task survey. In which they were asked to rate the ease of providing feedback, perceived conversation quality and final output quality of FeedbackGPT compared to ChatGPT. Then they were asked two questions for each feedback barrier where they rated their experience with FeedbackGPT. All of the post-task surveys were measured by a set of five-point Likert scales. The detailed questionnaire is available in the supplementary materials.

\subsection{Co-writing task}
We wanted participants to interact with the system to complete a task that would involve multiple turns, and where participants would have to provide feedback to complete the task successfully. We found co-writing task with the topic focused on the users personal experiences to be the best fit for our purpose. In our setup, there was an inherent knowledge gap between the user and the model which could only be bridged by user providing feedback to the agent. Furthermore, unlike coding and image editing which require special interfaces, co-writing task does not require special interface elements and is standard across different platforms such as Gemini and ChatGPT. The three topics we chose we as follows (detailed task descriptions are in Appendix \ref{appendix: task description}:
\begin{itemize}
    \item Fellowship application for transformative AI usecases.
    \item Proposal for neighborhood improvement plan
    \item Pitch for STEM leadership development recruitment
\end{itemize}

\subsection{Measures}
\label{sec:measures}
We used three families of measures capturing (1) the objective quality of user feedback, (2) participants' subjective experience of each system, and (3) their perceived ability of FeedbackGPT's affordances to overcome the four feedback barriers. We additionally collected pre-task attitudes and demographics for contextualization.

\subsubsection{Feedback Quality and Barriers(Log-based)}
\paragraph{Unit of analysis.}
We define a \emph{feedback turn} as any user message after the first system response that contains evaluative, corrective, or guiding content intended to shape the system's subsequent output (e.g., ``this section is too vague, please add concrete examples'' or ``keep the introduction but shorten the methods''). For each participant, we extracted all feedback turns and the preceding model response from both the ChatGPT and FeedbackGPT episodes.

\paragraph{Coding scheme - Feedback Quality.}
Using the definition of high-quality feedback (Section~\ref{sec: high-quality-feedback}), we manually annotated each feedback turn along three binary dimensions for goal-referenced, actionable and articulated, and the total number of feedback turns for progressive. Each dimension was coded as present (1) or absent (0) for a given feedback turn.\footnote{Turns could simultaneously express multiple dimensions, e.g., a message can be both highly relevant and highly specific.}. The detailed codebook is in supplementary materials. First author and a second coder independently annotated a randomly sampled subset of 20\% of the feedback turns to assess reliability, yielding a Cohen's $\kappa$ of 0.895 for goal-referenced, 0.878 for actionable and 1 for articulated. Disagreements were discussed and resolved. Following which the remaining data was coded by first author.

\paragraph{Coding scheme - Feedback Barriers.}
In parallel, the first author coded feedback barrier present in each model response using definitions of feedback barriers(\ref{sec:barriers}). Following the same process as above, a second coder coded a random 20\% sample independently. Cohen's $\kappa$: Common Ground=1, Verifiability=0.95, Communication=1, Informativeness=1.

\paragraph{Derived metrics.}
For each participant $p$ and system condition $s \in \{\text{ChatGPT}, \text{FeedbackGPT}\}$, we computed the proportions for each dimension $d$ (goal-referenced, actionable and articulated):
\begin{itemize}
    \item \textbf{dimension proportion:}
    \[
        \text{$d$}_{p,s} = \frac{\#\{\text{feedback turns coded as $d$ for }p,s\}}{\#\{\text{feedback turns for }p,s\}} \times 100.
    \]
\end{itemize}
These normalized rates allow us to compare feedback quality across systems while controlling for differences in the number of turns.

\textbf{Progressive feedback} was computed as:
\[
    \text{Vol}_{p,s} = \frac{\sum_{\text{feedback turns } t \text{ for } p,s} \text{characters}(t)}{\#\{\text{feedback turns for }p,s\}},
\]
i.e., the mean number of characters per feedback turn in that episode.

\subsubsection{Effect of Feedback Barriers}
To assess whether FeedbackGPT's scaffolds reduced the four feedback barriers(\ref{sec:barriers}), we combined post-task survey scales with qualitative analysis of open-ended responses and interviews. In post-task survey, participants answered barrier-focused items (5-point Likert scale) about their experience with \emph{FeedbackGPT}).

For each barrier, we constructed a two-item subscale. For each subscale we computed the participant-level mean of the two items. Internal consistency for each two-item subscale was acceptable ($0.701\leq\alpha\geq0.852$). In addition, we treated three global items as standalone indicators which capture overall ease of feedback and output preference between systems. The details are in the supplementary materials.

\paragraph{Open-ended interviews.}
To contextualize the survey ratings, participants participated in a brief post-study interview about their experience with both systems and the scaffolds.

\paragraph{Per-episode satisfaction and productivity.}
After each task, participants also completed three items (Output satisfaction, Conversation satisfaction and Perceived productivity) for the system they had just used (FeedbackGPT or ChatGPT), using a 5-point Likert scale from \emph{Strongly disagree} (1) to \emph{Strongly agree} (5).

\subsection{Participants}
We recruited participants via mailing lists and internal channels at several large international technology companies and universities. Inclusion criteria were: (1) fluent English speakers and (2) prior familiarity with conversational agents (e.g., ChatGPT).

In total, 22 participants started the study and 2 were excluded due to incomplete tasks and failed attention checks. Our analyses are based on the remaining 20 participants who completed the full protocol (two tasks, both systems, post-task survey, and interview).

The study was conducted either in person or virtually (video call with screen sharing), with each session lasting approximately one hour. All participants volunteered for the study. All participants provided informed consent, and the study was approved by the company's internal review committee.

Of the 20 participants, 6 identified as women, 13 as men, and 1 as non-binary or third gender. The median education level was a Bachelor's degree. The median annual household income fell between \$50{,}000 and \$100{,}000, and the median age was 25--34 years.

\subsection{Analysis}
Our primary goal in Study~2 is to compare how the baseline ChatGPT interface and the scaffolded FeedbackGPT interface affect (1) the objective quality of user feedback and (2) participants' subjective experience of giving feedback and steering the AI. Each participant used both systems in a randomized order; All inferential analyses treat \emph{system} (ChatGPT vs. FeedbackGPT) as a within-subject factor.

\paragraph{User-level analyses.}
For feedback quality metrics (Relevance, Specificity, Clarity proportions, and Volume) and participant-level survey composites (Perceived Interaction Quality, barrier subscales, and global items such as ease of giving feedback), we first computed per-participant scores for each system as described above. We then compared systems using the Wilcoxon signed-rank test, a non-parametric alternative to the paired $t$-test, appropriate given the small sample size ($N=
20$) and the non-normal distribution of several measures.

We report descriptive statistics as means and standard errors (e.g., $M \pm SE$) and provide exact $p$-values for Wilcoxon tests. Following common practice in HCI and psychology~\cite{cohen1992powerprimer}, we consider results with $p < .05$ statistically significant and treat $0.05 \leq p < 0.10$ as \emph{marginal}, indicating suggestive trends that we interpret cautiously.

\paragraph{Turn-level analyses.}
For outcomes defined at the turn or episode level (e.g., binary indicators of whether a feedback turn exhibits a given property), we used Generalized Estimating Equations (GEE) to account for the non-independence of observations nested within participants. We modeled participant ID as the clustering variable with an exchangeable working correlation structure and used:
\begin{itemize}
    \item a logit link for binary outcomes (e.g., whether a turn is coded as relevant), and
    \item an identity link for continuous outcomes (e.g., the number of characters per episode).
\end{itemize}

\paragraph{Qualitative analysis.}
Qualitative analyses of open-ended survey responses and post-task interviews followed the reflexive thematic analysis procedure described in Section~\ref{sec:thematic-analysis}. We use these themes to triangulate and interpret the quantitative findings, especially with respect to how participants experienced the feedback barriers and the affordances designed to address them.

%% file: sections/07-results.tex
\section{Results}

\subsection{Manipulation Check}
To ensure a fair comparison, we tested whether participants were exposed to a similar degree of feedback barriers in both conditions. For each barrier, we computed per-participant rates (proportion of turns in which the barrier was present) and compared systems using both descriptive statistics and a GEE logistic model with system (FeedbackGPT vs.\ ChatGPT) as a predictor.

Barrier rates were low for Relation and Disambiguation and moderate for Verifiability. On average, participants encountered Relation violations on $1.45\% \pm 1.08$ of turns with ChatGPT and $3.50\% \pm 2.64$ of turns with FeedbackGPT; Verifiability failures on $54.33\% \pm 8.83$ (ChatGPT) vs.\ $47.55\% \pm 7.84$ (FeedbackGPT); Disambiguation issues on $1.94\% \pm 1.17$ vs.\ $5.85\% \pm 2.75$; and Informativeness failures on $4.38\% \pm 2.18$ vs.\ $2.25\% \pm 1.27$. None of these per-user differences were statistically significant (all Wilcoxon $p \ge .173$).

Consistent with these descriptive patterns, GEE models found no significant system effect on barrier presence for any barrier: Relation (Odds Ratio (OR) = 1.17, $p = .878$), Verifiability (OR = 1.14, $p = .715$), Disambiguation (OR = 3.85, $p = .071$), and Informativeness (OR = 0.56, $p = .399$). Taken together, these results suggest that participants experienced comparable levels of model-side failures in both conditions, supporting a fair test of the scaffolds' effects.

\subsection{Scaffolds to lower feedback barriers increase feedback quality}
We next examine whether FeedbackGPT increased user feedback quality along our four dimensions: goal-referenced, actionable, articulate, and progressive. For the first three, we analyze proportions whereas for progressiveness, we analyze mean number of characters per feedback turn as described in Sec~\ref{sec:measures}.
\begin{figure*}[ht]
    \centering
    \includegraphics[width=0.95\textwidth]{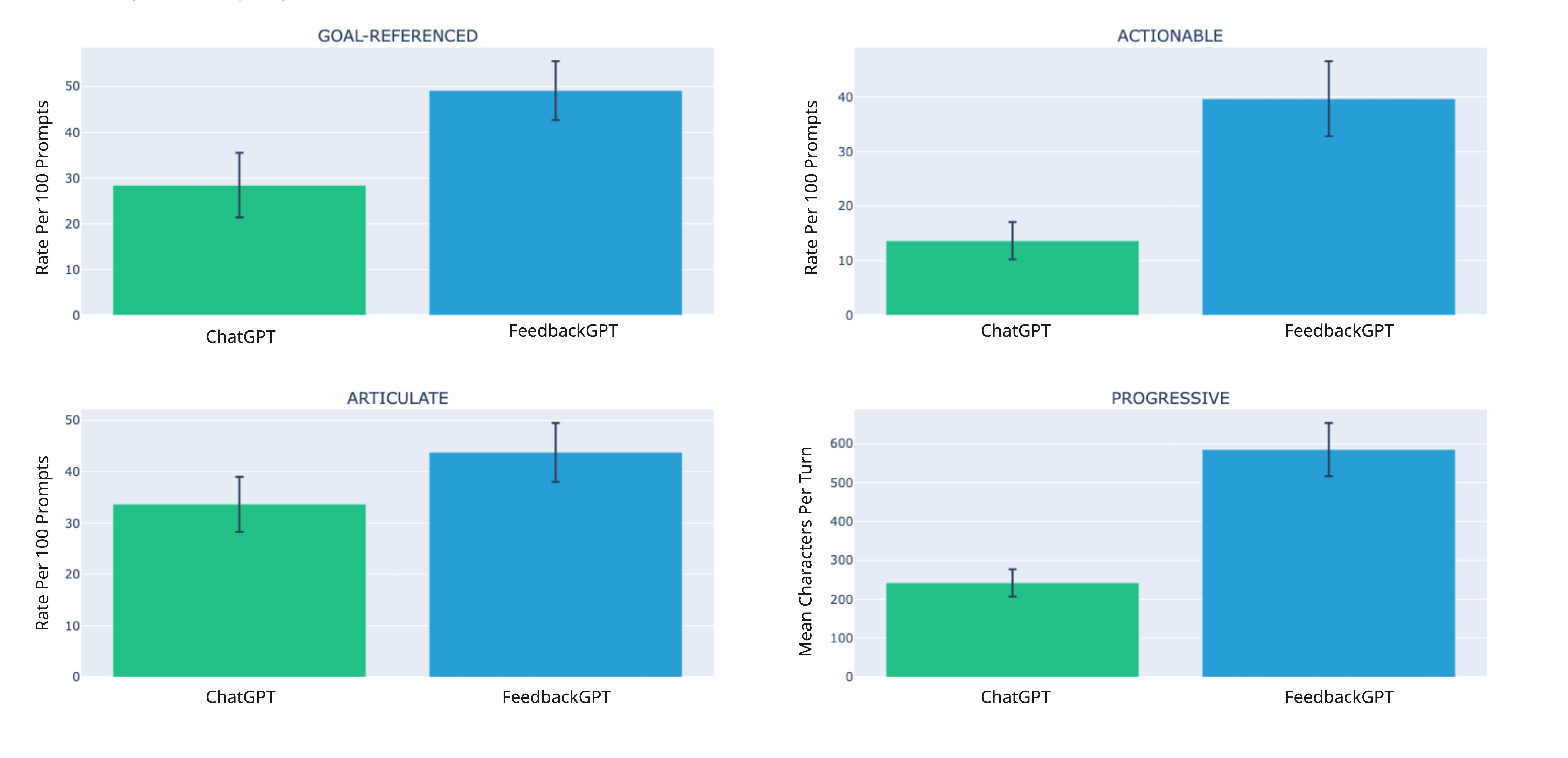}
    \caption{Scatterplot showing the distribution of user feedbacks. The x axis are the ChatGPT rate of giving high quality feedback and the Y axis are FeedbackGPT rate of giving high quality feedback per 100 feedbacks. We see FeedbackGPT made feedback more goal-referenced, actionable and progressive. However, users still struggle with producing articulate feedback.}
    \Description[Scatterplot showing the distribution of user feedbacks.]{The x axis are the ChatGPT rate of giving high quality feedback and the Y axis are FeedbackGPT rate of giving high quality feedback per 100 feedbacks. We see FeedbackGPT made feedback more goal-referenced, actionable and progressive. However, users still struggle with producing articulate feedback.}
    \label{fig:results}
\end{figure*}
\subsubsection*{Goal-referenced --- Scaffolds helped participants stay on-goal.}
FeedbackGPT (FG) made participant's feedback more goal-referenced than ChatGPT (CG). On average, participants’ relevant feedback proportion rose from $32.49\% \pm 7.07$ (CG) to $58.43\% \pm 6.56$ (FG), a $+25.94\%$  gain (SE=10.08); Wilcoxon signed-rank: n=20, W=40.5, $p<.05$. This objective gain aligns with post-task survey data, where participants strongly agreed that FeedbackGPT enabled with \textbf{Goal-referenced feedback (keeping goals/criteria in view)} ($\bar{x}=3.60$, $p<.01$). Furthermore, we found evidence of 2 participants restarting their conversations with ChatGPT but there was no evidence of restarts with FeedbackGPT.

Participants tied these gains to features that kept task goals ``in view.'' For example, anchored comments and comparison helped participants sustain intent across turns: ``With FeedbackGPT I maintained the conversation state longer'' (P9) and ``I compared a few prompts to see which response I liked better'' (P11). Inline controls also increased trust that changes would be localized: ``Marking good/bad gave me confidence the model won’t change the parts I want to keep'' (P15); ``The inline comment feature… lets me point to the specific part that’s wrong and ask for a fix'' (P18).

\subsubsection{Actionable --- Scaffolds enabled precise and detailed feedback.}
The normalized specificity rate rose from $65.86\% \pm 4.58$ (CG) to $90.62\% \pm 2.60$ (FG), a $+24.76$ increase (SE=4.34); Wilcoxon: n=20, W=1.0, $p<0.01$. This was corroborated by post-task ratings, where participants agreed that FeedbackGPT helped with \emph{Specificity (saying exactly what to change)} ($\bar{x}=3.48$, $p<.01$).

Interviews point to two mechanisms driving this. First, locus-specific annotation enabled precise edits without lengthy explanations: ``I could add multiple comments to specific areas instead of having to… manually copy things and explain everything'' (P8). Second, scaffolds that encouraged \emph{input refinement and short back-and-forth exchanges} elicited more detailed inputs before generation. As P6 explained, ``Huddle was basically prompting me for details... I would go provide it.'' This proactive engagement from the system encouraged users to supply concrete details, contrasting with their usual pattern where the model often ``fill[s] the gaps itself'' (P11).

\subsubsection{Articulate --- Vague feedback persisted despite perceived improvements.}
We observe no significant difference in clarity: $77.59\% \pm 6.59$ (CG) vs $84.85\% \pm 4.77$ (FG); $\Delta=+7.26$ (SE=8.21); Wilcoxon: n=20, W=35.0, $p=0.463$. This result reveals a notable tension. Qualitatively, some participants felt FeedbackGPT helped them increase the quality of their inputs: ``The auto evaluate and prompt refinement improved my prompt quality'' (P6; P15). This perception is strongly supported by post-task ratings, where users felt FeedbackGPT significantly improved \emph{Communication (expressing feedback)} ($\bar{x}=3.69$, $p<0.01$). However, in practice, participants still provided vague and ambiguous inputs like ``Need one more version'' and ``trim to 950 words.'' This disconnect—a perceived ease of expression that did not translate to objectively clearer feedback—correlates with self-reported ratings for \emph{conversation quality} that showed no significant improvement (FG=3.70 vs. CG=3.63).

\subsubsection{Progressive --- Scaffolds encouraged richer feedback at the cost of effort.}
Participants contributed significantly more feedback volume with FeedbackGPT. Measured as characters per turn, participants wrote $585.06 \pm 68.55$ characters with FG vs $242.33 \pm 35.24$ with CG; Wilcoxon: n=20, W=32.0, $p<.01$. This aligns with post-task survey data showing users felt FeedbackGPT was superior for \emph{Informativeness (supplying needed info)} ($\bar{x}=3.93$, $p<0.01$).

The interviews reveal this increase in engagement came with an effort trade-off. Participants felt more in control and productive—``My favorite way to interact was to comment and regenerate'' (P2)—but also noted the cost: ``There was more cognitive load because I was interacting with comments but it allowed me to be more productive and in control'' (P2). This pattern of a \emph{productive but more effortful workflow} explains why objective gains in feedback quality and volume did not translate to significantly higher subjective ratings of \emph{perceived productivity} (FG=3.65 vs. CG=3.63) or \emph{output quality} (FG=3.91 vs. CG=3.87). Many participants found this trade-off acceptable because the interaction felt more like a \emph{human-like collaboration}. As P3 explained, ``It feels more like interacting with a human because you can point somewhere and they know what you’re talking about.''. Ultimately, while the scaffolds successfully lowered barriers to providing high-quality feedback, the increased cognitive load tempered users’ subjective ratings of the overall experience. Nonetheless, participants confirmed it was ``easier to give feedback in FeedbackGPT'' ($\bar{x}=3.69$, $p<0.01$) and that it ``improved my ability to guide the AI'' ($\bar{x}=3.63$, $p<0.01$).

%% file: sections/08-discussion.tex
\section{Discussion}
Our two studies demonstrate that the feedback loop in modern human-AI collaboration is systematically broken, not because either humans or models “fail” in isolation, but because of a reciprocal collapse of cooperative principles. Formative study identified four \textbf{Feedback Barriers}---Common Ground, Verifiability, Communication and Informativeness---that emerged from a combination of model limitations and user-side cognitive constraints. Second study showed that, \textbf{FeedbackGPT}, a system equipped with lightweight, model and task agnostic scaffolds, can mitigate the barriers in practice: enabling users to provide significantly more goal-referenced, actionable and progressive feedback.

At the same time, our findings also revealed a critical trade-off. While the scaffolds empowered users to be more effective collaborators, they also increased the perceived cognitive-load: 9/20 users felt that ``burden to provide the right feedback'' remained squarely on them. This underscores that simply providing tools is not enough; the path to effective human-AI collaboration requires deeper rethinking of the interaction paradigm itself. Moreover, while individual scaffolds may be subsumed into future commercial CA interfaces, our enduring contribution is to articulate the design desiderata---and the underlying Feedback Barriers they target---that should guide which affordances are built and how they are composed, rather than added ad hoc. Below, we interpret our findings, propose design implications for the next generation of Conversational Agents, and issue a call for action to the broader Artificial Intelligence community to address challenges at the model level.

\subsection{From Monologue to Dialogue: Interpreting the Impact of Scaffolding}
Our results suggest that the scaffolds in FeedbackGPT began to shift the interaction from a transactional, command-based monologue to a more collaborative dialogue. In the baseline condition (ChatGPT), users treated the CA as a tool to be instructed, and when instructions failed, they resorted to low-effort repair strategies like re-prompting or abandonment. The interaction was one-directional and brittle.

The scaffolds created a sense of a shared conversational space. Features like inline comments allowed users to anchor their feedback to a specific locus, making their intent more precise and reducing the need for lengthy, decontextualized explanations. As P8 noted, this meant they could ``add multiple comments to specific areas instead of having to... manually copy things and explain everything''.  This aligns with principles of external cognition~\cite{scaife1996external}, where visible, persistent artifacts offload cognitive burden and support a shared frame of reference. However, this increased engagement came at a cost. Participants reported a higher cognitive load, even as they felt more productive and in control---``the burden to provide the right feedback was on me''(P19;P17,P2). This ``productive friction'' is a key finding: effective collaboration requires mental effort. While our scaffolds lowered the barrier to providing feedback, the act of formulating good feedback remains a cognitively demanding task. This explains why objective gains in feedback quality did not always translate to higher subjective ratings of ease or productivity. It suggests that future systems must not only provide tools for feedback but also more actively share the cognitive load of the collaborative process itself~\cite{Westbrook2015Cognitive}. Taken together, these findings suggest that the value of scaffolding is not limited to improving the interactional experience itself, but may also shape how users engage with and take ownership over AI outputs downstream.

The effect of scaffolds on downstream performance was evident in our interviews. Participants consistently described a stronger sense of authorship and increased perceived quality: ``I feel I own the FeedbackGPT output more because I have better control.'' (P16; P3, P5, P19), and ``...saw my prompts quality increasing which also increased the output quality'' (P20; P5, P7). While we did not observe a significant difference in self-reported output quality, we expect that as systems become better at incorporating user feedback, and users become more comfortable giving it, these interactional changes will translate into measurable downstream gains. Future work could operationalize such gains through process-oriented metrics, such as the proportion of user-generated content in the final artifact, the number of turns required to reach a stable output, or objective task performance in domains where ground truth exists.

\subsection{Design Implications for Truly Collaborative Agents}
Our findings, particularly the qualitative feedback from participants, point toward three key design principles for moving beyond mere assistance to true collaboration.
\begin{itemize}
    \item \textbf{Design for Proactivity and Mixed Initiative.} Participants expressed a clear desire for the AI to be more than a passive respondent; they wanted a collaborator that could take initiative by suggesting its own refinements (P6), offering a high-level plan before generating a full response (P5, P20), or even ``persuade'' them towards a better idea (P15). This calls for a return to the HCI principles of mixed-initiative interaction, where the system can proactively guide the user, ask clarifying questions, and share responsibility for the task's success. The goal is not just to react to feedback, but to co-create the solution.
    \item \textbf{Foster Reciprocity and Mutual Grounding.} The interaction still felt ``one-directional'' to many participants. As P3 noted, ``I leave comments for a model, but like the model doesn't leave comments for me.'' Users desire a reciprocal dialogue where the AI also provides its own feedback and insights. P19 summarized this perfectly: ``the conversation is two way you know, as much as i want to provide feedback I also want to receive feedback.'' Future systems should incorporate mechanisms for explicit grounding, where the model confirms its understanding. As P8 envisioned, a comment should become ``a conversation with friends in that comment,'' allowing for back-and-forth clarifications.
    \item \textbf{Make the Value of Feedback Interpretable.} Users are more likely to provide high-quality feedback if they understand how it is being used and see its benefit. Current systems with simple thumbs-up/down controls offer no transparency. Organizational science shows that people give more feedback when its benefits are clear \cite{abi2022just,Anseel2015FeedbackSeekingMetaAnalysis}. By showing users how their feedback is leveraged beyond the immediate turn, we can help them become partners in AI development, thereby increasing trust, alignment, and safety.
\end{itemize}

\subsection{A call for action for the broader Artificial Intelligence community}
\label{sec:call-for-action}
While our work demonstrates the power of interface and interaction design, fundamentally overcoming the Feedback Barriers requires parallel advances in core LLM capabilities. We urge the AI community to address the model-side drivers of these barriers:
\begin{itemize}
    \item \textbf{To Overcome the Barrier of Common Ground,} models need more than just longer context windows \cite{liu2025comprehensivesurveylongcontext}; they need \textbf{context-robust attention and memory}. The ``lost-in-the-middle'' effect \cite{liu-etal-2024-lost} is a critical failure of shared reference that burdens users with constant re-contextualization. Research into mitigating this effect \cite{10.5555/3737916.3739859,Chung_2025}, along with developing more faithful chain-of-thought reasoning \cite{turpin2023language,Lee2025FaithfulSelfExplanation} and explicit planning capabilities \cite{wei2025plangenllms}, is essential for maintaining a stable shared goal that updates with the conversation.
    \item \textbf{To Overcome the Barrier of Verifiability,} the burden of fact-checking must shift away from the user \cite{parshakov2025usersfavorllmgeneratedcontent}. Hallucinations remain a central barrier to trust \cite{Sun2024AIHallucination,kalai2025languagemodelshallucinate}, and while Retrieval Augmented Generation (RAG) \cite{rag} helps \cite{niu2023ragtruth,shuster2021retrievalaugmentationreduceshallucination}, its effectiveness is sensitive to retrieval quality \cite{10.1145/3637528.3671470}. We need models that are verifiable by design. This requires progress in three key areas: \textbf{(i) calibration} \cite{geng2023survey,Steyvers2025LLMKnowledge}, so models can express reliable confidence scores, aiding paralinguistic features; \textbf{(ii) principled abstention} \cite{wen2025knowlimitssurveyabstention,wang2025learningaskllmagents,wu-etal-2024-need}, so models can ask for help when uncertain; and \textbf{(iii) reliable inline citation} \cite{zhang2025verifiabledesignaligninglanguage,zhang2024longciteenablingllmsgenerate, 10.1145/3544548.3580847}, to ground claims in evidence.
    \item \textbf{To Overcome the Barriers of Communication and Informativeness,} models must be trained for \textbf{multi-turn, collaborative dialogue} \cite{laban2025llmslostmultiturnconversation,zhang2025surveymultiturninteractioncapabilities}, not just single-turn instruction following \cite{shen2023trickle, ouyang2022training}. This involves developing benchmarks and training schemes that reward proactivity, seeking clarification, and self-critique, mirroring what participants want \cite{chakraborty2025t1,sirdeshmukh2025multichallengerealisticmultiturnconversation,chen2024learning,gou2024criticlargelanguagemodels, lin2024criticbench,shi2025argumentativeexperiencereducingconfirmation}. Furthermore, models must learn to adapt their communication style based on user cues, leveraging research into personalization \cite{zhang2025personalizationlargelanguagemodels,zhao-etal-2025-personalens} and para-linguistics  \cite{Scherer1973VoiceConfidence,Guyer2021ParalinguisticConfidence,schroeder2017humanizing,10.5555/3721488.3721577,Klein2025SocialCuesMetaAnalysis} to deliver information at the right level of detail, just as humans do \cite{hattie2007power, shute2008formative}.  

\end{itemize}

\subsection{Utilizing user feedback from scaffolded interfaces in RLHF}
As discussed earlier (Sec~\ref{sec:problems-in-feedback}), current RLHF and preference optimization pipelines mostly learn from coarse, turn-level signals (e.g., thumbs up/down, whole-response comparisons), which constrains the granularity of reward models and the kinds of collaborative behaviors they can support~\cite{ouyang2022training,rafailov2024directpreferenceoptimizationlanguage,winata2025preference,xu2025largereasoningmodelssurvey}. Scaffolded interfaces like FeedbackGPT can instead yield structured, high-quality preference data: span-level highlights and comments, huddle transcripts, and prompt-evaluation scores naturally form rich preference tuples (e.g., pre- vs. post-feedback responses, competing revisions that better satisfy marked constraints) that can be fed into RLHF and DPO-style training, including online and continual preference learning~\cite{rafailov2024directpreferenceoptimizationlanguage,park-etal-2024-disentangling,xu2025largereasoningmodelssurvey}. Moreover, logs of when users open huddles, request explanations, or repeatedly flag the same error expose “missed opportunities’’ for the model to ask clarifying questions or self-critique; these are precisely the states targeted by recent work on proactive and collaborative LLMs~\cite{wu2025collabllm,zhang2025modeling,10.1145/3715097,mysore-etal-2025-prototypical}.

%% file: sections/09-conclusion.tex
\section{Limitations}
We have identified several limitations as a part of this empirical study. First, our sample is not representative of the population. Although we try to get diverse participants with diverse use-cases there were imbalances in our sample particularly with gender and the education levels. Furthermore, because of the nature and the high cognitive demand of the second study, there was a selection bias with participants who volunteered and completed the task.

Second, our task focuses on co-writing to make our findings generalizable however other interfaces with their task specific interfaces may elicit different behaviors from users.

Third, our scaffolds are not exhaustive but rather one instantiation of the design desiderata. Different feature combinations might elicit different behaviors from users. Furthermore, we have not studied how different actions interact with each other or with the user to elicit different types of feedback and behaviors from both the user and the model. We leave this exploration to future work.

Finally, we also acknowledge that users feedback behaviors will change over time as they get more accustomed to the interfaces and model behaviors. Given the short term nature of our study, we cannot predict how user behaviors will evolve. We leave exploration of longitudinal studies on user feedback behavior as future work.

\section{Conclusion}
As Conversational Agents become integral collaborators in complex, multi-turn tasks, their effectiveness is critically dependent on high-quality user feedback—a resource that is often sparse and difficult to elicit. Across a formative study and a controlled comparison of a scaffolded system (FeedbackGPT) with a baseline (ChatGPT), we empirically show: (1) that naturally occurring feedback is impeded by four Feedback Barriers (Common Ground, Verifiability, Communication, and Informativeness), arising from reciprocal user and model breakdowns; and (2) that lightweight, model-agnostic scaffolds that operationalize persistent shared reference, reciprocal interaction, and verifiable reasoning can significantly increase the relevance, specificity, and volume of user feedback. These findings underscore a key insight: the path to high-quality feedback lies not only in more powerful models but also in interaction designs that make giving feedback cognitively affordable, socially reciprocal, and visibly impactful.

Looking forward, our work points to a dual agenda for the field. On the design side, practitioners must integrate scaffolds that help users disseminate, engage with, and evaluate feedback in ways that sustain cooperative norms. On the model side, advances in memory, calibration, and multi-turn training are essential to reduce the user's burden and enable truly reciprocal dialogue. More broadly, bringing rich human feedback back into the loop is not merely a matter of usability. It is a fundamental requirement for building more trustworthy, adaptive, and socially aligned AI systems, turning users from passive consumers into active partners in the development of safe and effective AI.

%% file: sections/10-appendix.tex
\section{Appendix}
\subsection{Study 1: Participant table}
See Table\ref{tab:participants}
\begin{table*}[ht]
\small
\centering
\resizebox{\linewidth}{!}{
\begin{tabular}{l c c c c}
\toprule
\textbf{Participant} & \textbf{Age Range} & \textbf{Gender} & \textbf{AI Usage} & \textbf{Educational Background} \\
\midrule
P1  & 18--24 & Male       & Coding, Homework                          & High-School graduate \\
P2  & 25--34 & Male       & Coding, Information Seeking, Research     & Doctorate \\
P3  & 18--24 & Male       & Decision Making, Writing, Planning        & Bachelors \\
P4  & 25--34 & Male       & Image Generation, Coding                  & Doctorate \\
P5  & 25--34 & Female     & Essay Writing, Coding                     & Bachelors \\
P6  & 25--34 & Female     & Automation, Coding                        & Bachelors \\
P7  & 18--24 & Female     & Cooking, Automation, Research             & High-School graduate \\
P8  & 25--34 & Male       & Learning, Image Generation                & Doctorate \\
P9  & 18--24 & Male       & Image Generation, Information Seeking     & Bachelors \\
P10 & 18--24 & Male       & Writing                                   & Bachelors \\
P11 & 18--24 & Non-Binary & Learning, Coding, Decision Making         & High-School graduate \\\hline
P12 & 35--44 & Female & Messaging, Information Seeking, Cooking, Decision Making         & Bachelors \\
P13 & 18--24 & Female & Information Seeking, Decision Making         & High-School graduate \\
P14 & 75--84 & Male & Information Seeking, Email, Decision Making         & High-School graduate \\
P15 & 45--54 & Male & Information Seeking, Decision Making         & Bachelors \\
P16 & 45--54 & Female & Information Seeking, Decision Making         & Bachelors \\
\bottomrule
\end{tabular}
}
\caption{Participant demographics and reported AI use-case expertise. (P12-P16 are inexperienced users)}
\label{tab:participants}
\end{table*}

\subsection{Detailed task descriptions}
\label{appendix: task description}
\subsubsection{Task 1: \$250K STEM Fellowship Application}
\textbf{Instructions for participants:}  
You are applying for a \$250,000 STEM Fellowship aimed at individuals with bold, creative visions for using artificial intelligence in transformative ways in their respective fields.

\textbf{Requirements (500–1000 words):}
\begin{itemize}
    \item \textbf{Who you are:} Share your background, values, and what personally motivates you.
    \item \textbf{Why you are uniquely qualified:} Highlight your personal strengths, experiences, and perspectives.
    \item \textbf{Your AI vision:} What innovative AI project or initiative would you pursue?
    \item \textbf{Impact of the fellowship:} Explain how it will advance your personal journey and create a wider positive social impact.
\end{itemize}

\textbf{Important:}
\begin{itemize}
    \item Must be 500–1000 words (applications outside this range will be disqualified).
    \item Strive for an authentic, imaginative, and thematic essay that blends clarity with personal narrative.
    \item You will need to deliver a talk for the first minute of the essay you have drafted.
\end{itemize}

\subsubsection{Task 2: \$50K Neighborhood Improvement Grant}
\textbf{Instructions for participants:}  
You are proposing a local solution for your community via a \$50,000 Neighborhood Improvement Grant.

\textbf{Requirements (500–1000 words):}
\begin{itemize}
    \item \textbf{Neighborhood \& problem statement:} Describe your neighborhood and a pressing issue it faces. Why is this problem urgent?
    \item \textbf{Proposed solution:} What concrete action or project are you proposing (e.g., a public space, safety improvement, communal benefit)?
    \item \textbf{Your role \& qualifications:} Your personal connection, experience, or past involvement in community improvement.
    \item \textbf{Grant usage \& sustainability:} How will funds be spent? How will the project continue after funding ends?
    \item \textbf{Community impact:} Who will benefit and how will this enhance neighborhood well-being?
\end{itemize}

\textbf{Important:}
\begin{itemize}
    \item Must be 500–1000 words, with strict disqualification if under or over.
    \item You will need to deliver a talk for the first minute of the essay you have drafted.
\end{itemize}

\subsubsection{Task 3: \$10K Competition — 5-Minute Personal Pitch}
\textbf{Instructions for participants:}  
You are in a competition to win \$10,000 for participants who display readiness to tackle the unknown and can be leaders in their respective fields. This fund is aimed to develop leaders who can drive change and boost the economy.

\textbf{Requirements:}
\begin{itemize}
    \item 5-minute pitch (\(\sim\)500–1000 words total).
    \item Submit both:
    \begin{itemize}
        \item The transcript.
        \item A 1-minute audio recording of you speaking a portion of it.
    \end{itemize}
\end{itemize}

\textbf{In this pitch, aim to convey:}
\begin{itemize}
    \item What you do and why it matters.
    \item What unique perspective or expertise you offer.
    \item A clear ask or next step (e.g., working with you, supporting your work).
    \item Why you are positioned to be the leader of your field, with examples from your past.
    \item The impact you will deliver if chosen, and why you want to be a leader.
\end{itemize}

\textbf{Important:}
\begin{itemize}
    \item Time/word constraint strict—5 minutes (\(\sim\)500–1000 words).
    \item You will need to deliver a talk for the first minute of the pitch.
\end{itemize}

%% file: 00-main.bbl

\providecommand{\CNFX}[1]{{\em{\textrm{(#1)}}}}
\begin{thebibliography}{155}


\ifx \showCODEN    \undefined \def \showCODEN     #1{\unskip}     \fi
\ifx \showISBNx    \undefined \def \showISBNx     #1{\unskip}     \fi
\ifx \showISBNxiii \undefined \def \showISBNxiii  #1{\unskip}     \fi
\ifx \showISSN     \undefined \def \showISSN      #1{\unskip}     \fi
\ifx \showLCCN     \undefined \def \showLCCN      #1{\unskip}     \fi
\ifx \shownote     \undefined \def \shownote      #1{#1}          \fi
\ifx \showarticletitle \undefined \def \showarticletitle #1{#1}   \fi
\ifx \showURL      \undefined \def \showURL       {\relax}        \fi
\providecommand\bibfield[2]{#2}
\providecommand\bibinfo[2]{#2}
\providecommand\natexlab[1]{#1}
\providecommand\showeprint[2][]{arXiv:#2}

\bibitem[Abi-Esber et~al\mbox{.}(2022)]%
        {abi2022just}
\bibfield{author}{\bibinfo{person}{Nicole Abi-Esber}, \bibinfo{person}{Jennifer~E Abel}, \bibinfo{person}{Juliana Schroeder}, {and} \bibinfo{person}{Francesca Gino}.} \bibinfo{year}{2022}\natexlab{}.
\newblock \showarticletitle{“Just letting you know…” Underestimating others’ desire for constructive feedback.}
\newblock \bibinfo{journal}{\emph{Journal of Personality and Social Psychology}} \bibinfo{volume}{123}, \bibinfo{number}{6} (\bibinfo{year}{2022}), \bibinfo{pages}{1362}.
\newblock


\bibitem[Ahn et~al\mbox{.}(2024)]%
        {ahn2021trust}
\bibfield{author}{\bibinfo{person}{Daehwan Ahn}, \bibinfo{person}{Abdullah Almaatouq}, \bibinfo{person}{Monisha Gulabani}, {and} \bibinfo{person}{Kartik Hosanagar}.} \bibinfo{year}{2024}\natexlab{}.
\newblock \showarticletitle{Impact of Model Interpretability and Outcome Feedback on Trust in AI}. In \bibinfo{booktitle}{\emph{Proceedings of the 2024 CHI Conference on Human Factors in Computing Systems}} (Honolulu, HI, USA) \emph{(\bibinfo{series}{CHI '24})}. \bibinfo{publisher}{Association for Computing Machinery}, \bibinfo{address}{New York, NY, USA}, Article \bibinfo{articleno}{27}, \bibinfo{numpages}{25}~pages.
\newblock
\showISBNx{9798400703300}
\href{https://doi.org/10.1145/3613904.3642780}{doi:\nolinkurl{10.1145/3613904.3642780}}


\bibitem[Allen et~al\mbox{.}(1999)]%
        {allen1999mixedinitiative}
\bibfield{author}{\bibinfo{person}{J.E. Allen}, \bibinfo{person}{C.I. Guinn}, {and} \bibinfo{person}{E. Horvtz}.} \bibinfo{year}{1999}\natexlab{}.
\newblock \showarticletitle{Mixed-initiative interaction}.
\newblock \bibinfo{journal}{\emph{IEEE Intelligent Systems and their Applications}} \bibinfo{volume}{14}, \bibinfo{number}{5} (\bibinfo{year}{1999}), \bibinfo{pages}{14--23}.
\newblock
\href{https://doi.org/10.1109/5254.796083}{doi:\nolinkurl{10.1109/5254.796083}}


\bibitem[Amershi et~al\mbox{.}(2014)]%
        {amershi2014power}
\bibfield{author}{\bibinfo{person}{Saleema Amershi}, \bibinfo{person}{Maya Cakmak}, \bibinfo{person}{William~Bradley Knox}, {and} \bibinfo{person}{Todd Kulesza}.} \bibinfo{year}{2014}\natexlab{}.
\newblock \showarticletitle{Power to the People: The Role of Humans in Interactive Machine Learning}.
\newblock \bibinfo{journal}{\emph{AI Magazine}} \bibinfo{volume}{35}, \bibinfo{number}{4} (\bibinfo{date}{Dec.} \bibinfo{year}{2014}), \bibinfo{pages}{105--120}.
\newblock
\href{https://doi.org/10.1609/aimag.v35i4.2513}{doi:\nolinkurl{10.1609/aimag.v35i4.2513}}


\bibitem[Anderson(2003)]%
        {Anderson2003PsychologyOfDoingNothing}
\bibfield{author}{\bibinfo{person}{Christopher~J. Anderson}.} \bibinfo{year}{2003}\natexlab{}.
\newblock \showarticletitle{The Psychology of Doing Nothing: Forms of Decision Avoidance Result from Reason and Emotion}.
\newblock \bibinfo{journal}{\emph{Psychological Bulletin}}  \bibinfo{volume}{129} (\bibinfo{year}{2003}), \bibinfo{pages}{139--167}.
\newblock
\urldef\tempurl%
\url{https://ssrn.com/abstract=895727}
\showURL{%
\tempurl}


\bibitem[Anseel et~al\mbox{.}(2015)]%
        {Anseel2015FeedbackSeekingMetaAnalysis}
\bibfield{author}{\bibinfo{person}{F. Anseel}, \bibinfo{person}{A.~S. Beatty}, \bibinfo{person}{W. Shen}, \bibinfo{person}{F. Lievens}, {and} \bibinfo{person}{P.~R. Sackett}.} \bibinfo{year}{2015}\natexlab{}.
\newblock \showarticletitle{How Are We Doing After 30 Years? A Meta-Analytic Review of the Antecedents and Outcomes of Feedback-Seeking Behavior}.
\newblock \bibinfo{journal}{\emph{Journal of Management}} \bibinfo{volume}{41}, \bibinfo{number}{1} (\bibinfo{year}{2015}), \bibinfo{pages}{318--348}.
\newblock
\href{https://doi.org/10.1177/0149206313484521}{doi:\nolinkurl{10.1177/0149206313484521}}


\bibitem[Arora et~al\mbox{.}(2025)]%
        {doi:10.1177/00222429241276529}
\bibfield{author}{\bibinfo{person}{Neeraj Arora}, \bibinfo{person}{Ishita Chakraborty}, {and} \bibinfo{person}{Yohei Nishimura}.} \bibinfo{year}{2025}\natexlab{}.
\newblock \showarticletitle{AI–Human Hybrids for Marketing Research: Leveraging Large Language Models (LLMs) as Collaborators}.
\newblock \bibinfo{journal}{\emph{Journal of Marketing}} \bibinfo{volume}{89}, \bibinfo{number}{2} (\bibinfo{year}{2025}), \bibinfo{pages}{43--70}.
\newblock
\showeprint{https://doi.org/10.1177/00222429241276529}
\href{https://doi.org/10.1177/00222429241276529}{doi:\nolinkurl{10.1177/00222429241276529}}


\bibitem[Ashery et~al\mbox{.}(2025)]%
        {Ashery2025Conventions}
\bibfield{author}{\bibinfo{person}{Ariel~Flint Ashery}, \bibinfo{person}{Luca~Maria Aiello}, {and} \bibinfo{person}{Andrea Baronchelli}.} \bibinfo{year}{2025}\natexlab{}.
\newblock \showarticletitle{Emergent social conventions and collective bias in {LLM} populations}.
\newblock \bibinfo{journal}{\emph{Science Advances}} \bibinfo{volume}{11}, \bibinfo{number}{20} (\bibinfo{year}{2025}), \bibinfo{pages}{eadu9368}.
\newblock
\href{https://doi.org/10.1126/sciadv.adu9368}{doi:\nolinkurl{10.1126/sciadv.adu9368}}


\bibitem[Bacchelli and Bird(2013)]%
        {bacchelli2013expectations}
\bibfield{author}{\bibinfo{person}{Alberto Bacchelli} {and} \bibinfo{person}{Christian Bird}.} \bibinfo{year}{2013}\natexlab{}.
\newblock \showarticletitle{Expectations, outcomes, and challenges of modern code review}. In \bibinfo{booktitle}{\emph{2013 35th International Conference on Software Engineering (ICSE)}}. \bibinfo{pages}{712--721}.
\newblock
\href{https://doi.org/10.1109/ICSE.2013.6606617}{doi:\nolinkurl{10.1109/ICSE.2013.6606617}}


\bibitem[Black and Wiliam(1998)]%
        {Black01031998}
\bibfield{author}{\bibinfo{person}{Paul Black} {and} \bibinfo{person}{Dylan Wiliam}.} \bibinfo{year}{1998}\natexlab{}.
\newblock \showarticletitle{Assessment and Classroom Learning}.
\newblock \bibinfo{journal}{\emph{Assessment in Education: Principles, Policy \& Practice}} \bibinfo{volume}{5}, \bibinfo{number}{1} (\bibinfo{year}{1998}), \bibinfo{pages}{7--74}.
\newblock
\showeprint{https://doi.org/10.1080/0969595980050102}
\href{https://doi.org/10.1080/0969595980050102}{doi:\nolinkurl{10.1080/0969595980050102}}


\bibitem[Braun and Clarke(2023)]%
        {Braun25012023}
\bibfield{author}{\bibinfo{person}{Virginia Braun} {and} \bibinfo{person}{Victoria Clarke}.} \bibinfo{year}{2023}\natexlab{}.
\newblock \showarticletitle{Toward good practice in thematic analysis: Avoiding common problems and be(com)ing a knowing researcher}.
\newblock \bibinfo{journal}{\emph{International Journal of Transgender Health}} \bibinfo{volume}{24}, \bibinfo{number}{1} (\bibinfo{year}{2023}), \bibinfo{pages}{1--6}.
\newblock
\showeprint{https://doi.org/10.1080/26895269.2022.2129597}
\href{https://doi.org/10.1080/26895269.2022.2129597}{doi:\nolinkurl{10.1080/26895269.2022.2129597}}


\bibitem[Carless and Boud(2018)]%
        {carless2018developing}
\bibfield{author}{\bibinfo{person}{David Carless} {and} \bibinfo{person}{David Boud}.} \bibinfo{year}{2018}\natexlab{}.
\newblock \showarticletitle{The development of student feedback literacy: enabling uptake of feedback}.
\newblock \bibinfo{journal}{\emph{Assessment \& Evaluation in Higher Education}} \bibinfo{volume}{43}, \bibinfo{number}{8} (\bibinfo{year}{2018}), \bibinfo{pages}{1315--1325}.
\newblock
\showeprint{https://doi.org/10.1080/02602938.2018.1463354}
\href{https://doi.org/10.1080/02602938.2018.1463354}{doi:\nolinkurl{10.1080/02602938.2018.1463354}}


\bibitem[Chakraborty et~al\mbox{.}(2025)]%
        {chakraborty2025t1}
\bibfield{author}{\bibinfo{person}{Amartya Chakraborty}, \bibinfo{person}{Paresh Dashore}, \bibinfo{person}{Nadia Bathaee}, \bibinfo{person}{Anmol Jain}, \bibinfo{person}{Anirban Das}, \bibinfo{person}{Shi-Xiong Zhang}, \bibinfo{person}{Sambit Sahu}, \bibinfo{person}{Milind Naphade}, {and} \bibinfo{person}{Genta~Indra Winata}.} \bibinfo{year}{2025}\natexlab{}.
\newblock \showarticletitle{T1: A Tool-Oriented Conversational Dataset for Multi-Turn Agentic Planning}.
\newblock \bibinfo{journal}{\emph{arXiv preprint arXiv:2505.16986}} (\bibinfo{year}{2025}).
\newblock


\bibitem[Chang et~al\mbox{.}(2023)]%
        {10.1145/3544548.3580847}
\bibfield{author}{\bibinfo{person}{Joseph~Chee Chang}, \bibinfo{person}{Amy~X. Zhang}, \bibinfo{person}{Jonathan Bragg}, \bibinfo{person}{Andrew Head}, \bibinfo{person}{Kyle Lo}, \bibinfo{person}{Doug Downey}, {and} \bibinfo{person}{Daniel~S. Weld}.} \bibinfo{year}{2023}\natexlab{}.
\newblock \showarticletitle{CiteSee: Augmenting Citations in Scientific Papers with Persistent and Personalized Historical Context}. In \bibinfo{booktitle}{\emph{Proceedings of the 2023 CHI Conference on Human Factors in Computing Systems}} (Hamburg, Germany) \emph{(\bibinfo{series}{CHI '23})}. \bibinfo{publisher}{Association for Computing Machinery}, \bibinfo{address}{New York, NY, USA}, Article \bibinfo{articleno}{737}, \bibinfo{numpages}{15}~pages.
\newblock
\showISBNx{9781450394215}
\href{https://doi.org/10.1145/3544548.3580847}{doi:\nolinkurl{10.1145/3544548.3580847}}


\bibitem[Chen et~al\mbox{.}(2017)]%
        {chen2017survey}
\bibfield{author}{\bibinfo{person}{Hongshen Chen}, \bibinfo{person}{Xiaorui Liu}, \bibinfo{person}{Dawei Yin}, {and} \bibinfo{person}{Jiliang Tang}.} \bibinfo{year}{2017}\natexlab{}.
\newblock \showarticletitle{A survey on dialogue systems: Recent advances and new frontiers}.
\newblock \bibinfo{journal}{\emph{Acm Sigkdd Explorations Newsletter}} \bibinfo{volume}{19}, \bibinfo{number}{2} (\bibinfo{year}{2017}), \bibinfo{pages}{25--35}.
\newblock


\bibitem[Chen and Pu(2007)]%
        {10.1145/1297231.1297263}
\bibfield{author}{\bibinfo{person}{Li Chen} {and} \bibinfo{person}{Pearl Pu}.} \bibinfo{year}{2007}\natexlab{}.
\newblock \showarticletitle{The evaluation of a hybrid critiquing system with preference-based recommendations organization}. In \bibinfo{booktitle}{\emph{Proceedings of the 2007 ACM Conference on Recommender Systems}} (Minneapolis, MN, USA) \emph{(\bibinfo{series}{RecSys '07})}. \bibinfo{publisher}{Association for Computing Machinery}, \bibinfo{address}{New York, NY, USA}, \bibinfo{pages}{169–172}.
\newblock
\showISBNx{9781595937308}
\href{https://doi.org/10.1145/1297231.1297263}{doi:\nolinkurl{10.1145/1297231.1297263}}


\bibitem[Chen et~al\mbox{.}(2025)]%
        {chen2024learning}
\bibfield{author}{\bibinfo{person}{Maximillian Chen}, \bibinfo{person}{Ruoxi Sun}, \bibinfo{person}{Tomas Pfister}, {and} \bibinfo{person}{Sercan Arik}.} \bibinfo{year}{2025}\natexlab{}.
\newblock \showarticletitle{Learning to Clarify: Multi-turn Conversations with Action-Based Contrastive Self-Training}. In \bibinfo{booktitle}{\emph{International Conference on Learning Representations}}, \bibfield{editor}{\bibinfo{person}{Y.~Yue}, \bibinfo{person}{A.~Garg}, \bibinfo{person}{N.~Peng}, \bibinfo{person}{F.~Sha}, {and} \bibinfo{person}{R.~Yu}} (Eds.), Vol.~\bibinfo{volume}{2025}. \bibinfo{pages}{32244–32279}.
\newblock
\urldef\tempurl%
\url{https://proceedings.iclr.cc/paper_files/paper/2025/file/4ffd05ca3cf3985f4572af015b4cfc1e-Paper-Conference.pdf}
\showURL{%
\tempurl}


\bibitem[Chung et~al\mbox{.}(2025)]%
        {Chung_2025}
\bibfield{author}{\bibinfo{person}{Yeounoh Chung}, \bibinfo{person}{Gaurav~T. Kakkar}, \bibinfo{person}{Yu Gan}, \bibinfo{person}{Brenton Milne}, {and} \bibinfo{person}{Fatma Özcan}.} \bibinfo{year}{2025}\natexlab{}.
\newblock \showarticletitle{Is Long Context All You Need? Leveraging LLM’s Extended Context for NL2SQL}.
\newblock \bibinfo{journal}{\emph{Proceedings of the VLDB Endowment}} \bibinfo{volume}{18}, \bibinfo{number}{8} (\bibinfo{date}{April} \bibinfo{year}{2025}), \bibinfo{pages}{2735–2747}.
\newblock
\showISSN{2150-8097}
\href{https://doi.org/10.14778/3742728.3742761}{doi:\nolinkurl{10.14778/3742728.3742761}}


\bibitem[Clark and Brennan(1991)]%
        {clark1991grounding}
\bibfield{author}{\bibinfo{person}{Herbert~H Clark} {and} \bibinfo{person}{Susan~E Brennan}.} \bibinfo{year}{1991}\natexlab{}.
\newblock \showarticletitle{Grounding in communication.}
\newblock In \bibinfo{booktitle}{\emph{Perspectives on socially shared cognition.}} \bibinfo{publisher}{American Psychological Association}, \bibinfo{pages}{127--149}.
\newblock


\bibitem[Cohen(1992)]%
        {cohen1992powerprimer}
\bibfield{author}{\bibinfo{person}{Jacob Cohen}.} \bibinfo{year}{1992}\natexlab{}.
\newblock \showarticletitle{A power primer}.
\newblock \bibinfo{journal}{\emph{Psychological Bulletin}} \bibinfo{volume}{112}, \bibinfo{number}{1} (\bibinfo{year}{1992}), \bibinfo{pages}{155--159}.
\newblock
\href{https://doi.org/10.1037//0033-2909.112.1.155}{doi:\nolinkurl{10.1037//0033-2909.112.1.155}}


\bibitem[Cohn et~al\mbox{.}(2025)]%
        {cohn2025cotalhumaninthelooppromptengineering}
\bibfield{author}{\bibinfo{person}{Clayton Cohn}, \bibinfo{person}{Ashwin~T S}, \bibinfo{person}{Naveeduddin Mohammed}, {and} \bibinfo{person}{Gautam Biswas}.} \bibinfo{year}{2025}\natexlab{}.
\newblock \bibinfo{title}{CoTAL: Human-in-the-Loop Prompt Engineering for Generalizable Formative Assessment Scoring}.
\newblock
\showeprint[arxiv]{2504.02323}~[cs.CL]
\urldef\tempurl%
\url{https://arxiv.org/abs/2504.02323}
\showURL{%
\tempurl}


\bibitem[Dai et~al\mbox{.}(2025)]%
        {dai2025critic}
\bibfield{author}{\bibinfo{person}{David~Wei Dai}, \bibinfo{person}{Hua Zhu}, {and} \bibinfo{person}{Guanliang Chen}.} \bibinfo{year}{2025}\natexlab{}.
\newblock \showarticletitle{How does interaction with LLM powered chatbots shape human understanding of culture? The need for Critical Interactional Competence (CritIC)}.
\newblock \bibinfo{journal}{\emph{Annual Review of Applied Linguistics}}  \bibinfo{volume}{45} (\bibinfo{year}{2025}), \bibinfo{pages}{28--49}.
\newblock
\href{https://doi.org/10.1017/S0267190525000054}{doi:\nolinkurl{10.1017/S0267190525000054}}


\bibitem[Deng et~al\mbox{.}(2025)]%
        {10.1145/3715097}
\bibfield{author}{\bibinfo{person}{Yang Deng}, \bibinfo{person}{Lizi Liao}, \bibinfo{person}{Wenqiang Lei}, \bibinfo{person}{Grace~Hui Yang}, \bibinfo{person}{Wai Lam}, {and} \bibinfo{person}{Tat-Seng Chua}.} \bibinfo{year}{2025}\natexlab{}.
\newblock \showarticletitle{Proactive Conversational AI: A Comprehensive Survey of Advancements and Opportunities}.
\newblock \bibinfo{journal}{\emph{ACM Trans. Inf. Syst.}} \bibinfo{volume}{43}, \bibinfo{number}{3}, Article \bibinfo{articleno}{67} (\bibinfo{date}{March} \bibinfo{year}{2025}), \bibinfo{numpages}{45}~pages.
\newblock
\showISSN{1046-8188}
\href{https://doi.org/10.1145/3715097}{doi:\nolinkurl{10.1145/3715097}}


\bibitem[Deshpande et~al\mbox{.}(2025)]%
        {sirdeshmukh2025multichallengerealisticmultiturnconversation}
\bibfield{author}{\bibinfo{person}{Kaustubh Deshpande}, \bibinfo{person}{Ved Sirdeshmukh}, \bibinfo{person}{Johannes~Baptist Mols}, \bibinfo{person}{Lifeng Jin}, \bibinfo{person}{Ed-Yeremai Hernandez-Cardona}, \bibinfo{person}{Dean Lee}, \bibinfo{person}{Jeremy Kritz}, \bibinfo{person}{Willow~E. Primack}, \bibinfo{person}{Summer Yue}, {and} \bibinfo{person}{Chen Xing}.} \bibinfo{year}{2025}\natexlab{}.
\newblock \showarticletitle{{M}ulti{C}hallenge: A Realistic Multi-Turn Conversation Evaluation Benchmark Challenging to Frontier {LLM}s}. In \bibinfo{booktitle}{\emph{Findings of the Association for Computational Linguistics: ACL 2025}}, \bibfield{editor}{\bibinfo{person}{Wanxiang Che}, \bibinfo{person}{Joyce Nabende}, \bibinfo{person}{Ekaterina Shutova}, {and} \bibinfo{person}{Mohammad~Taher Pilehvar}} (Eds.). \bibinfo{publisher}{Association for Computational Linguistics}, \bibinfo{address}{Vienna, Austria}, \bibinfo{pages}{18632--18702}.
\newblock
\showISBNx{979-8-89176-256-5}
\href{https://doi.org/10.18653/v1/2025.findings-acl.958}{doi:\nolinkurl{10.18653/v1/2025.findings-acl.958}}


\bibitem[Dietvorst et~al\mbox{.}(2015)]%
        {dietvorst2015algorithm}
\bibfield{author}{\bibinfo{person}{Berkeley~J Dietvorst}, \bibinfo{person}{Joseph~P Simmons}, {and} \bibinfo{person}{Cade Massey}.} \bibinfo{year}{2015}\natexlab{}.
\newblock \showarticletitle{Algorithm aversion: People erroneously avoid algorithms after seeing them err}.
\newblock \bibinfo{journal}{\emph{Journal of Experimental Psychology: General}} \bibinfo{volume}{144}, \bibinfo{number}{1} (\bibinfo{year}{2015}), \bibinfo{pages}{114--126}.
\newblock
\href{https://doi.org/10.1037/xge0000033}{doi:\nolinkurl{10.1037/xge0000033}}


\bibitem[Don-Yehiya et~al\mbox{.}(2025)]%
        {DonYehiya2024NaturallyOF}
\bibfield{author}{\bibinfo{person}{Shachar Don-Yehiya}, \bibinfo{person}{Leshem Choshen}, {and} \bibinfo{person}{Omri Abend}.} \bibinfo{year}{2025}\natexlab{}.
\newblock \bibinfo{title}{Naturally Occurring Feedback is Common, Extractable and Useful}.
\newblock
\showeprint[arxiv]{2407.10944}~[cs.CL]
\urldef\tempurl%
\url{https://arxiv.org/abs/2407.10944}
\showURL{%
\tempurl}


\bibitem[Dudley and Kristensson(2018)]%
        {10.1145/3185517}
\bibfield{author}{\bibinfo{person}{John~J. Dudley} {and} \bibinfo{person}{Per~Ola Kristensson}.} \bibinfo{year}{2018}\natexlab{}.
\newblock \showarticletitle{A Review of User Interface Design for Interactive Machine Learning}.
\newblock \bibinfo{journal}{\emph{ACM Trans. Interact. Intell. Syst.}} \bibinfo{volume}{8}, \bibinfo{number}{2}, Article \bibinfo{articleno}{8} (\bibinfo{date}{June} \bibinfo{year}{2018}), \bibinfo{numpages}{37}~pages.
\newblock
\showISSN{2160-6455}
\href{https://doi.org/10.1145/3185517}{doi:\nolinkurl{10.1145/3185517}}


\bibitem[Durante et~al\mbox{.}(2024)]%
        {durante2024agentaisurveyinghorizons}
\bibfield{author}{\bibinfo{person}{Zane Durante}, \bibinfo{person}{Qiuyuan Huang}, \bibinfo{person}{Naoki Wake}, \bibinfo{person}{Ran Gong}, \bibinfo{person}{Jae~Sung Park}, \bibinfo{person}{Bidipta Sarkar}, \bibinfo{person}{Rohan Taori}, \bibinfo{person}{Yusuke Noda}, \bibinfo{person}{Demetri Terzopoulos}, \bibinfo{person}{Yejin Choi}, \bibinfo{person}{Katsushi Ikeuchi}, \bibinfo{person}{Hoi Vo}, \bibinfo{person}{Li Fei-Fei}, {and} \bibinfo{person}{Jianfeng Gao}.} \bibinfo{year}{2024}\natexlab{}.
\newblock \bibinfo{title}{Agent AI: Surveying the Horizons of Multimodal Interaction}.
\newblock
\showeprint[arxiv]{2401.03568}~[cs.AI]
\urldef\tempurl%
\url{https://arxiv.org/abs/2401.03568}
\showURL{%
\tempurl}


\bibitem[Fails and Olsen~Jr(2003)]%
        {fails2003interactive}
\bibfield{author}{\bibinfo{person}{Jerry~Alan Fails} {and} \bibinfo{person}{Dan~R Olsen~Jr}.} \bibinfo{year}{2003}\natexlab{}.
\newblock \showarticletitle{Interactive machine learning}. In \bibinfo{booktitle}{\emph{Proceedings of the 8th International Conference on Intelligent User Interfaces}}. ACM, \bibinfo{pages}{39--45}.
\newblock
\href{https://doi.org/10.1145/604045.604056}{doi:\nolinkurl{10.1145/604045.604056}}


\bibitem[Fan et~al\mbox{.}(2024)]%
        {10.1145/3637528.3671470}
\bibfield{author}{\bibinfo{person}{Wenqi Fan}, \bibinfo{person}{Yujuan Ding}, \bibinfo{person}{Liangbo Ning}, \bibinfo{person}{Shijie Wang}, \bibinfo{person}{Hengyun Li}, \bibinfo{person}{Dawei Yin}, \bibinfo{person}{Tat-Seng Chua}, {and} \bibinfo{person}{Qing Li}.} \bibinfo{year}{2024}\natexlab{}.
\newblock \showarticletitle{A Survey on RAG Meeting LLMs: Towards Retrieval-Augmented Large Language Models}. In \bibinfo{booktitle}{\emph{Proceedings of the 30th ACM SIGKDD Conference on Knowledge Discovery and Data Mining}} (Barcelona, Spain) \emph{(\bibinfo{series}{KDD '24})}. \bibinfo{publisher}{Association for Computing Machinery}, \bibinfo{address}{New York, NY, USA}, \bibinfo{pages}{6491–6501}.
\newblock
\showISBNx{9798400704901}
\href{https://doi.org/10.1145/3637528.3671470}{doi:\nolinkurl{10.1145/3637528.3671470}}


\bibitem[Fernandes et~al\mbox{.}(2023)]%
        {fernandes-etal-2023-bridging}
\bibfield{author}{\bibinfo{person}{Patrick Fernandes}, \bibinfo{person}{Aman Madaan}, \bibinfo{person}{Emmy Liu}, \bibinfo{person}{Ant{\'o}nio Farinhas}, \bibinfo{person}{Pedro~Henrique Martins}, \bibinfo{person}{Amanda Bertsch}, \bibinfo{person}{Jos{\'e} G.~C. de Souza}, \bibinfo{person}{Shuyan Zhou}, \bibinfo{person}{Tongshuang Wu}, \bibinfo{person}{Graham Neubig}, {and} \bibinfo{person}{Andr{\'e} F.~T. Martins}.} \bibinfo{year}{2023}\natexlab{}.
\newblock \showarticletitle{Bridging the Gap: A Survey on Integrating (Human) Feedback for Natural Language Generation}.
\newblock \bibinfo{journal}{\emph{Transactions of the Association for Computational Linguistics}}  \bibinfo{volume}{11} (\bibinfo{year}{2023}), \bibinfo{pages}{1643--1668}.
\newblock
\href{https://doi.org/10.1162/tacl_a_00626}{doi:\nolinkurl{10.1162/tacl_a_00626}}


\bibitem[Gao et~al\mbox{.}(2024)]%
        {Gao_2024}
\bibfield{author}{\bibinfo{person}{Jie Gao}, \bibinfo{person}{Simret~Araya Gebreegziabher}, \bibinfo{person}{Kenny Tsu~Wei Choo}, \bibinfo{person}{Toby Jia-Jun Li}, \bibinfo{person}{Simon~Tangi Perrault}, {and} \bibinfo{person}{Thomas~W Malone}.} \bibinfo{year}{2024}\natexlab{}.
\newblock \showarticletitle{A Taxonomy for Human-LLM Interaction Modes: An Initial Exploration}. In \bibinfo{booktitle}{\emph{Extended Abstracts of the CHI Conference on Human Factors in Computing Systems}} \emph{(\bibinfo{series}{CHI ’24})}. \bibinfo{publisher}{ACM}, \bibinfo{pages}{1–11}.
\newblock
\href{https://doi.org/10.1145/3613905.3650786}{doi:\nolinkurl{10.1145/3613905.3650786}}


\bibitem[Ge et~al\mbox{.}(2023)]%
        {ge2023should}
\bibfield{author}{\bibinfo{person}{Yubin Ge}, \bibinfo{person}{Ziang Xiao}, \bibinfo{person}{Jana Diesner}, \bibinfo{person}{Heng Ji}, \bibinfo{person}{Karrie Karahalios}, {and} \bibinfo{person}{Hari Sundaram}.} \bibinfo{year}{2023}\natexlab{}.
\newblock \showarticletitle{What should i ask: A knowledge-driven approach for follow-up questions generation in conversational surveys}. In \bibinfo{booktitle}{\emph{Proceedings of the 37th Pacific Asia Conference on Language, Information and Computation}}. \bibinfo{pages}{113--124}.
\newblock


\bibitem[Geng et~al\mbox{.}(2024)]%
        {geng2023survey}
\bibfield{author}{\bibinfo{person}{Jiahui Geng}, \bibinfo{person}{Fengyu Cai}, \bibinfo{person}{Yuxia Wang}, \bibinfo{person}{Heinz Koeppl}, \bibinfo{person}{Preslav Nakov}, {and} \bibinfo{person}{Iryna Gurevych}.} \bibinfo{year}{2024}\natexlab{}.
\newblock \showarticletitle{A Survey of Confidence Estimation and Calibration in Large Language Models}. In \bibinfo{booktitle}{\emph{Proceedings of the 2024 Conference of the North American Chapter of the Association for Computational Linguistics: Human Language Technologies (Volume 1: Long Papers)}}, \bibfield{editor}{\bibinfo{person}{Kevin Duh}, \bibinfo{person}{Helena Gomez}, {and} \bibinfo{person}{Steven Bethard}} (Eds.). \bibinfo{publisher}{Association for Computational Linguistics}, \bibinfo{address}{Mexico City, Mexico}, \bibinfo{pages}{6577--6595}.
\newblock
\href{https://doi.org/10.18653/v1/2024.naacl-long.366}{doi:\nolinkurl{10.18653/v1/2024.naacl-long.366}}


\bibitem[Gou et~al\mbox{.}(2024)]%
        {gou2024criticlargelanguagemodels}
\bibfield{author}{\bibinfo{person}{Zhibin Gou}, \bibinfo{person}{Zhihong Shao}, \bibinfo{person}{Yeyun Gong}, \bibinfo{person}{yelong shen}, \bibinfo{person}{Yujiu Yang}, \bibinfo{person}{Nan Duan}, {and} \bibinfo{person}{Weizhu Chen}.} \bibinfo{year}{2024}\natexlab{}.
\newblock \showarticletitle{{CRITIC}: Large Language Models Can Self-Correct with Tool-Interactive Critiquing}. In \bibinfo{booktitle}{\emph{The Twelfth International Conference on Learning Representations}}.
\newblock
\urldef\tempurl%
\url{https://openreview.net/forum?id=Sx038qxjek}
\showURL{%
\tempurl}


\bibitem[Grice(1975)]%
        {grice1975logic}
\bibfield{author}{\bibinfo{person}{H.~P. Grice}.} \bibinfo{year}{1975}\natexlab{}.
\newblock \bibinfo{booktitle}{\emph{Logic and Conversation}}.
\newblock \bibinfo{publisher}{Brill}, \bibinfo{address}{Leiden, The Netherlands}, \bibinfo{pages}{41 -- 58}.
\newblock
\showISBNx{9789004368811}
\href{https://doi.org/10.1163/9789004368811_003}{doi:\nolinkurl{10.1163/9789004368811_003}}


\bibitem[Gunning and Aha(2019)]%
        {gunning2019darpa}
\bibfield{author}{\bibinfo{person}{David Gunning} {and} \bibinfo{person}{David~W Aha}.} \bibinfo{year}{2019}\natexlab{}.
\newblock \showarticletitle{DARPA's explainable artificial intelligence (XAI) program}.
\newblock \bibinfo{journal}{\emph{AI Magazine}} \bibinfo{volume}{40}, \bibinfo{number}{2} (\bibinfo{year}{2019}), \bibinfo{pages}{44--58}.
\newblock


\bibitem[Guyer et~al\mbox{.}(2021)]%
        {Guyer2021ParalinguisticConfidence}
\bibfield{author}{\bibinfo{person}{J.~J. Guyer}, \bibinfo{person}{P. Briñol}, \bibinfo{person}{T.~I. Vaughan-Johnston}, \bibinfo{person}{L.~R. Fabrigar}, \bibinfo{person}{L. Moreno}, {and} \bibinfo{person}{R.~E. Petty}.} \bibinfo{year}{2021}\natexlab{}.
\newblock \showarticletitle{Paralinguistic Features Communicated through Voice can Affect Appraisals of Confidence and Evaluative Judgments}.
\newblock \bibinfo{journal}{\emph{Journal of Nonverbal Behavior}} \bibinfo{volume}{45}, \bibinfo{number}{4} (\bibinfo{year}{2021}), \bibinfo{pages}{479--504}.
\newblock
\showeprint{2021-07-06}
\href{https://doi.org/10.1007/s10919-021-00374-2}{doi:\nolinkurl{10.1007/s10919-021-00374-2}}


\bibitem[Hattie and Timperley(2007)]%
        {hattie2007power}
\bibfield{author}{\bibinfo{person}{John Hattie} {and} \bibinfo{person}{Helen Timperley}.} \bibinfo{year}{2007}\natexlab{}.
\newblock \showarticletitle{The power of feedback}.
\newblock \bibinfo{journal}{\emph{Review of educational research}} \bibinfo{volume}{77}, \bibinfo{number}{1} (\bibinfo{year}{2007}), \bibinfo{pages}{81--112}.
\newblock


\bibitem[He et~al\mbox{.}(2016)]%
        {He2016InteractiveRecommenders}
\bibfield{author}{\bibinfo{person}{Chen He}, \bibinfo{person}{Denis Parra}, {and} \bibinfo{person}{Katrien Verbert}.} \bibinfo{year}{2016}\natexlab{}.
\newblock \showarticletitle{Interactive recommender systems: A survey of the state of the art and future research challenges and opportunities}.
\newblock \bibinfo{journal}{\emph{Expert Systems with Applications}}  \bibinfo{volume}{56} (\bibinfo{year}{2016}), \bibinfo{pages}{9--27}.
\newblock
\href{https://doi.org/10.1016/j.eswa.2016.02.013}{doi:\nolinkurl{10.1016/j.eswa.2016.02.013}}


\bibitem[Henderson et~al\mbox{.}(2025)]%
        {Henderson2025Comparing}
\bibfield{author}{\bibinfo{person}{M. Henderson}, \bibinfo{person}{M. Bearman}, \bibinfo{person}{J. Chung}, \bibinfo{person}{T. Fawns}, \bibinfo{person}{S. Buckingham~Shum}, \bibinfo{person}{K.~E. Matthews}, {and} \bibinfo{person}{J. de Mello~Heredia}.} \bibinfo{year}{2025}\natexlab{}.
\newblock \showarticletitle{Comparing Generative AI and teacher feedback: student perceptions of usefulness and trustworthiness}.
\newblock \bibinfo{journal}{\emph{Assessment \& Evaluation in Higher Education}} (\bibinfo{year}{2025}), \bibinfo{pages}{1--16}.
\newblock
\href{https://doi.org/10.1080/02602938.2025.2502582}{doi:\nolinkurl{10.1080/02602938.2025.2502582}}


\bibitem[Honeycutt et~al\mbox{.}(2020)]%
        {honeycutt2020soliciting}
\bibfield{author}{\bibinfo{person}{Donald Honeycutt}, \bibinfo{person}{Mahsan Nourani}, {and} \bibinfo{person}{Eric Ragan}.} \bibinfo{year}{2020}\natexlab{}.
\newblock \showarticletitle{Soliciting Human-in-the-Loop User Feedback for Interactive Machine Learning Reduces User Trust and Impressions of Model Accuracy}.
\newblock \bibinfo{journal}{\emph{Proceedings of the AAAI Conference on Human Computation and Crowdsourcing}} \bibinfo{volume}{8}, \bibinfo{number}{1} (\bibinfo{date}{Oct.} \bibinfo{year}{2020}), \bibinfo{pages}{63–72}.
\newblock
\href{https://doi.org/10.1609/hcomp.v8i1.7464}{doi:\nolinkurl{10.1609/hcomp.v8i1.7464}}


\bibitem[Horvitz(1999)]%
        {horvitz1999principles}
\bibfield{author}{\bibinfo{person}{Eric Horvitz}.} \bibinfo{year}{1999}\natexlab{}.
\newblock \showarticletitle{Principles of mixed-initiative user interfaces}. In \bibinfo{booktitle}{\emph{Proceedings of the SIGCHI Conference on Human Factors in Computing Systems}}. ACM, \bibinfo{pages}{159--166}.
\newblock
\href{https://doi.org/10.1145/302979.303030}{doi:\nolinkurl{10.1145/302979.303030}}


\bibitem[Horvitz et~al\mbox{.}(1999)]%
        {horvitz1999continual}
\bibfield{author}{\bibinfo{person}{Eric Horvitz}, \bibinfo{person}{Tim Paek}, {and} \bibinfo{person}{Mani Subramani}.} \bibinfo{year}{1999}\natexlab{}.
\newblock \showarticletitle{Continual computation policies for utility-directed preference assessment}. In \bibinfo{booktitle}{\emph{Proceedings of the Fifteenth Conference on Uncertainty in Artificial Intelligence}}. Morgan Kaufmann Publishers Inc., \bibinfo{pages}{269--278}.
\newblock


\bibitem[Jannach et~al\mbox{.}(2021)]%
        {10.1145/3453154}
\bibfield{author}{\bibinfo{person}{Dietmar Jannach}, \bibinfo{person}{Ahtsham Manzoor}, \bibinfo{person}{Wanling Cai}, {and} \bibinfo{person}{Li Chen}.} \bibinfo{year}{2021}\natexlab{}.
\newblock \showarticletitle{A Survey on Conversational Recommender Systems}.
\newblock \bibinfo{journal}{\emph{ACM Comput. Surv.}} \bibinfo{volume}{54}, \bibinfo{number}{5}, Article \bibinfo{articleno}{105} (\bibinfo{date}{May} \bibinfo{year}{2021}), \bibinfo{numpages}{36}~pages.
\newblock
\showISSN{0360-0300}
\href{https://doi.org/10.1145/3453154}{doi:\nolinkurl{10.1145/3453154}}


\bibitem[Jiang et~al\mbox{.}(2025b)]%
        {10.1145/3747588}
\bibfield{author}{\bibinfo{person}{Juyong Jiang}, \bibinfo{person}{Fan Wang}, \bibinfo{person}{Jiasi Shen}, \bibinfo{person}{Sungju Kim}, {and} \bibinfo{person}{Sunghun Kim}.} \bibinfo{year}{2025}\natexlab{b}.
\newblock \showarticletitle{A Survey on Large Language Models for Code Generation}.
\newblock \bibinfo{journal}{\emph{ACM Trans. Softw. Eng. Methodol.}} (\bibinfo{date}{July} \bibinfo{year}{2025}).
\newblock
\showISSN{1049-331X}
\href{https://doi.org/10.1145/3747588}{doi:\nolinkurl{10.1145/3747588}}
\newblock
\shownote{Just Accepted}.


\bibitem[Jiang et~al\mbox{.}(2025a)]%
        {jiang2025onewayinfluencebidirectionalopinion}
\bibfield{author}{\bibinfo{person}{Yuyang Jiang}, \bibinfo{person}{Longjie Guo}, \bibinfo{person}{Yuchen Wu}, \bibinfo{person}{Aylin Caliskan}, \bibinfo{person}{Tanu Mitra}, {and} \bibinfo{person}{Hua Shen}.} \bibinfo{year}{2025}\natexlab{a}.
\newblock \bibinfo{title}{Beyond One-Way Influence: Bidirectional Opinion Dynamics in Multi-Turn Human-LLM Interactions}.
\newblock
\showeprint[arxiv]{2510.20039}~[cs.HC]
\urldef\tempurl%
\url{https://arxiv.org/abs/2510.20039}
\showURL{%
\tempurl}


\bibitem[Jin et~al\mbox{.}(2025)]%
        {jin2025erarealworldhumaninteraction}
\bibfield{author}{\bibinfo{person}{Chuanyang Jin}, \bibinfo{person}{Jing Xu}, \bibinfo{person}{Bo Liu}, \bibinfo{person}{Leitian Tao}, \bibinfo{person}{Olga Golovneva}, \bibinfo{person}{Tianmin Shu}, \bibinfo{person}{Wenting Zhao}, \bibinfo{person}{Xian Li}, {and} \bibinfo{person}{Jason Weston}.} \bibinfo{year}{2025}\natexlab{}.
\newblock \bibinfo{title}{The Era of Real-World Human Interaction: RL from User Conversations}.
\newblock
\showeprint[arxiv]{2509.25137}~[cs.AI]
\urldef\tempurl%
\url{https://arxiv.org/abs/2509.25137}
\showURL{%
\tempurl}


\bibitem[Joko et~al\mbox{.}(2025)]%
        {Joko2025WildClaimsIA}
\bibfield{author}{\bibinfo{person}{Hideaki Joko}, \bibinfo{person}{Shakiba Amirshahi}, \bibinfo{person}{Charles L.~A. Clarke}, {and} \bibinfo{person}{Faegheh Hasibi}.} \bibinfo{year}{2025}\natexlab{}.
\newblock \bibinfo{title}{WildClaims: Information Access Conversations in the Wild(Chat)}.
\newblock
\showeprint[arxiv]{2509.17442}~[cs.IR]
\urldef\tempurl%
\url{https://arxiv.org/abs/2509.17442}
\showURL{%
\tempurl}


\bibitem[Ju et~al\mbox{.}(2025)]%
        {ju2025collaborating}
\bibfield{author}{\bibinfo{person}{Wendy Ju} {et~al\mbox{.}}} \bibinfo{year}{2025}\natexlab{}.
\newblock \showarticletitle{Collaborating with AI Agents in the Field: A Study of Situated Human--AI Interaction}. In \bibinfo{booktitle}{\emph{Proceedings of the 2025 CHI Conference on Human Factors in Computing Systems}}. ACM.
\newblock


\bibitem[Kalai et~al\mbox{.}(2025)]%
        {kalai2025languagemodelshallucinate}
\bibfield{author}{\bibinfo{person}{Adam~Tauman Kalai}, \bibinfo{person}{Ofir Nachum}, \bibinfo{person}{Santosh~S. Vempala}, {and} \bibinfo{person}{Edwin Zhang}.} \bibinfo{year}{2025}\natexlab{}.
\newblock \bibinfo{title}{Why Language Models Hallucinate}.
\newblock
\showeprint[arxiv]{2509.04664}~[cs.CL]
\urldef\tempurl%
\url{https://arxiv.org/abs/2509.04664}
\showURL{%
\tempurl}


\bibitem[Kang et~al\mbox{.}(2024)]%
        {kang-etal-2024-large}
\bibfield{author}{\bibinfo{person}{Dongjin Kang}, \bibinfo{person}{Sunghwan Kim}, \bibinfo{person}{Taeyoon Kwon}, \bibinfo{person}{Seungjun Moon}, \bibinfo{person}{Hyunsouk Cho}, \bibinfo{person}{Youngjae Yu}, \bibinfo{person}{Dongha Lee}, {and} \bibinfo{person}{Jinyoung Yeo}.} \bibinfo{year}{2024}\natexlab{}.
\newblock \showarticletitle{Can Large Language Models be Good Emotional Supporter? Mitigating Preference Bias on Emotional Support Conversation}. In \bibinfo{booktitle}{\emph{Proceedings of the 62nd Annual Meeting of the Association for Computational Linguistics (Volume 1: Long Papers)}}, \bibfield{editor}{\bibinfo{person}{Lun-Wei Ku}, \bibinfo{person}{Andre Martins}, {and} \bibinfo{person}{Vivek Srikumar}} (Eds.). \bibinfo{publisher}{Association for Computational Linguistics}, \bibinfo{address}{Bangkok, Thailand}, \bibinfo{pages}{15232--15261}.
\newblock
\href{https://doi.org/10.18653/v1/2024.acl-long.813}{doi:\nolinkurl{10.18653/v1/2024.acl-long.813}}


\bibitem[Kim et~al\mbox{.}(2025)]%
        {DBLP:journals/corr/abs-2503-00858}
\bibfield{author}{\bibinfo{person}{Yoonsu Kim}, \bibinfo{person}{Brandon Chin}, \bibinfo{person}{Kihoon Son}, \bibinfo{person}{Seoyoung Kim}, {and} \bibinfo{person}{Juho Kim}.} \bibinfo{year}{2025}\natexlab{}.
\newblock \showarticletitle{Applying the Gricean Maxims to a Human-LLM Interaction Cycle: Design Insights from a Participatory Approach}.
\newblock \bibinfo{journal}{\emph{CoRR}}  \bibinfo{volume}{abs/2503.00858} (\bibinfo{date}{March} \bibinfo{year}{2025}).
\newblock
\urldef\tempurl%
\url{https://doi.org/10.48550/arXiv.2503.00858}
\showURL{%
\tempurl}


\bibitem[Kirk et~al\mbox{.}(2023)]%
        {kirk-etal-2023-past}
\bibfield{author}{\bibinfo{person}{Hannah~Rose Kirk}, \bibinfo{person}{Andrew~M. Bean}, \bibinfo{person}{Bertie Vidgen}, \bibinfo{person}{Paul R{\"o}ttger}, {and} \bibinfo{person}{Scott~A. Hale}.} \bibinfo{year}{2023}\natexlab{}.
\newblock \showarticletitle{The Past, Present and Better Future of Feedback Learning in Large Language Models for Subjective Human Preferences and Values}. In \bibinfo{booktitle}{\emph{Proceedings of the 2023 Conference on Empirical Methods in Natural Language Processing}}, \bibfield{editor}{\bibinfo{person}{Houda Bouamor}, \bibinfo{person}{Juan Pino}, {and} \bibinfo{person}{Kalika Bali}} (Eds.). \bibinfo{publisher}{Association for Computational Linguistics}, \bibinfo{address}{Singapore}, \bibinfo{pages}{2409--2430}.
\newblock
\href{https://doi.org/10.18653/v1/2023.emnlp-main.148}{doi:\nolinkurl{10.18653/v1/2023.emnlp-main.148}}


\bibitem[Kirk et~al\mbox{.}(2025)]%
        {kirk2025socioaffective}
\bibfield{author}{\bibinfo{person}{Hannah~Rose Kirk}, \bibinfo{person}{Iason Gabriel}, \bibinfo{person}{Chris Summerfield}, \bibinfo{person}{Bertie Vidgen}, {and} \bibinfo{person}{Scott~A. Hale}.} \bibinfo{year}{2025}\natexlab{}.
\newblock \showarticletitle{Why human--AI relationships need socioaffective alignment}.
\newblock \bibinfo{journal}{\emph{Humanities and Social Sciences Communications}}  \bibinfo{volume}{12} (\bibinfo{year}{2025}), \bibinfo{pages}{728}.
\newblock
\href{https://doi.org/10.1057/s41599-025-04532-5}{doi:\nolinkurl{10.1057/s41599-025-04532-5}}


\bibitem[Kirkeby-Hinrup and Stenseke(2025)]%
        {KirkebyHinrup_Stenseke_2025}
\bibfield{author}{\bibinfo{person}{Asger Kirkeby-Hinrup} {and} \bibinfo{person}{Jakob Stenseke}.} \bibinfo{year}{2025}\natexlab{}.
\newblock \showarticletitle{The psychology of LLM interactions: the uncanny valley and other minds}.
\newblock \bibinfo{journal}{\emph{Journal of Psychology and AI}} \bibinfo{volume}{1}, \bibinfo{number}{1} (\bibinfo{year}{2025}).
\newblock
\href{https://doi.org/10.1080/29974100.2025.2457627}{doi:\nolinkurl{10.1080/29974100.2025.2457627}}


\bibitem[Kirsh and Maglio(1994)]%
        {kirsh1994distinguishing}
\bibfield{author}{\bibinfo{person}{David Kirsh} {and} \bibinfo{person}{Paul Maglio}.} \bibinfo{year}{1994}\natexlab{}.
\newblock \showarticletitle{On distinguishing epistemic from pragmatic action}.
\newblock \bibinfo{journal}{\emph{Cognitive Science}} \bibinfo{volume}{18}, \bibinfo{number}{4} (\bibinfo{year}{1994}), \bibinfo{pages}{513--549}.
\newblock


\bibitem[Klein(2025)]%
        {Klein2025SocialCuesMetaAnalysis}
\bibfield{author}{\bibinfo{person}{S.~H. Klein}.} \bibinfo{year}{2025}\natexlab{}.
\newblock \showarticletitle{The effects of human-like social cues on social responses towards text-based conversational agents—a meta-analysis}.
\newblock \bibinfo{journal}{\emph{Humanities and Social Sciences Communications}}  \bibinfo{volume}{12} (\bibinfo{year}{2025}), \bibinfo{pages}{1322}.
\newblock
\href{https://doi.org/10.1057/s41599-025-05618-w}{doi:\nolinkurl{10.1057/s41599-025-05618-w}}


\bibitem[Kluger and DeNisi(1996)]%
        {kluger1996effect}
\bibfield{author}{\bibinfo{person}{Avraham~N Kluger} {and} \bibinfo{person}{Angelo DeNisi}.} \bibinfo{year}{1996}\natexlab{}.
\newblock \showarticletitle{The effect of feedback interventions on performance: A historical review, a meta-analysis, and a preliminary feedback intervention theory}.
\newblock \bibinfo{journal}{\emph{Psychological Bulletin}} \bibinfo{volume}{119}, \bibinfo{number}{2} (\bibinfo{year}{1996}), \bibinfo{pages}{254--284}.
\newblock


\bibitem[Kontogiorgos and Shah(2025)]%
        {10.5555/3721488.3721577}
\bibfield{author}{\bibinfo{person}{Dimosthenis Kontogiorgos} {and} \bibinfo{person}{Julie Shah}.} \bibinfo{year}{2025}\natexlab{}.
\newblock \showarticletitle{Questioning the Robot: Using Human Non-verbal Cues to Estimate the Need for Explanations}. In \bibinfo{booktitle}{\emph{Proceedings of the 2025 ACM/IEEE International Conference on Human-Robot Interaction}} (Melbourne, Australia) \emph{(\bibinfo{series}{HRI '25})}. \bibinfo{publisher}{IEEE Press}, \bibinfo{pages}{717–728}.
\newblock


\bibitem[Krause and Vossen(2024)]%
        {krause-vossen-2024-gricean-maxims}
\bibfield{author}{\bibinfo{person}{Lea Krause} {and} \bibinfo{person}{Piek~T.J.M. Vossen}.} \bibinfo{year}{2024}\natexlab{}.
\newblock \showarticletitle{The {G}ricean Maxims in {NLP} - A Survey}. In \bibinfo{booktitle}{\emph{Proceedings of the 17th International Natural Language Generation Conference}}, \bibfield{editor}{\bibinfo{person}{Saad Mahamood}, \bibinfo{person}{Nguyen~Le Minh}, {and} \bibinfo{person}{Daphne Ippolito}} (Eds.). \bibinfo{publisher}{Association for Computational Linguistics}, \bibinfo{address}{Tokyo, Japan}, \bibinfo{pages}{470--485}.
\newblock
\href{https://doi.org/10.18653/v1/2024.inlg-main.39}{doi:\nolinkurl{10.18653/v1/2024.inlg-main.39}}


\bibitem[Kulesza et~al\mbox{.}(2015)]%
        {10.1145/2678025.2701399}
\bibfield{author}{\bibinfo{person}{Todd Kulesza}, \bibinfo{person}{Margaret Burnett}, \bibinfo{person}{Weng-Keen Wong}, {and} \bibinfo{person}{Simone Stumpf}.} \bibinfo{year}{2015}\natexlab{}.
\newblock \showarticletitle{Principles of Explanatory Debugging to Personalize Interactive Machine Learning}. In \bibinfo{booktitle}{\emph{Proceedings of the 20th International Conference on Intelligent User Interfaces}} (Atlanta, Georgia, USA) \emph{(\bibinfo{series}{IUI '15})}. \bibinfo{publisher}{Association for Computing Machinery}, \bibinfo{address}{New York, NY, USA}, \bibinfo{pages}{126–137}.
\newblock
\showISBNx{9781450333061}
\href{https://doi.org/10.1145/2678025.2701399}{doi:\nolinkurl{10.1145/2678025.2701399}}


\bibitem[Kulesza et~al\mbox{.}(2013)]%
        {6645235}
\bibfield{author}{\bibinfo{person}{Todd Kulesza}, \bibinfo{person}{Simone Stumpf}, \bibinfo{person}{Margaret Burnett}, \bibinfo{person}{Sherry Yang}, \bibinfo{person}{Irwin Kwan}, {and} \bibinfo{person}{Weng-Keen Wong}.} \bibinfo{year}{2013}\natexlab{}.
\newblock \showarticletitle{Too much, too little, or just right? Ways explanations impact end users' mental models}. In \bibinfo{booktitle}{\emph{2013 IEEE Symposium on Visual Languages and Human Centric Computing}}. \bibinfo{pages}{3--10}.
\newblock
\href{https://doi.org/10.1109/VLHCC.2013.6645235}{doi:\nolinkurl{10.1109/VLHCC.2013.6645235}}


\bibitem[Kurniawan et~al\mbox{.}(2023)]%
        {Kurniawan2022PathOfLeastResistance}
\bibfield{author}{\bibinfo{person}{Indah Kurniawan}, \bibinfo{person}{Alexander Soutschek}, {and} \bibinfo{person}{Matthew Apps}.} \bibinfo{year}{2023}\natexlab{}.
\newblock \showarticletitle{Taking the path of least resistance now, but not later: Pushing cognitive effort into the future reduces effort discounting}.
\newblock \bibinfo{journal}{\emph{Psychonomic Bulletin \& Review}} \bibinfo{volume}{30}, \bibinfo{number}{1} (\bibinfo{year}{2023}), \bibinfo{pages}{1--15}.
\newblock
\href{https://doi.org/10.3758/s13423-022-02198-7}{doi:\nolinkurl{10.3758/s13423-022-02198-7}}


\bibitem[Laban et~al\mbox{.}(2025)]%
        {laban2025llmslostmultiturnconversation}
\bibfield{author}{\bibinfo{person}{Philippe Laban}, \bibinfo{person}{Hiroaki Hayashi}, \bibinfo{person}{Yingbo Zhou}, {and} \bibinfo{person}{Jennifer Neville}.} \bibinfo{year}{2025}\natexlab{}.
\newblock \bibinfo{title}{LLMs Get Lost In Multi-Turn Conversation}.
\newblock
\showeprint[arxiv]{2505.06120}~[cs.CL]
\urldef\tempurl%
\url{https://arxiv.org/abs/2505.06120}
\showURL{%
\tempurl}


\bibitem[Lai et~al\mbox{.}(2023)]%
        {10.1145/3610206}
\bibfield{author}{\bibinfo{person}{Vivian Lai}, \bibinfo{person}{Yiming Zhang}, \bibinfo{person}{Chacha Chen}, \bibinfo{person}{Q.~Vera Liao}, {and} \bibinfo{person}{Chenhao Tan}.} \bibinfo{year}{2023}\natexlab{}.
\newblock \showarticletitle{Selective Explanations: Leveraging Human Input to Align Explainable AI}.
\newblock \bibinfo{journal}{\emph{Proc. ACM Hum.-Comput. Interact.}} \bibinfo{volume}{7}, \bibinfo{number}{CSCW2}, Article \bibinfo{articleno}{357} (\bibinfo{date}{Oct.} \bibinfo{year}{2023}), \bibinfo{numpages}{35}~pages.
\newblock
\href{https://doi.org/10.1145/3610206}{doi:\nolinkurl{10.1145/3610206}}


\bibitem[Lee et~al\mbox{.}(2025a)]%
        {10.1145/3746058.3758376}
\bibfield{author}{\bibinfo{person}{Daniel Lee}, \bibinfo{person}{Nikhil Sharma}, \bibinfo{person}{Donghoon Shin}, \bibinfo{person}{DaEun Choi}, \bibinfo{person}{Harsh Sharma}, \bibinfo{person}{Jeonghwan Kim}, {and} \bibinfo{person}{Heng Ji}.} \bibinfo{year}{2025}\natexlab{a}.
\newblock \showarticletitle{ThematicPlane: Bridging Tacit User Intent and Latent Spaces for Image Generation}. In \bibinfo{booktitle}{\emph{Adjunct Proceedings of the 38th Annual ACM Symposium on User Interface Software and Technology}} \emph{(\bibinfo{series}{UIST Adjunct '25})}. \bibinfo{publisher}{Association for Computing Machinery}, \bibinfo{address}{New York, NY, USA}, Article \bibinfo{articleno}{120}, \bibinfo{numpages}{3}~pages.
\newblock
\showISBNx{9798400720369}
\href{https://doi.org/10.1145/3746058.3758376}{doi:\nolinkurl{10.1145/3746058.3758376}}


\bibitem[Lee et~al\mbox{.}(2024)]%
        {lee2023rlaif}
\bibfield{author}{\bibinfo{person}{Harrison Lee}, \bibinfo{person}{Samrat Phatale}, \bibinfo{person}{Hassan Mansoor}, \bibinfo{person}{Thomas Mesnard}, \bibinfo{person}{Johan Ferret}, \bibinfo{person}{Kellie Lu}, \bibinfo{person}{Colton Bishop}, \bibinfo{person}{Ethan Hall}, \bibinfo{person}{Victor Carbune}, \bibinfo{person}{Abhinav Rastogi}, {and} \bibinfo{person}{Sushant Prakash}.} \bibinfo{year}{2024}\natexlab{}.
\newblock \showarticletitle{RLAIF vs. RLHF: scaling reinforcement learning from human feedback with AI feedback}. In \bibinfo{booktitle}{\emph{Proceedings of the 41st International Conference on Machine Learning}} (Vienna, Austria) \emph{(\bibinfo{series}{ICML'24})}. \bibinfo{publisher}{JMLR.org}, Article \bibinfo{articleno}{1071}, \bibinfo{numpages}{28}~pages.
\newblock


\bibitem[Lee and See(2004)]%
        {lee2004trust}
\bibfield{author}{\bibinfo{person}{John~D Lee} {and} \bibinfo{person}{Kelly~A See}.} \bibinfo{year}{2004}\natexlab{}.
\newblock \showarticletitle{Trust in automation: designing for appropriate reliance}.
\newblock \bibinfo{journal}{\emph{Human Factors}} \bibinfo{volume}{46}, \bibinfo{number}{1} (\bibinfo{year}{2004}), \bibinfo{pages}{50--80}.
\newblock


\bibitem[Lee et~al\mbox{.}(2023)]%
        {Lee2022EvaluatingHM}
\bibfield{author}{\bibinfo{person}{Mina Lee}, \bibinfo{person}{Megha Srivastava}, \bibinfo{person}{Amelia Hardy}, \bibinfo{person}{John Thickstun}, \bibinfo{person}{Esin Durmus}, \bibinfo{person}{Ashwin Paranjape}, \bibinfo{person}{Ines Gerard-Ursin}, \bibinfo{person}{Xiang~Lisa Li}, \bibinfo{person}{Faisal Ladhak}, \bibinfo{person}{Frieda Rong}, \bibinfo{person}{Rose~E Wang}, \bibinfo{person}{Minae Kwon}, \bibinfo{person}{Joon~Sung Park}, \bibinfo{person}{Hancheng Cao}, \bibinfo{person}{Tony Lee}, \bibinfo{person}{Rishi Bommasani}, \bibinfo{person}{Michael~S. Bernstein}, {and} \bibinfo{person}{Percy Liang}.} \bibinfo{year}{2023}\natexlab{}.
\newblock \showarticletitle{Evaluating Human-Language Model Interaction}.
\newblock \bibinfo{journal}{\emph{Transactions on Machine Learning Research}} (\bibinfo{year}{2023}).
\newblock
\showISSN{2835-8856}
\urldef\tempurl%
\url{https://openreview.net/forum?id=hjDYJUn9l1}
\showURL{%
\tempurl}


\bibitem[Lee et~al\mbox{.}(2025b)]%
        {Lee2025FaithfulSelfExplanation}
\bibfield{author}{\bibinfo{person}{S. Lee}, \bibinfo{person}{X. Wang}, \bibinfo{person}{A. Zhang}, {et~al\mbox{.}}} \bibinfo{year}{2025}\natexlab{b}.
\newblock \showarticletitle{Toward faithful and human-aligned self-explanation of deep models}.
\newblock \bibinfo{journal}{\emph{npj Artificial Intelligence}}  \bibinfo{volume}{1} (\bibinfo{year}{2025}), \bibinfo{pages}{21}.
\newblock
\href{https://doi.org/10.1038/s44387-025-00023-9}{doi:\nolinkurl{10.1038/s44387-025-00023-9}}


\bibitem[Lewis et~al\mbox{.}(2020)]%
        {rag}
\bibfield{author}{\bibinfo{person}{Patrick Lewis}, \bibinfo{person}{Ethan Perez}, \bibinfo{person}{Aleksandra Piktus}, \bibinfo{person}{Fabio Petroni}, \bibinfo{person}{Vladimir Karpukhin}, \bibinfo{person}{Naman Goyal}, \bibinfo{person}{Heinrich K\"{u}ttler}, \bibinfo{person}{Mike Lewis}, \bibinfo{person}{Wen-tau Yih}, \bibinfo{person}{Tim Rockt\"{a}schel}, \bibinfo{person}{Sebastian Riedel}, {and} \bibinfo{person}{Douwe Kiela}.} \bibinfo{year}{2020}\natexlab{}.
\newblock \showarticletitle{Retrieval-augmented generation for knowledge-intensive NLP tasks}. In \bibinfo{booktitle}{\emph{Proceedings of the 34th International Conference on Neural Information Processing Systems}} (Vancouver, BC, Canada) \emph{(\bibinfo{series}{NIPS '20})}. \bibinfo{publisher}{Curran Associates Inc.}, \bibinfo{address}{Red Hook, NY, USA}, Article \bibinfo{articleno}{793}, \bibinfo{numpages}{16}~pages.
\newblock
\showISBNx{9781713829546}


\bibitem[Li et~al\mbox{.}(2025)]%
        {li2025we}
\bibfield{author}{\bibinfo{person}{Jingshu Li}, \bibinfo{person}{Tianqi Song}, \bibinfo{person}{Beichen Xue}, {and} \bibinfo{person}{Yi-Chieh Lee}.} \bibinfo{year}{2025}\natexlab{}.
\newblock \showarticletitle{We Shape {AI}, and Thereafter {AI} Shape Us: Humans Align with {AI} through Social Influences}. In \bibinfo{booktitle}{\emph{ICLR 2025 Workshop on Bidirectional Human-AI Alignment}}.
\newblock
\urldef\tempurl%
\url{https://openreview.net/forum?id=64rCWVC78p}
\showURL{%
\tempurl}


\bibitem[Lin et~al\mbox{.}({[n.\,d.]})]%
        {lin2023beyond}
\bibfield{author}{\bibinfo{person}{Zhiyu Lin}, \bibinfo{person}{Upol Ehsan}, \bibinfo{person}{Rohan Agarwal}, \bibinfo{person}{Samihan Dani}, \bibinfo{person}{Vidushi Vashishth}, {and} \bibinfo{person}{Mark~O. Riedl}.} \bibinfo{year}{[n.\,d.]}\natexlab{}.
\newblock \showarticletitle{Beyond Prompts: Exploring the Design Space of Mixed-Initiative Co-Creativity Systems}.
\newblock \bibinfo{journal}{\emph{International Conference on Computational Creativity}} (\bibinfo{year}{[n.\,d.]}).
\newblock
\urldef\tempurl%
\url{https://par.nsf.gov/biblio/10434407}
\showURL{%
\tempurl}


\bibitem[Lin et~al\mbox{.}(2024)]%
        {lin2024criticbench}
\bibfield{author}{\bibinfo{person}{Zicheng Lin}, \bibinfo{person}{Zhibin Gou}, \bibinfo{person}{Tian Liang}, \bibinfo{person}{Ruilin Luo}, \bibinfo{person}{Haowei Liu}, {and} \bibinfo{person}{Yujiu Yang}.} \bibinfo{year}{2024}\natexlab{}.
\newblock \showarticletitle{{C}ritic{B}ench: Benchmarking {LLM}s for Critique-Correct Reasoning}. In \bibinfo{booktitle}{\emph{Findings of the Association for Computational Linguistics: ACL 2024}}, \bibfield{editor}{\bibinfo{person}{Lun-Wei Ku}, \bibinfo{person}{Andre Martins}, {and} \bibinfo{person}{Vivek Srikumar}} (Eds.). \bibinfo{publisher}{Association for Computational Linguistics}, \bibinfo{address}{Bangkok, Thailand}, \bibinfo{pages}{1552--1587}.
\newblock
\href{https://doi.org/10.18653/v1/2024.findings-acl.91}{doi:\nolinkurl{10.18653/v1/2024.findings-acl.91}}


\bibitem[Liu et~al\mbox{.}(2025)]%
        {liu2025comprehensivesurveylongcontext}
\bibfield{author}{\bibinfo{person}{Jiaheng Liu}, \bibinfo{person}{Dawei Zhu}, \bibinfo{person}{Zhiqi Bai}, \bibinfo{person}{Yancheng He}, \bibinfo{person}{Huanxuan Liao}, \bibinfo{person}{Haoran Que}, \bibinfo{person}{Zekun Wang}, \bibinfo{person}{Chenchen Zhang}, \bibinfo{person}{Ge Zhang}, \bibinfo{person}{Jiebin Zhang}, \bibinfo{person}{Yuanxing Zhang}, \bibinfo{person}{Zhuo Chen}, \bibinfo{person}{Hangyu Guo}, \bibinfo{person}{Shilong Li}, \bibinfo{person}{Ziqiang Liu}, \bibinfo{person}{Yong Shan}, \bibinfo{person}{Yifan Song}, \bibinfo{person}{Jiayi Tian}, \bibinfo{person}{Wenhao Wu}, \bibinfo{person}{Zhejian Zhou}, \bibinfo{person}{Ruijie Zhu}, \bibinfo{person}{Junlan Feng}, \bibinfo{person}{Yang Gao}, \bibinfo{person}{Shizhu He}, \bibinfo{person}{Zhoujun Li}, \bibinfo{person}{Tianyu Liu}, \bibinfo{person}{Fanyu Meng}, \bibinfo{person}{Wenbo Su}, \bibinfo{person}{Yingshui Tan}, \bibinfo{person}{Zili Wang}, \bibinfo{person}{Jian Yang}, \bibinfo{person}{Wei Ye}, \bibinfo{person}{Bo Zheng},
  \bibinfo{person}{Wangchunshu Zhou}, \bibinfo{person}{Wenhao Huang}, \bibinfo{person}{Sujian Li}, {and} \bibinfo{person}{Zhaoxiang Zhang}.} \bibinfo{year}{2025}\natexlab{}.
\newblock \bibinfo{title}{A Comprehensive Survey on Long Context Language Modeling}.
\newblock
\showeprint[arxiv]{2503.17407}~[cs.CL]
\urldef\tempurl%
\url{https://arxiv.org/abs/2503.17407}
\showURL{%
\tempurl}


\bibitem[Liu et~al\mbox{.}(2024)]%
        {liu-etal-2024-lost}
\bibfield{author}{\bibinfo{person}{Nelson~F. Liu}, \bibinfo{person}{Kevin Lin}, \bibinfo{person}{John Hewitt}, \bibinfo{person}{Ashwin Paranjape}, \bibinfo{person}{Michele Bevilacqua}, \bibinfo{person}{Fabio Petroni}, {and} \bibinfo{person}{Percy Liang}.} \bibinfo{year}{2024}\natexlab{}.
\newblock \showarticletitle{Lost in the Middle: How Language Models Use Long Contexts}.
\newblock \bibinfo{journal}{\emph{Transactions of the Association for Computational Linguistics}}  \bibinfo{volume}{12} (\bibinfo{year}{2024}), \bibinfo{pages}{157--173}.
\newblock
\href{https://doi.org/10.1162/tacl_a_00638}{doi:\nolinkurl{10.1162/tacl_a_00638}}


\bibitem[Locke and Latham(2013)]%
        {locke2013new}
\bibfield{author}{\bibinfo{person}{Edwin~A Locke} {and} \bibinfo{person}{Gary~P Latham}.} \bibinfo{year}{2013}\natexlab{}.
\newblock \bibinfo{booktitle}{\emph{New developments in goal setting and task performance}}. Vol.~\bibinfo{volume}{24}.
\newblock \bibinfo{publisher}{Routledge New York}.
\newblock


\bibitem[Miehling et~al\mbox{.}(2024)]%
        {miehling2024languagemodelsdialogueconversational}
\bibfield{author}{\bibinfo{person}{Erik Miehling}, \bibinfo{person}{Manish Nagireddy}, \bibinfo{person}{Prasanna Sattigeri}, \bibinfo{person}{Elizabeth~M. Daly}, \bibinfo{person}{David Piorkowski}, {and} \bibinfo{person}{John~T. Richards}.} \bibinfo{year}{2024}\natexlab{}.
\newblock \showarticletitle{Language Models in Dialogue: Conversational Maxims for Human-{AI} Interactions}. In \bibinfo{booktitle}{\emph{Findings of the Association for Computational Linguistics: EMNLP 2024}}, \bibfield{editor}{\bibinfo{person}{Yaser Al-Onaizan}, \bibinfo{person}{Mohit Bansal}, {and} \bibinfo{person}{Yun-Nung Chen}} (Eds.). \bibinfo{publisher}{Association for Computational Linguistics}, \bibinfo{address}{Miami, Florida, USA}, \bibinfo{pages}{14420--14437}.
\newblock
\href{https://doi.org/10.18653/v1/2024.findings-emnlp.843}{doi:\nolinkurl{10.18653/v1/2024.findings-emnlp.843}}


\bibitem[Miller(2019)]%
        {miller2019explanation}
\bibfield{author}{\bibinfo{person}{Tim Miller}.} \bibinfo{year}{2019}\natexlab{}.
\newblock \showarticletitle{Explanation in artificial intelligence: Insights from the social sciences}.
\newblock \bibinfo{journal}{\emph{Artificial Intelligence}}  \bibinfo{volume}{267} (\bibinfo{year}{2019}), \bibinfo{pages}{1--38}.
\newblock


\bibitem[Min et~al\mbox{.}(2025)]%
        {Min_2025}
\bibfield{author}{\bibinfo{person}{Taywon Min}, \bibinfo{person}{Haeone Lee}, \bibinfo{person}{Yongchan Kwon}, {and} \bibinfo{person}{Kimin Lee}.} \bibinfo{year}{2025}\natexlab{}.
\newblock \showarticletitle{Understanding Impact of Human Feedback via Influence Functions}. In \bibinfo{booktitle}{\emph{Proceedings of the 63rd Annual Meeting of the Association for Computational Linguistics (Volume 1: Long Papers)}}. \bibinfo{publisher}{Association for Computational Linguistics}, \bibinfo{pages}{27471–27500}.
\newblock
\href{https://doi.org/10.18653/v1/2025.acl-long.1333}{doi:\nolinkurl{10.18653/v1/2025.acl-long.1333}}


\bibitem[Mosqueira-Rey et~al\mbox{.}(2023)]%
        {MosqueiraRey2023HITL}
\bibfield{author}{\bibinfo{person}{Eduardo Mosqueira-Rey}, \bibinfo{person}{Elena Hern{\'a}ndez-Pereira}, \bibinfo{person}{David Alonso-R{\'i}os}, \bibinfo{person}{Jos{\'e} Bobes-Bascar{\'a}n}, {and} \bibinfo{person}{{\'A}ngel Fern{\'a}ndez-Leal}.} \bibinfo{year}{2023}\natexlab{}.
\newblock \showarticletitle{Human-in-the-loop machine learning: a state of the art}.
\newblock \bibinfo{journal}{\emph{Artificial Intelligence Review}}  \bibinfo{volume}{56} (\bibinfo{year}{2023}), \bibinfo{pages}{3005--3054}.
\newblock
\href{https://doi.org/10.1007/s10462-022-10246-w}{doi:\nolinkurl{10.1007/s10462-022-10246-w}}


\bibitem[Mysore et~al\mbox{.}(2025)]%
        {mysore-etal-2025-prototypical}
\bibfield{author}{\bibinfo{person}{Sheshera Mysore}, \bibinfo{person}{Debarati Das}, \bibinfo{person}{Hancheng Cao}, {and} \bibinfo{person}{Bahareh Sarrafzadeh}.} \bibinfo{year}{2025}\natexlab{}.
\newblock \showarticletitle{Prototypical Human-{AI} Collaboration Behaviors from {LLM}-Assisted Writing in the Wild}. In \bibinfo{booktitle}{\emph{Proceedings of the 2025 Conference on Empirical Methods in Natural Language Processing}}, \bibfield{editor}{\bibinfo{person}{Christos Christodoulopoulos}, \bibinfo{person}{Tanmoy Chakraborty}, \bibinfo{person}{Carolyn Rose}, {and} \bibinfo{person}{Violet Peng}} (Eds.). \bibinfo{publisher}{Association for Computational Linguistics}, \bibinfo{address}{Suzhou, China}, \bibinfo{pages}{16819--16846}.
\newblock
\showISBNx{979-8-89176-332-6}
\href{https://doi.org/10.18653/v1/2025.emnlp-main.852}{doi:\nolinkurl{10.18653/v1/2025.emnlp-main.852}}


\bibitem[Nass et~al\mbox{.}(1994)]%
        {nass1994computers}
\bibfield{author}{\bibinfo{person}{Clifford Nass}, \bibinfo{person}{Jonathan Steuer}, {and} \bibinfo{person}{Ellen~R Tauber}.} \bibinfo{year}{1994}\natexlab{}.
\newblock \showarticletitle{Computers are social actors}. In \bibinfo{booktitle}{\emph{Proceedings of the SIGCHI conference on Human factors in computing systems}}. ACM, \bibinfo{pages}{72--78}.
\newblock


\bibitem[Nicol and Macfarlane‐Dick(2006)]%
        {Nicol01042006}
\bibfield{author}{\bibinfo{person}{David~J. Nicol} {and} \bibinfo{person}{Debra Macfarlane‐Dick}.} \bibinfo{year}{2006}\natexlab{}.
\newblock \showarticletitle{Formative assessment and self‐regulated learning: a model and seven principles of good feedback practice}.
\newblock \bibinfo{journal}{\emph{Studies in Higher Education}} \bibinfo{volume}{31}, \bibinfo{number}{2} (\bibinfo{year}{2006}), \bibinfo{pages}{199–218}.
\newblock
\href{https://doi.org/10.1080/03075070600572090}{doi:\nolinkurl{10.1080/03075070600572090}}


\bibitem[Niu et~al\mbox{.}(2024)]%
        {niu2023ragtruth}
\bibfield{author}{\bibinfo{person}{Cheng Niu}, \bibinfo{person}{Yuanhao Wu}, \bibinfo{person}{Juno Zhu}, \bibinfo{person}{Siliang Xu}, \bibinfo{person}{KaShun Shum}, \bibinfo{person}{Randy Zhong}, \bibinfo{person}{Juntong Song}, {and} \bibinfo{person}{Tong Zhang}.} \bibinfo{year}{2024}\natexlab{}.
\newblock \showarticletitle{{RAGT}ruth: A Hallucination Corpus for Developing Trustworthy Retrieval-Augmented Language Models}. In \bibinfo{booktitle}{\emph{Proceedings of the 62nd Annual Meeting of the Association for Computational Linguistics (Volume 1: Long Papers)}}, \bibfield{editor}{\bibinfo{person}{Lun-Wei Ku}, \bibinfo{person}{Andre Martins}, {and} \bibinfo{person}{Vivek Srikumar}} (Eds.). \bibinfo{publisher}{Association for Computational Linguistics}, \bibinfo{address}{Bangkok, Thailand}, \bibinfo{pages}{10862--10878}.
\newblock
\href{https://doi.org/10.18653/v1/2024.acl-long.585}{doi:\nolinkurl{10.18653/v1/2024.acl-long.585}}


\bibitem[Norman(1993)]%
        {norman1993things}
\bibfield{author}{\bibinfo{person}{Donald~A Norman}.} \bibinfo{year}{1993}\natexlab{}.
\newblock \bibinfo{booktitle}{\emph{Things that make us smart: Defending human attributes in the age of the machine}}.
\newblock \bibinfo{publisher}{Basic Books}.
\newblock


\bibitem[OpenAI(2022)]%
        {chatgpt}
\bibfield{author}{\bibinfo{person}{OpenAI}.} \bibinfo{year}{2022}\natexlab{}.
\newblock \bibinfo{title}{ChatGPT}.
\newblock
\urldef\tempurl%
\url{https://openai.com/blog/chatgpt}
\showURL{%
\tempurl}


\bibitem[Ou et~al\mbox{.}(2022)]%
        {ou2022artemis}
\bibfield{author}{\bibinfo{person}{Jianzhe Ou}, \bibinfo{person}{Krzysztof~Z Gajos}, \bibinfo{person}{Wojciech Matusik}, \bibinfo{person}{Wilmot Li}, {and} \bibinfo{person}{Maneesh Agrawala}.} \bibinfo{year}{2022}\natexlab{}.
\newblock \showarticletitle{ARTEMIS: A Human-AI Loop for Creative 3D Modeling}. In \bibinfo{booktitle}{\emph{Proceedings of the 35th Annual ACM Symposium on User Interface Software and Technology (UIST '22)}}. \bibinfo{pages}{75--88}.
\newblock
\href{https://doi.org/10.1145/3526113.3545685}{doi:\nolinkurl{10.1145/3526113.3545685}}


\bibitem[Ouyang et~al\mbox{.}(2022)]%
        {ouyang2022training}
\bibfield{author}{\bibinfo{person}{Long Ouyang}, \bibinfo{person}{Jeff Wu}, \bibinfo{person}{Xu Jiang}, \bibinfo{person}{Diogo Almeida}, \bibinfo{person}{Carroll~L Wainwright}, \bibinfo{person}{Pamela Mishkin}, \bibinfo{person}{Chong Zhang}, \bibinfo{person}{Sandhini Agarwal}, \bibinfo{person}{Katarina Slama}, \bibinfo{person}{Alex Ray}, {et~al\mbox{.}}} \bibinfo{year}{2022}\natexlab{}.
\newblock \showarticletitle{{Training Language Models to Follow Instructions with Human Feedback}}. In \bibinfo{booktitle}{\emph{Advances in Neural Information Processing Systems {NeurIPS}}}.
\newblock
\urldef\tempurl%
\url{https://arxiv.org/abs/2203.02155}
\showURL{%
\tempurl}


\bibitem[Panfili et~al\mbox{.}(2021)]%
        {Panfili_2021}
\bibfield{author}{\bibinfo{person}{Laura Panfili}, \bibinfo{person}{Steve Duman}, \bibinfo{person}{Andrew Nave}, \bibinfo{person}{Katherine~Phelps Ridgeway}, \bibinfo{person}{Nathan Eversole}, {and} \bibinfo{person}{Ruhi Sarikaya}.} \bibinfo{year}{2021}\natexlab{}.
\newblock \showarticletitle{Human-AI interactions through a Gricean lens}.
\newblock \bibinfo{journal}{\emph{Proceedings of the Linguistic Society of America}} \bibinfo{volume}{6}, \bibinfo{number}{1} (\bibinfo{date}{March} \bibinfo{year}{2021}), \bibinfo{pages}{288}.
\newblock
\showISSN{2473-8689}
\href{https://doi.org/10.3765/plsa.v6i1.4971}{doi:\nolinkurl{10.3765/plsa.v6i1.4971}}


\bibitem[Park et~al\mbox{.}(2024a)]%
        {park2024pragmatic}
\bibfield{author}{\bibinfo{person}{Dojun Park}, \bibinfo{person}{Jiwoo Lee}, \bibinfo{person}{Hyeyun Jeong}, \bibinfo{person}{Seohyun Park}, {and} \bibinfo{person}{Sungeun Lee}.} \bibinfo{year}{2024}\natexlab{a}.
\newblock \showarticletitle{Pragmatic Competence Evaluation of Large Language Models for the {K}orean Language}. In \bibinfo{booktitle}{\emph{Proceedings of the 38th Pacific Asia Conference on Language, Information and Computation}}, \bibfield{editor}{\bibinfo{person}{Nathaniel Oco}, \bibinfo{person}{Shirley~N. Dita}, \bibinfo{person}{Ariane~Macalinga Borlongan}, {and} \bibinfo{person}{Jong-Bok Kim}} (Eds.). \bibinfo{publisher}{Tokyo University of Foreign Studies}, \bibinfo{address}{Tokyo, Japan}, \bibinfo{pages}{256--266}.
\newblock
\urldef\tempurl%
\url{https://aclanthology.org/2024.paclic-1.25/}
\showURL{%
\tempurl}


\bibitem[Park et~al\mbox{.}(2024b)]%
        {park-etal-2024-disentangling}
\bibfield{author}{\bibinfo{person}{Ryan Park}, \bibinfo{person}{Rafael Rafailov}, \bibinfo{person}{Stefano Ermon}, {and} \bibinfo{person}{Chelsea Finn}.} \bibinfo{year}{2024}\natexlab{b}.
\newblock \showarticletitle{Disentangling Length from Quality in Direct Preference Optimization}. In \bibinfo{booktitle}{\emph{Findings of the Association for Computational Linguistics: ACL 2024}}, \bibfield{editor}{\bibinfo{person}{Lun-Wei Ku}, \bibinfo{person}{Andre Martins}, {and} \bibinfo{person}{Vivek Srikumar}} (Eds.). \bibinfo{publisher}{Association for Computational Linguistics}, \bibinfo{address}{Bangkok, Thailand}, \bibinfo{pages}{4998--5017}.
\newblock
\href{https://doi.org/10.18653/v1/2024.findings-acl.297}{doi:\nolinkurl{10.18653/v1/2024.findings-acl.297}}


\bibitem[Parshakov et~al\mbox{.}(2025)]%
        {parshakov2025usersfavorllmgeneratedcontent}
\bibfield{author}{\bibinfo{person}{Petr Parshakov}, \bibinfo{person}{Iuliia Naidenova}, \bibinfo{person}{Sofia Paklina}, \bibinfo{person}{Nikita Matkin}, {and} \bibinfo{person}{Cornel Nesseler}.} \bibinfo{year}{2025}\natexlab{}.
\newblock \bibinfo{booktitle}{\emph{Users Favor LLM-Generated Content -- Until They Know It's AI}}.
\newblock \bibinfo{type}{Papers} 2503.16458. \bibinfo{institution}{arXiv.org}.
\newblock
\href{https://doi.org/None}{doi:\nolinkurl{None}}


\bibitem[Peter et~al\mbox{.}(2025)]%
        {peter2025anthropomorphic}
\bibfield{author}{\bibinfo{person}{Sandra Peter}, \bibinfo{person}{Kai Riemer}, {and} \bibinfo{person}{Jevin~D. West}.} \bibinfo{year}{2025}\natexlab{}.
\newblock \showarticletitle{The benefits and dangers of anthropomorphic conversational agents}.
\newblock \bibinfo{journal}{\emph{Proceedings of the National Academy of Sciences of the United States of America}} \bibinfo{volume}{122}, \bibinfo{number}{22} (\bibinfo{year}{2025}), \bibinfo{pages}{e2415898122}.
\newblock
\href{https://doi.org/10.1073/pnas.2415898122}{doi:\nolinkurl{10.1073/pnas.2415898122}}


\bibitem[Pfeuffer et~al\mbox{.}(2023)]%
        {Pfeuffer2023XILADR}
\bibfield{author}{\bibinfo{person}{Nicolas Pfeuffer}, \bibinfo{person}{Lorenz Baum}, \bibinfo{person}{Wolfgang Stammer}, \bibinfo{person}{Benjamin~M. Abdel-Karim}, \bibinfo{person}{Patrick Schramowski}, \bibinfo{person}{Andreas~M. Bucher}, \bibinfo{person}{Christian H{\"u}gel}, \bibinfo{person}{Gernot Rohde}, \bibinfo{person}{Kristian Kersting}, {and} \bibinfo{person}{Oliver Hinz}.} \bibinfo{year}{2023}\natexlab{}.
\newblock \showarticletitle{Explanatory Interactive Machine Learning - Establishing an Action Design Research Process for Machine Learning Projects}.
\newblock \bibinfo{journal}{\emph{Business \& Information Systems Engineering}} \bibinfo{volume}{65}, \bibinfo{number}{6} (\bibinfo{year}{2023}), \bibinfo{pages}{677--701}.
\newblock
\urldef\tempurl%
\url{https://aisel.aisnet.org/bise/vol65/iss6/4}
\showURL{%
\tempurl}


\bibitem[Rafailov et~al\mbox{.}(2023)]%
        {rafailov2024directpreferenceoptimizationlanguage}
\bibfield{author}{\bibinfo{person}{Rafael Rafailov}, \bibinfo{person}{Archit Sharma}, \bibinfo{person}{Eric Mitchell}, \bibinfo{person}{Stefano Ermon}, \bibinfo{person}{Christopher~D. Manning}, {and} \bibinfo{person}{Chelsea Finn}.} \bibinfo{year}{2023}\natexlab{}.
\newblock \showarticletitle{Direct preference optimization: your language model is secretly a reward model}. In \bibinfo{booktitle}{\emph{Proceedings of the 37th International Conference on Neural Information Processing Systems}} (New Orleans, LA, USA) \emph{(\bibinfo{series}{NIPS '23})}. \bibinfo{publisher}{Curran Associates Inc.}, \bibinfo{address}{Red Hook, NY, USA}, Article \bibinfo{articleno}{2338}, \bibinfo{numpages}{14}~pages.
\newblock


\bibitem[Saad et~al\mbox{.}(2025)]%
        {10.5555/3709347.3743817}
\bibfield{author}{\bibinfo{person}{Fardin Saad}, \bibinfo{person}{Pradeep~K. Murukannaiah}, {and} \bibinfo{person}{Munindar~P. Singh}.} \bibinfo{year}{2025}\natexlab{}.
\newblock \showarticletitle{Gricean Norms as a Basis for Effective Collaboration}. In \bibinfo{booktitle}{\emph{Proceedings of the 24th International Conference on Autonomous Agents and Multiagent Systems}} (Detroit, MI, USA) \emph{(\bibinfo{series}{AAMAS '25})}. \bibinfo{publisher}{International Foundation for Autonomous Agents and Multiagent Systems}, \bibinfo{address}{Richland, SC}, \bibinfo{pages}{1812–1820}.
\newblock
\showISBNx{9798400714269}


\bibitem[Sadler(1989)]%
        {sadler1989formative}
\bibfield{author}{\bibinfo{person}{D.~Royce Sadler}.} \bibinfo{year}{1989}\natexlab{}.
\newblock \showarticletitle{Formative assessment and the design of instructional systems}.
\newblock \bibinfo{journal}{\emph{Instructional Science}} \bibinfo{volume}{18}, \bibinfo{number}{2} (\bibinfo{year}{1989}), \bibinfo{pages}{119--144}.
\newblock
\href{https://doi.org/10.1007/BF00117714}{doi:\nolinkurl{10.1007/BF00117714}}


\bibitem[Sahoo et~al\mbox{.}(2025)]%
        {sahoo2025systematicsurveypromptengineering}
\bibfield{author}{\bibinfo{person}{Pranab Sahoo}, \bibinfo{person}{Ayush~Kumar Singh}, \bibinfo{person}{Sriparna Saha}, \bibinfo{person}{Vinija Jain}, \bibinfo{person}{Samrat Mondal}, {and} \bibinfo{person}{Aman Chadha}.} \bibinfo{year}{2025}\natexlab{}.
\newblock \bibinfo{title}{A Systematic Survey of Prompt Engineering in Large Language Models: Techniques and Applications}.
\newblock
\showeprint[arxiv]{2402.07927}~[cs.AI]
\urldef\tempurl%
\url{https://arxiv.org/abs/2402.07927}
\showURL{%
\tempurl}


\bibitem[Scaife and Rogers(1996)]%
        {scaife1996external}
\bibfield{author}{\bibinfo{person}{Mike Scaife} {and} \bibinfo{person}{Yvonne Rogers}.} \bibinfo{year}{1996}\natexlab{}.
\newblock \showarticletitle{External cognition: How do graphical representations work?}
\newblock \bibinfo{journal}{\emph{International Journal of Human-Computer Studies}} \bibinfo{volume}{45}, \bibinfo{number}{2} (\bibinfo{year}{1996}), \bibinfo{pages}{185--213}.
\newblock


\bibitem[Scherer et~al\mbox{.}(1973)]%
        {Scherer1973VoiceConfidence}
\bibfield{author}{\bibinfo{person}{Klaus~R. Scherer}, \bibinfo{person}{Harvey London}, {and} \bibinfo{person}{Jared~J. Wolf}.} \bibinfo{year}{1973}\natexlab{}.
\newblock \showarticletitle{The voice of confidence: Paralinguistic cues and audience evaluation}.
\newblock \bibinfo{journal}{\emph{Journal of Research in Personality}} \bibinfo{volume}{7}, \bibinfo{number}{1} (\bibinfo{year}{1973}), \bibinfo{pages}{31--44}.
\newblock
\showISSN{0092-6566}
\href{https://doi.org/10.1016/0092-6566(73)90030-5}{doi:\nolinkurl{10.1016/0092-6566(73)90030-5}}


\bibitem[Schroeder et~al\mbox{.}(2017)]%
        {schroeder2017humanizing}
\bibfield{author}{\bibinfo{person}{Juliana Schroeder}, \bibinfo{person}{Michael Kardas}, {and} \bibinfo{person}{Nicholas Epley}.} \bibinfo{year}{2017}\natexlab{}.
\newblock \showarticletitle{The humanizing voice: Speech reveals, and text conceals, a more thoughtful mind in the midst of disagreement}.
\newblock \bibinfo{journal}{\emph{Psychological science}} \bibinfo{volume}{28}, \bibinfo{number}{12} (\bibinfo{year}{2017}), \bibinfo{pages}{1745--1762}.
\newblock


\bibitem[Seeger et~al\mbox{.}(2021)]%
        {Seeger2021TextingWH}
\bibfield{author}{\bibinfo{person}{Anna-Maria Seeger}, \bibinfo{person}{Jella Pfeiffer}, {and} \bibinfo{person}{Armin Heinzl}.} \bibinfo{year}{2021}\natexlab{}.
\newblock \showarticletitle{Texting with humanlike conversational agents: Designing for anthropomorphism}.
\newblock \bibinfo{journal}{\emph{Journal of the Association for Information systems}} \bibinfo{volume}{22}, \bibinfo{number}{4} (\bibinfo{year}{2021}), \bibinfo{pages}{8}.
\newblock


\bibitem[Setlur and Tory(2022)]%
        {10.1145/3491102.3501972}
\bibfield{author}{\bibinfo{person}{Vidya Setlur} {and} \bibinfo{person}{Melanie Tory}.} \bibinfo{year}{2022}\natexlab{}.
\newblock \showarticletitle{How do you Converse with an Analytical Chatbot? Revisiting Gricean Maxims for Designing Analytical Conversational Behavior}. In \bibinfo{booktitle}{\emph{Proceedings of the 2022 CHI Conference on Human Factors in Computing Systems}} (New Orleans, LA, USA) \emph{(\bibinfo{series}{CHI '22})}. \bibinfo{publisher}{Association for Computing Machinery}, \bibinfo{address}{New York, NY, USA}, Article \bibinfo{articleno}{29}, \bibinfo{numpages}{17}~pages.
\newblock
\showISBNx{9781450391573}
\href{https://doi.org/10.1145/3491102.3501972}{doi:\nolinkurl{10.1145/3491102.3501972}}


\bibitem[Sharma et~al\mbox{.}(2024b)]%
        {sharma2025understandingsycophancylanguagemodels}
\bibfield{author}{\bibinfo{person}{Mrinank Sharma}, \bibinfo{person}{Meg Tong}, \bibinfo{person}{Tomek Korbak}, \bibinfo{person}{David Duvenaud}, \bibinfo{person}{Amanda Askell}, \bibinfo{person}{Sam Bowman}, \bibinfo{person}{Esin DURMUS}, \bibinfo{person}{Zac Hatfield-Dodds}, \bibinfo{person}{Scott Johnston}, \bibinfo{person}{Shauna Kravec}, \bibinfo{person}{Timothy Maxwell}, \bibinfo{person}{Sam McCandlish}, \bibinfo{person}{Kamal Ndousse}, \bibinfo{person}{Oliver Rausch}, \bibinfo{person}{Nicholas Schiefer}, \bibinfo{person}{Da Yan}, \bibinfo{person}{Miranda Zhang}, {and} \bibinfo{person}{Ethan Perez}.} \bibinfo{year}{2024}\natexlab{b}.
\newblock \showarticletitle{Towards Understanding Sycophancy in Language Models}. In \bibinfo{booktitle}{\emph{International Conference on Learning Representations}}, \bibfield{editor}{\bibinfo{person}{B.~Kim}, \bibinfo{person}{Y.~Yue}, \bibinfo{person}{S.~Chaudhuri}, \bibinfo{person}{K.~Fragkiadaki}, \bibinfo{person}{M.~Khan}, {and} \bibinfo{person}{Y.~Sun}} (Eds.), Vol.~\bibinfo{volume}{2024}. \bibinfo{pages}{110–144}.
\newblock
\urldef\tempurl%
\url{https://proceedings.iclr.cc/paper_files/paper/2024/file/0105f7972202c1d4fb817da9f21a9663-Paper-Conference.pdf}
\showURL{%
\tempurl}


\bibitem[Sharma et~al\mbox{.}(2024a)]%
        {generative-echo-chamber}
\bibfield{author}{\bibinfo{person}{Nikhil Sharma}, \bibinfo{person}{Q.~Vera Liao}, {and} \bibinfo{person}{Ziang Xiao}.} \bibinfo{year}{2024}\natexlab{a}.
\newblock \showarticletitle{Generative Echo Chamber? Effect of LLM-Powered Search Systems on Diverse Information Seeking}. In \bibinfo{booktitle}{\emph{Proceedings of the 2024 CHI Conference on Human Factors in Computing Systems}} (Honolulu, HI, USA) \emph{(\bibinfo{series}{CHI '24})}. \bibinfo{publisher}{Association for Computing Machinery}, \bibinfo{address}{New York, NY, USA}, Article \bibinfo{articleno}{1033}, \bibinfo{numpages}{17}~pages.
\newblock
\showISBNx{9798400703300}
\href{https://doi.org/10.1145/3613904.3642459}{doi:\nolinkurl{10.1145/3613904.3642459}}


\bibitem[Sharma et~al\mbox{.}(2025)]%
        {faux-polyglot}
\bibfield{author}{\bibinfo{person}{Nikhil Sharma}, \bibinfo{person}{Kenton Murray}, {and} \bibinfo{person}{Ziang Xiao}.} \bibinfo{year}{2025}\natexlab{}.
\newblock \showarticletitle{Faux Polyglot: A Study on Information Disparity in Multilingual Large Language Models}. In \bibinfo{booktitle}{\emph{Proceedings of the 2025 Conference of the Nations of the Americas Chapter of the Association for Computational Linguistics: Human Language Technologies (Volume 1: Long Papers)}}. \bibinfo{publisher}{Association for Computational Linguistics}, \bibinfo{pages}{8090–8107}.
\newblock
\href{https://doi.org/10.18653/v1/2025.naacl-long.411}{doi:\nolinkurl{10.18653/v1/2025.naacl-long.411}}


\bibitem[Shen et~al\mbox{.}(2025)]%
        {10.1145/3706599.3716291}
\bibfield{author}{\bibinfo{person}{Hua Shen}, \bibinfo{person}{Tiffany Knearem}, \bibinfo{person}{Reshmi Ghosh}, \bibinfo{person}{Michael~Xieyang Liu}, \bibinfo{person}{Andr\'{e}s Monroy-Hern\'{a}ndez}, \bibinfo{person}{Tongshuang Wu}, \bibinfo{person}{Diyi Yang}, \bibinfo{person}{Yun Huang}, \bibinfo{person}{Tanushree Mitra}, \bibinfo{person}{Yang Li}, {and} \bibinfo{person}{Marti Hearst}.} \bibinfo{year}{2025}\natexlab{}.
\newblock \showarticletitle{Bidirectional Human-AI Alignment: Emerging Challenges and Opportunities}. In \bibinfo{booktitle}{\emph{Proceedings of the Extended Abstracts of the CHI Conference on Human Factors in Computing Systems}} \emph{(\bibinfo{series}{CHI EA '25})}. \bibinfo{publisher}{Association for Computing Machinery}, \bibinfo{address}{New York, NY, USA}, Article \bibinfo{articleno}{857}, \bibinfo{numpages}{6}~pages.
\newblock
\showISBNx{9798400713958}
\href{https://doi.org/10.1145/3706599.3716291}{doi:\nolinkurl{10.1145/3706599.3716291}}


\bibitem[Shen et~al\mbox{.}(2024)]%
        {shen2023trickle}
\bibfield{author}{\bibinfo{person}{Lingfeng Shen}, \bibinfo{person}{Sihao Chen}, \bibinfo{person}{Linfeng Song}, \bibinfo{person}{Lifeng Jin}, \bibinfo{person}{Baolin Peng}, \bibinfo{person}{Haitao Mi}, \bibinfo{person}{Daniel Khashabi}, {and} \bibinfo{person}{Dong Yu}.} \bibinfo{year}{2024}\natexlab{}.
\newblock \showarticletitle{The Trickle-down Impact of Reward Inconsistency on {RLHF}}. In \bibinfo{booktitle}{\emph{The Twelfth International Conference on Learning Representations}}.
\newblock
\urldef\tempurl%
\url{https://openreview.net/forum?id=MeHmwCDifc}
\showURL{%
\tempurl}


\bibitem[Shi et~al\mbox{.}(2025)]%
        {shi2025argumentativeexperiencereducingconfirmation}
\bibfield{author}{\bibinfo{person}{Li Shi}, \bibinfo{person}{Houjiang Liu}, \bibinfo{person}{Yian Wong}, \bibinfo{person}{Utkarsh Mujumdar}, \bibinfo{person}{Dan Zhang}, \bibinfo{person}{Jacek Gwizdka}, {and} \bibinfo{person}{Matthew Lease}.} \bibinfo{year}{2025}\natexlab{}.
\newblock \bibinfo{title}{Argumentative Experience: Reducing Confirmation Bias on Controversial Issues through LLM-Generated Multi-Persona Debates}.
\newblock
\showeprint[arxiv]{2412.04629}~[cs.HC]
\urldef\tempurl%
\url{https://arxiv.org/abs/2412.04629}
\showURL{%
\tempurl}


\bibitem[Shi et~al\mbox{.}(2024)]%
        {shi2025wildfeedbackaligningllmsinsitu}
\bibfield{author}{\bibinfo{person}{Taiwei Shi}, \bibinfo{person}{Zhuoer Wang}, \bibinfo{person}{Longqi Yang}, \bibinfo{person}{Ying-Chun Lin}, \bibinfo{person}{Zexue He}, \bibinfo{person}{Mengting Wan}, \bibinfo{person}{Pei Zhou}, \bibinfo{person}{Sujay~Kumar Jauhar}, \bibinfo{person}{Xiaofeng Xu}, \bibinfo{person}{Xia Song}, {and} \bibinfo{person}{Jennifer Neville}.} \bibinfo{year}{2024}\natexlab{}.
\newblock \showarticletitle{WildFeedback: Aligning {LLM}s With In-situ User Interactions And Feedback}. In \bibinfo{booktitle}{\emph{NeurIPS 2024 Workshop on Behavioral Machine Learning}}.
\newblock
\urldef\tempurl%
\url{https://openreview.net/forum?id=07QCozT1pi}
\showURL{%
\tempurl}


\bibitem[Shuster et~al\mbox{.}(2021)]%
        {shuster2021retrievalaugmentationreduceshallucination}
\bibfield{author}{\bibinfo{person}{Kurt Shuster}, \bibinfo{person}{Spencer Poff}, \bibinfo{person}{Moya Chen}, \bibinfo{person}{Douwe Kiela}, {and} \bibinfo{person}{Jason Weston}.} \bibinfo{year}{2021}\natexlab{}.
\newblock \showarticletitle{Retrieval Augmentation Reduces Hallucination in Conversation}. In \bibinfo{booktitle}{\emph{Findings of the Association for Computational Linguistics: EMNLP 2021}}, \bibfield{editor}{\bibinfo{person}{Marie-Francine Moens}, \bibinfo{person}{Xuanjing Huang}, \bibinfo{person}{Lucia Specia}, {and} \bibinfo{person}{Scott Wen-tau Yih}} (Eds.). \bibinfo{publisher}{Association for Computational Linguistics}, \bibinfo{address}{Punta Cana, Dominican Republic}, \bibinfo{pages}{3784--3803}.
\newblock
\href{https://doi.org/10.18653/v1/2021.findings-emnlp.320}{doi:\nolinkurl{10.18653/v1/2021.findings-emnlp.320}}


\bibitem[Shute(2008)]%
        {shute2008formative}
\bibfield{author}{\bibinfo{person}{Valerie~J. Shute}.} \bibinfo{year}{2008}\natexlab{}.
\newblock \showarticletitle{Focus on Formative Feedback}.
\newblock \bibinfo{journal}{\emph{Review of Educational Research}} \bibinfo{volume}{78}, \bibinfo{number}{1} (\bibinfo{year}{2008}), \bibinfo{pages}{153--189}.
\newblock
\showeprint{https://doi.org/10.3102/0034654307313795}
\href{https://doi.org/10.3102/0034654307313795}{doi:\nolinkurl{10.3102/0034654307313795}}


\bibitem[Simard et~al\mbox{.}(2017)]%
        {simard2017machine}
\bibfield{author}{\bibinfo{person}{Patrice~Y. Simard}, \bibinfo{person}{Saleema Amershi}, \bibinfo{person}{David~Maxwell Chickering}, \bibinfo{person}{Alicia~Edelman Pelton}, \bibinfo{person}{Soroush Ghorashi}, \bibinfo{person}{Christopher Meek}, \bibinfo{person}{Gonzalo~A. Ramos}, \bibinfo{person}{Jina Suh}, \bibinfo{person}{Johan Verwey}, \bibinfo{person}{Mo Wang}, {and} \bibinfo{person}{John Wernsing}.} \bibinfo{year}{2017}\natexlab{}.
\newblock \showarticletitle{Machine Teaching: {A} New Paradigm for Building Machine Learning Systems}.
\newblock \bibinfo{journal}{\emph{CoRR}}  \bibinfo{volume}{abs/1707.06742} (\bibinfo{year}{2017}).
\newblock
\showeprint[arXiv]{1707.06742}
\urldef\tempurl%
\url{http://arxiv.org/abs/1707.06742}
\showURL{%
\tempurl}


\bibitem[Steyvers et~al\mbox{.}(2025)]%
        {Steyvers2025LLMKnowledge}
\bibfield{author}{\bibinfo{person}{M. Steyvers}, \bibinfo{person}{H. Tejeda}, \bibinfo{person}{A. Kumar}, {et~al\mbox{.}}} \bibinfo{year}{2025}\natexlab{}.
\newblock \showarticletitle{What large language models know and what people think they know}.
\newblock \bibinfo{journal}{\emph{Nature Machine Intelligence}}  \bibinfo{volume}{7} (\bibinfo{year}{2025}), \bibinfo{pages}{221--231}.
\newblock
\href{https://doi.org/10.1038/s42256-024-00976-7}{doi:\nolinkurl{10.1038/s42256-024-00976-7}}


\bibitem[Stumpf et~al\mbox{.}(2009)]%
        {Stumpf2009InteractingMeaningfully}
\bibfield{author}{\bibinfo{person}{Simone Stumpf}, \bibinfo{person}{Vidya Rajaram}, \bibinfo{person}{Lida Li}, \bibinfo{person}{Weng-Keen Wong}, \bibinfo{person}{Margaret Burnett}, \bibinfo{person}{Thomas Dietterich}, \bibinfo{person}{Erin Sullivan}, {and} \bibinfo{person}{Jonathan Herlocker}.} \bibinfo{year}{2009}\natexlab{}.
\newblock \showarticletitle{Interacting meaningfully with machine learning systems: Three experiments}.
\newblock \bibinfo{journal}{\emph{International Journal of Human-Computer Studies}} \bibinfo{volume}{67}, \bibinfo{number}{8} (\bibinfo{year}{2009}), \bibinfo{pages}{639--662}.
\newblock
\href{https://doi.org/10.1016/j.ijhcs.2009.03.004}{doi:\nolinkurl{10.1016/j.ijhcs.2009.03.004}}


\bibitem[Sun et~al\mbox{.}(2024)]%
        {Sun2024AIHallucination}
\bibfield{author}{\bibinfo{person}{Y. Sun}, \bibinfo{person}{D. Sheng}, \bibinfo{person}{Z. Zhou}, {et~al\mbox{.}}} \bibinfo{year}{2024}\natexlab{}.
\newblock \showarticletitle{AI hallucination: towards a comprehensive classification of distorted information in artificial intelligence-generated content}.
\newblock \bibinfo{journal}{\emph{Humanities and Social Sciences Communications}}  \bibinfo{volume}{11} (\bibinfo{year}{2024}), \bibinfo{pages}{1278}.
\newblock
\href{https://doi.org/10.1057/s41599-024-03811-x}{doi:\nolinkurl{10.1057/s41599-024-03811-x}}


\bibitem[Tao et~al\mbox{.}(2024)]%
        {Tao_2024}
\bibfield{author}{\bibinfo{person}{Yan Tao}, \bibinfo{person}{Olga Viberg}, \bibinfo{person}{Ryan~S Baker}, {and} \bibinfo{person}{René~F Kizilcec}.} \bibinfo{year}{2024}\natexlab{}.
\newblock \showarticletitle{Cultural bias and cultural alignment of large language models}.
\newblock \bibinfo{journal}{\emph{PNAS Nexus}} \bibinfo{volume}{3}, \bibinfo{number}{9} (\bibinfo{date}{Sept.} \bibinfo{year}{2024}).
\newblock
\showISSN{2752-6542}
\href{https://doi.org/10.1093/pnasnexus/pgae346}{doi:\nolinkurl{10.1093/pnasnexus/pgae346}}


\bibitem[Tohidi et~al\mbox{.}(2006)]%
        {tohidi2006getting}
\bibfield{author}{\bibinfo{person}{Maryam Tohidi}, \bibinfo{person}{William Buxton}, \bibinfo{person}{Ronald Baecker}, {and} \bibinfo{person}{Abigail Sellen}.} \bibinfo{year}{2006}\natexlab{}.
\newblock \showarticletitle{Getting the right design and the design right}. In \bibinfo{booktitle}{\emph{Proceedings of the SIGCHI conference on Human Factors in computing systems}}. \bibinfo{pages}{1243--1252}.
\newblock


\bibitem[Topping(2009)]%
        {Topping01012009}
\bibfield{author}{\bibinfo{person}{Keith~J. Topping}.} \bibinfo{year}{2009}\natexlab{}.
\newblock \showarticletitle{Peer Assessment}.
\newblock \bibinfo{journal}{\emph{Theory Into Practice}} \bibinfo{volume}{48}, \bibinfo{number}{1} (\bibinfo{year}{2009}), \bibinfo{pages}{20--27}.
\newblock
\showeprint{https://doi.org/10.1080/00405840802577569}
\href{https://doi.org/10.1080/00405840802577569}{doi:\nolinkurl{10.1080/00405840802577569}}


\bibitem[Turner et~al\mbox{.}(2022)]%
        {turner2022calibrating}
\bibfield{author}{\bibinfo{person}{Amy Turner}, \bibinfo{person}{Meena Kaushik}, \bibinfo{person}{Mu-Ti Huang}, {and} \bibinfo{person}{Srikar Varanasi}.} \bibinfo{year}{2022}\natexlab{}.
\newblock \showarticletitle{Calibrating trust in AI-assisted decision making}.
\newblock \bibinfo{journal}{\emph{Google Scholar Google Scholar Navigate to}} (\bibinfo{year}{2022}).
\newblock


\bibitem[Turpin et~al\mbox{.}(2023)]%
        {turpin2023language}
\bibfield{author}{\bibinfo{person}{Miles Turpin}, \bibinfo{person}{Julian Michael}, \bibinfo{person}{Ethan Perez}, {and} \bibinfo{person}{Samuel~R. Bowman}.} \bibinfo{year}{2023}\natexlab{}.
\newblock \showarticletitle{Language Models Don't Always Say What They Think: Unfaithful Explanations in Chain-of-Thought Prompting}. In \bibinfo{booktitle}{\emph{Thirty-seventh Conference on Neural Information Processing Systems}}.
\newblock
\urldef\tempurl%
\url{https://openreview.net/forum?id=bzs4uPLXvi}
\showURL{%
\tempurl}


\bibitem[Vaccaro and colleagues(2024)]%
        {vaccaro2024meta}
\bibfield{author}{\bibinfo{person}{Andrea Vaccaro} {and} \bibinfo{person}{colleagues}.} \bibinfo{year}{2024}\natexlab{}.
\newblock \showarticletitle{Humans working with AI often underperform humans working alone: A meta-analysis}.
\newblock \bibinfo{journal}{\emph{Nature Human Behaviour}} (\bibinfo{year}{2024}).
\newblock
\href{https://doi.org/10.1038/s41562-024-02024-1}{doi:\nolinkurl{10.1038/s41562-024-02024-1}}


\bibitem[Van~Pinxteren et~al\mbox{.}(2020)]%
        {Pinxteren2020HumanlikeCI}
\bibfield{author}{\bibinfo{person}{Michelle~M.E. Van~Pinxteren}, \bibinfo{person}{Mark Pluymaekers}, {and} \bibinfo{person}{Jos~G.A.M. Lemmink}.} \bibinfo{year}{2020}\natexlab{}.
\newblock \showarticletitle{Human-like communication in conversational agents: a literature review and research agenda}.
\newblock \bibinfo{journal}{\emph{Journal of Service Management}} \bibinfo{volume}{31}, \bibinfo{number}{2} (\bibinfo{year}{2020}), \bibinfo{pages}{203–225}.
\newblock
\showISSN{1757-5818}
\href{https://doi.org/10.1108/JOSM-06-2019-0175}{doi:\nolinkurl{10.1108/JOSM-06-2019-0175}}


\bibitem[Vats et~al\mbox{.}(2025)]%
        {vats2025surveyhumanaicollaborationlarge}
\bibfield{author}{\bibinfo{person}{Vanshika Vats}, \bibinfo{person}{Marzia~Binta Nizam}, \bibinfo{person}{Minghao Liu}, \bibinfo{person}{Ziyuan Wang}, \bibinfo{person}{Richard Ho}, \bibinfo{person}{Mohnish~Sai Prasad}, \bibinfo{person}{Vincent Titterton}, \bibinfo{person}{Sai~Venkat Malreddy}, \bibinfo{person}{Riya Aggarwal}, \bibinfo{person}{Yanwen Xu}, \bibinfo{person}{Lei Ding}, \bibinfo{person}{Jay Mehta}, \bibinfo{person}{Nathan Grinnell}, \bibinfo{person}{Li Liu}, \bibinfo{person}{Sijia Zhong}, \bibinfo{person}{Devanathan~Nallur Gandamani}, \bibinfo{person}{Xinyi Tang}, \bibinfo{person}{Rohan Ghosalkar}, \bibinfo{person}{Celeste Shen}, \bibinfo{person}{Rachel Shen}, \bibinfo{person}{Nafisa Hussain}, \bibinfo{person}{Kesav Ravichandran}, {and} \bibinfo{person}{James Davis}.} \bibinfo{year}{2025}\natexlab{}.
\newblock \bibinfo{title}{A Survey on Human-AI Collaboration with Large Foundation Models}.
\newblock
\showeprint[arxiv]{2403.04931}~[cs.AI]
\urldef\tempurl%
\url{https://arxiv.org/abs/2403.04931}
\showURL{%
\tempurl}


\bibitem[Wang et~al\mbox{.}(2024)]%
        {wang2024largelanguagemodelseducation}
\bibfield{author}{\bibinfo{person}{Shen Wang}, \bibinfo{person}{Tianlong Xu}, \bibinfo{person}{Hang Li}, \bibinfo{person}{Chaoli Zhang}, \bibinfo{person}{Joleen Liang}, \bibinfo{person}{Jiliang Tang}, \bibinfo{person}{Philip~S. Yu}, {and} \bibinfo{person}{Qingsong Wen}.} \bibinfo{year}{2024}\natexlab{}.
\newblock \showarticletitle{Large Language Models for Education: {A} Survey and Outlook}.
\newblock \bibinfo{journal}{\emph{CoRR}}  \bibinfo{volume}{abs/2403.18105} (\bibinfo{year}{2024}).
\newblock
\showeprint[arXiv]{2403.18105}
\href{https://doi.org/10.48550/ARXIV.2403.18105}{doi:\nolinkurl{10.48550/ARXIV.2403.18105}}


\bibitem[Wang et~al\mbox{.}(2025)]%
        {wang2025learningaskllmagents}
\bibfield{author}{\bibinfo{person}{Wenxuan Wang}, \bibinfo{person}{Shi Juluan}, \bibinfo{person}{Zixuan Ling}, \bibinfo{person}{Yuk-Kit Chan}, \bibinfo{person}{Chaozheng Wang}, \bibinfo{person}{Cheryl Lee}, \bibinfo{person}{Youliang Yuan}, \bibinfo{person}{Jen-tse Huang}, \bibinfo{person}{Wenxiang Jiao}, {and} \bibinfo{person}{Michael~R. Lyu}.} \bibinfo{year}{2025}\natexlab{}.
\newblock \showarticletitle{Learning to Ask: When {LLM} Agents Meet Unclear Instruction}. In \bibinfo{booktitle}{\emph{Proceedings of the 2025 Conference on Empirical Methods in Natural Language Processing}}, \bibfield{editor}{\bibinfo{person}{Christos Christodoulopoulos}, \bibinfo{person}{Tanmoy Chakraborty}, \bibinfo{person}{Carolyn Rose}, {and} \bibinfo{person}{Violet Peng}} (Eds.). \bibinfo{publisher}{Association for Computational Linguistics}, \bibinfo{address}{Suzhou, China}, \bibinfo{pages}{21773--21784}.
\newblock
\showISBNx{979-8-89176-332-6}
\href{https://doi.org/10.18653/v1/2025.emnlp-main.1104}{doi:\nolinkurl{10.18653/v1/2025.emnlp-main.1104}}


\bibitem[Wang et~al\mbox{.}(2021)]%
        {wang-etal-2021-putting}
\bibfield{author}{\bibinfo{person}{Zijie~J. Wang}, \bibinfo{person}{Dongjin Choi}, \bibinfo{person}{Shenyu Xu}, {and} \bibinfo{person}{Diyi Yang}.} \bibinfo{year}{2021}\natexlab{}.
\newblock \showarticletitle{Putting Humans in the Natural Language Processing Loop: A Survey}. In \bibinfo{booktitle}{\emph{Proceedings of the First Workshop on Bridging Human{--}Computer Interaction and Natural Language Processing}}, \bibfield{editor}{\bibinfo{person}{Su~Lin Blodgett}, \bibinfo{person}{Michael Madaio}, \bibinfo{person}{Brendan O'Connor}, \bibinfo{person}{Hanna Wallach}, {and} \bibinfo{person}{Qian Yang}} (Eds.). \bibinfo{publisher}{Association for Computational Linguistics}, \bibinfo{address}{Online}, \bibinfo{pages}{47--52}.
\newblock
\urldef\tempurl%
\url{https://aclanthology.org/2021.hcinlp-1.8/}
\showURL{%
\tempurl}


\bibitem[Wei et~al\mbox{.}(2025)]%
        {wei2025plangenllms}
\bibfield{author}{\bibinfo{person}{Hui Wei}, \bibinfo{person}{Zihao Zhang}, \bibinfo{person}{Shenghua He}, \bibinfo{person}{Tian Xia}, \bibinfo{person}{Shijia Pan}, {and} \bibinfo{person}{Fei Liu}.} \bibinfo{year}{2025}\natexlab{}.
\newblock \showarticletitle{{P}lan{G}en{LLM}s: A Modern Survey of {LLM} Planning Capabilities}. In \bibinfo{booktitle}{\emph{Proceedings of the 63rd Annual Meeting of the Association for Computational Linguistics (Volume 1: Long Papers)}}, \bibfield{editor}{\bibinfo{person}{Wanxiang Che}, \bibinfo{person}{Joyce Nabende}, \bibinfo{person}{Ekaterina Shutova}, {and} \bibinfo{person}{Mohammad~Taher Pilehvar}} (Eds.). \bibinfo{publisher}{Association for Computational Linguistics}, \bibinfo{address}{Vienna, Austria}, \bibinfo{pages}{19497--19521}.
\newblock
\showISBNx{979-8-89176-251-0}
\href{https://doi.org/10.18653/v1/2025.acl-long.958}{doi:\nolinkurl{10.18653/v1/2025.acl-long.958}}


\bibitem[Wen et~al\mbox{.}(2025)]%
        {wen2025knowlimitssurveyabstention}
\bibfield{author}{\bibinfo{person}{Bingbing Wen}, \bibinfo{person}{Jihan Yao}, \bibinfo{person}{Shangbin Feng}, \bibinfo{person}{Chenjun Xu}, \bibinfo{person}{Yulia Tsvetkov}, \bibinfo{person}{Bill Howe}, {and} \bibinfo{person}{Lucy~Lu Wang}.} \bibinfo{year}{2025}\natexlab{}.
\newblock \showarticletitle{Know Your Limits: A Survey of Abstention in Large Language Models}.
\newblock \bibinfo{journal}{\emph{Transactions of the Association for Computational Linguistics}}  \bibinfo{volume}{13} (\bibinfo{year}{2025}), \bibinfo{pages}{529--556}.
\newblock
\href{https://doi.org/10.1162/tacl_a_00754}{doi:\nolinkurl{10.1162/tacl_a_00754}}


\bibitem[Westbrook and Braver(2015)]%
        {Westbrook2015Cognitive}
\bibfield{author}{\bibinfo{person}{Andrew Westbrook} {and} \bibinfo{person}{Todd~S. Braver}.} \bibinfo{year}{2015}\natexlab{}.
\newblock \showarticletitle{Cognitive effort: A neuroeconomic approach}.
\newblock \bibinfo{journal}{\emph{Cognitive, Affective, \& Behavioral Neuroscience}}  \bibinfo{volume}{15} (\bibinfo{year}{2015}), \bibinfo{pages}{395--415}.
\newblock
\href{https://doi.org/10.3758/s13415-015-0334-y}{doi:\nolinkurl{10.3758/s13415-015-0334-y}}


\bibitem[White et~al\mbox{.}(2023)]%
        {white2023promptpatterncatalogenhance}
\bibfield{author}{\bibinfo{person}{Jules White}, \bibinfo{person}{Quchen Fu}, \bibinfo{person}{Sam Hays}, \bibinfo{person}{Michael Sandborn}, \bibinfo{person}{Carlos Olea}, \bibinfo{person}{Henry Gilbert}, \bibinfo{person}{Ashraf Elnashar}, \bibinfo{person}{Jesse Spencer-Smith}, {and} \bibinfo{person}{Douglas~C. Schmidt}.} \bibinfo{year}{2023}\natexlab{}.
\newblock \showarticletitle{A Prompt Pattern Catalog to Enhance Prompt Engineering with ChatGPT}. In \bibinfo{booktitle}{\emph{Proceedings of the 30th Conference on Pattern Languages of Programs}} (Monticello, IL, USA) \emph{(\bibinfo{series}{PLoP '23})}. \bibinfo{publisher}{The Hillside Group}, \bibinfo{address}{USA}, Article \bibinfo{articleno}{5}, \bibinfo{numpages}{31}~pages.
\newblock
\showISBNx{9781941652190}


\bibitem[Winata et~al\mbox{.}(2025)]%
        {winata2025preference}
\bibfield{author}{\bibinfo{person}{Genta~Indra Winata}, \bibinfo{person}{Hanyang Zhao}, \bibinfo{person}{Anirban Das}, \bibinfo{person}{Wenpin Tang}, \bibinfo{person}{David~D Yao}, \bibinfo{person}{Shi-Xiong Zhang}, {and} \bibinfo{person}{Sambit Sahu}.} \bibinfo{year}{2025}\natexlab{}.
\newblock \showarticletitle{Preference tuning with human feedback on language, speech, and vision tasks: A survey}.
\newblock \bibinfo{journal}{\emph{Journal of Artificial Intelligence Research}}  \bibinfo{volume}{82} (\bibinfo{year}{2025}), \bibinfo{pages}{2595--2661}.
\newblock


\bibitem[Wisniewski et~al\mbox{.}(2020)]%
        {Wisniewski2020FeedbackMetaAnalysis}
\bibfield{author}{\bibinfo{person}{Benedikt Wisniewski}, \bibinfo{person}{Klaus Zierer}, {and} \bibinfo{person}{John Hattie}.} \bibinfo{year}{2020}\natexlab{}.
\newblock \showarticletitle{The Power of Feedback Revisited: A Meta-Analysis of Educational Feedback Research}.
\newblock \bibinfo{journal}{\emph{Frontiers in Psychology}}  \bibinfo{volume}{10} (\bibinfo{year}{2020}), \bibinfo{pages}{3087}.
\newblock
\href{https://doi.org/10.3389/fpsyg.2019.03087}{doi:\nolinkurl{10.3389/fpsyg.2019.03087}}


\bibitem[Wu et~al\mbox{.}(2024)]%
        {wu-etal-2024-need}
\bibfield{author}{\bibinfo{person}{Cheng-Kuang Wu}, \bibinfo{person}{Zhi~Rui Tam}, \bibinfo{person}{Chao-Chung Wu}, \bibinfo{person}{Chieh-Yen Lin}, \bibinfo{person}{Hung-yi Lee}, {and} \bibinfo{person}{Yun-Nung Chen}.} \bibinfo{year}{2024}\natexlab{}.
\newblock \showarticletitle{{I} Need Help! Evaluating {LLM}{'}s Ability to Ask for Users' Support: A Case Study on Text-to-{SQL} Generation}. In \bibinfo{booktitle}{\emph{Proceedings of the 2024 Conference on Empirical Methods in Natural Language Processing}}, \bibfield{editor}{\bibinfo{person}{Yaser Al-Onaizan}, \bibinfo{person}{Mohit Bansal}, {and} \bibinfo{person}{Yun-Nung Chen}} (Eds.). \bibinfo{publisher}{Association for Computational Linguistics}, \bibinfo{address}{Miami, Florida, USA}, \bibinfo{pages}{2191--2199}.
\newblock
\href{https://doi.org/10.18653/v1/2024.emnlp-main.131}{doi:\nolinkurl{10.18653/v1/2024.emnlp-main.131}}


\bibitem[Wu et~al\mbox{.}(2025)]%
        {wu2025collabllm}
\bibfield{author}{\bibinfo{person}{Shirley Wu}, \bibinfo{person}{Michel Galley}, \bibinfo{person}{Baolin Peng}, \bibinfo{person}{Hao Cheng}, \bibinfo{person}{Gavin Li}, \bibinfo{person}{Yao Dou}, \bibinfo{person}{Weixin Cai}, \bibinfo{person}{James Zou}, \bibinfo{person}{Jure Leskovec}, {and} \bibinfo{person}{Jianfeng Gao}.} \bibinfo{year}{2025}\natexlab{}.
\newblock \showarticletitle{Collab{LLM}: From Passive Responders to Active Collaborators}. In \bibinfo{booktitle}{\emph{Forty-second International Conference on Machine Learning}}.
\newblock
\urldef\tempurl%
\url{https://openreview.net/forum?id=DmH4HHVb3y}
\showURL{%
\tempurl}


\bibitem[Xiao et~al\mbox{.}(2021)]%
        {xiao2021let}
\bibfield{author}{\bibinfo{person}{Ziang Xiao}, \bibinfo{person}{Sarah Mennicken}, \bibinfo{person}{Bernd Huber}, \bibinfo{person}{Adam Shonkoff}, {and} \bibinfo{person}{Jennifer Thom}.} \bibinfo{year}{2021}\natexlab{}.
\newblock \showarticletitle{Let me ask you this: How can a voice assistant elicit explicit user feedback?}
\newblock \bibinfo{journal}{\emph{Proceedings of the ACM on Human-Computer Interaction}} \bibinfo{volume}{5}, \bibinfo{number}{CSCW2} (\bibinfo{year}{2021}), \bibinfo{pages}{1--24}.
\newblock


\bibitem[Xiao et~al\mbox{.}(2020a)]%
        {xiao2020if}
\bibfield{author}{\bibinfo{person}{Ziang Xiao}, \bibinfo{person}{Michelle~X Zhou}, \bibinfo{person}{Wenxi Chen}, \bibinfo{person}{Huahai Yang}, {and} \bibinfo{person}{Changyan Chi}.} \bibinfo{year}{2020}\natexlab{a}.
\newblock \showarticletitle{If I hear you correctly: Building and evaluating interview chatbots with active listening skills}. In \bibinfo{booktitle}{\emph{Proceedings of the 2020 CHI Conference on Human Factors in Computing Systems}}. \bibinfo{pages}{1--14}.
\newblock


\bibitem[Xiao et~al\mbox{.}(2020b)]%
        {xiao2020tell}
\bibfield{author}{\bibinfo{person}{Ziang Xiao}, \bibinfo{person}{Michelle~X Zhou}, \bibinfo{person}{Q~Vera Liao}, \bibinfo{person}{Gloria Mark}, \bibinfo{person}{Changyan Chi}, \bibinfo{person}{Wenxi Chen}, {and} \bibinfo{person}{Huahai Yang}.} \bibinfo{year}{2020}\natexlab{b}.
\newblock \showarticletitle{Tell me about yourself: Using an AI-powered chatbot to conduct conversational surveys with open-ended questions}.
\newblock \bibinfo{journal}{\emph{ACM Transactions on Computer-Human Interaction (TOCHI)}} \bibinfo{volume}{27}, \bibinfo{number}{3} (\bibinfo{year}{2020}), \bibinfo{pages}{1--37}.
\newblock


\bibitem[Xu et~al\mbox{.}(2025)]%
        {xu2025largereasoningmodelssurvey}
\bibfield{author}{\bibinfo{person}{Fengli Xu}, \bibinfo{person}{Qianyue Hao}, \bibinfo{person}{Chenyang Shao}, \bibinfo{person}{Zefang Zong}, \bibinfo{person}{Yu Li}, \bibinfo{person}{Jingwei Wang}, \bibinfo{person}{Yunke Zhang}, \bibinfo{person}{Jingyi Wang}, \bibinfo{person}{Xiaochong Lan}, \bibinfo{person}{Jiahui Gong}, \bibinfo{person}{Tianjian Ouyang}, \bibinfo{person}{Fanjin Meng}, \bibinfo{person}{Yuwei Yan}, \bibinfo{person}{Qinglong Yang}, \bibinfo{person}{Yiwen Song}, \bibinfo{person}{Sijian Ren}, \bibinfo{person}{Xinyuan Hu}, \bibinfo{person}{Jie Feng}, \bibinfo{person}{Chen Gao}, {and} \bibinfo{person}{Yong Li}.} \bibinfo{year}{2025}\natexlab{}.
\newblock \showarticletitle{Toward large reasoning models: A survey of reinforced reasoning with large language models}.
\newblock \bibinfo{journal}{\emph{Patterns}} \bibinfo{volume}{6}, \bibinfo{number}{10} (\bibinfo{year}{2025}), \bibinfo{pages}{101370}.
\newblock
\showISSN{2666-3899}
\href{https://doi.org/10.1016/j.patter.2025.101370}{doi:\nolinkurl{10.1016/j.patter.2025.101370}}


\bibitem[Yang et~al\mbox{.}(2024)]%
        {yang2024human}
\bibfield{author}{\bibinfo{person}{Diyi Yang}, \bibinfo{person}{Sherry~Tongshuang Wu}, {and} \bibinfo{person}{Marti~A Hearst}.} \bibinfo{year}{2024}\natexlab{}.
\newblock \showarticletitle{Human-AI interaction in the age of LLMs}. In \bibinfo{booktitle}{\emph{Proceedings of the 2024 Conference of the North American Chapter of the Association for Computational Linguistics: Human Language Technologies (Volume 5: Tutorial Abstracts)}}. \bibinfo{pages}{34--38}.
\newblock


\bibitem[Zamfirescu-Pereira et~al\mbox{.}(2023)]%
        {10.1145/3544548.3581388}
\bibfield{author}{\bibinfo{person}{J.D. Zamfirescu-Pereira}, \bibinfo{person}{Richmond~Y. Wong}, \bibinfo{person}{Bjoern Hartmann}, {and} \bibinfo{person}{Qian Yang}.} \bibinfo{year}{2023}\natexlab{}.
\newblock \showarticletitle{Why Johnny Can’t Prompt: How Non-AI Experts Try (and Fail) to Design LLM Prompts}. In \bibinfo{booktitle}{\emph{Proceedings of the 2023 CHI Conference on Human Factors in Computing Systems}} (Hamburg, Germany) \emph{(\bibinfo{series}{CHI '23})}. \bibinfo{publisher}{Association for Computing Machinery}, \bibinfo{address}{New York, NY, USA}, Article \bibinfo{articleno}{437}, \bibinfo{numpages}{21}~pages.
\newblock
\showISBNx{9781450394215}
\href{https://doi.org/10.1145/3544548.3581388}{doi:\nolinkurl{10.1145/3544548.3581388}}


\bibitem[Zhang et~al\mbox{.}(2025c)]%
        {zhang2025surveymultiturninteractioncapabilities}
\bibfield{author}{\bibinfo{person}{Chen Zhang}, \bibinfo{person}{Xinyi Dai}, \bibinfo{person}{Yaxiong Wu}, \bibinfo{person}{Qu Yang}, \bibinfo{person}{Yasheng Wang}, \bibinfo{person}{Ruiming Tang}, {and} \bibinfo{person}{Yong Liu}.} \bibinfo{year}{2025}\natexlab{c}.
\newblock \bibinfo{title}{A Survey on Multi-Turn Interaction Capabilities of Large Language Models}.
\newblock
\showeprint[arxiv]{2501.09959}~[cs.CL]
\urldef\tempurl%
\url{https://arxiv.org/abs/2501.09959}
\showURL{%
\tempurl}


\bibitem[Zhang et~al\mbox{.}(2025a)]%
        {zhang2024longciteenablingllmsgenerate}
\bibfield{author}{\bibinfo{person}{Jiajie Zhang}, \bibinfo{person}{Yushi Bai}, \bibinfo{person}{Xin Lv}, \bibinfo{person}{Wanjun Gu}, \bibinfo{person}{Danqing Liu}, \bibinfo{person}{Minhao Zou}, \bibinfo{person}{Shulin Cao}, \bibinfo{person}{Lei Hou}, \bibinfo{person}{Yuxiao Dong}, \bibinfo{person}{Ling Feng}, {and} \bibinfo{person}{Juanzi Li}.} \bibinfo{year}{2025}\natexlab{a}.
\newblock \showarticletitle{{L}ong{C}ite: Enabling {LLM}s to Generate Fine-grained Citations in Long-Context {QA}}. In \bibinfo{booktitle}{\emph{Findings of the Association for Computational Linguistics: ACL 2025}}, \bibfield{editor}{\bibinfo{person}{Wanxiang Che}, \bibinfo{person}{Joyce Nabende}, \bibinfo{person}{Ekaterina Shutova}, {and} \bibinfo{person}{Mohammad~Taher Pilehvar}} (Eds.). \bibinfo{publisher}{Association for Computational Linguistics}, \bibinfo{address}{Vienna, Austria}, \bibinfo{pages}{5098--5122}.
\newblock
\showISBNx{979-8-89176-256-5}
\href{https://doi.org/10.18653/v1/2025.findings-acl.264}{doi:\nolinkurl{10.18653/v1/2025.findings-acl.264}}


\bibitem[Zhang et~al\mbox{.}(2025e)]%
        {zhang2025verifiabledesignaligninglanguage}
\bibfield{author}{\bibinfo{person}{Jingyu Zhang}, \bibinfo{person}{Marc Marone}, \bibinfo{person}{Tianjian Li}, \bibinfo{person}{Benjamin Van~Durme}, {and} \bibinfo{person}{Daniel Khashabi}.} \bibinfo{year}{2025}\natexlab{e}.
\newblock \showarticletitle{Verifiable by Design: Aligning Language Models to Quote from Pre-Training Data}. In \bibinfo{booktitle}{\emph{Proceedings of the 2025 Conference of the Nations of the Americas Chapter of the Association for Computational Linguistics: Human Language Technologies (Volume 1: Long Papers)}}, \bibfield{editor}{\bibinfo{person}{Luis Chiruzzo}, \bibinfo{person}{Alan Ritter}, {and} \bibinfo{person}{Lu~Wang}} (Eds.). \bibinfo{publisher}{Association for Computational Linguistics}, \bibinfo{address}{Albuquerque, New Mexico}, \bibinfo{pages}{3748--3768}.
\newblock
\showISBNx{979-8-89176-189-6}
\href{https://doi.org/10.18653/v1/2025.naacl-long.191}{doi:\nolinkurl{10.18653/v1/2025.naacl-long.191}}


\bibitem[Zhang and Norman(1994)]%
        {zhang1994representations}
\bibfield{author}{\bibinfo{person}{Jiajie Zhang} {and} \bibinfo{person}{Donald~A Norman}.} \bibinfo{year}{1994}\natexlab{}.
\newblock \showarticletitle{Representations in distributed cognitive tasks}.
\newblock \bibinfo{journal}{\emph{Cognitive Science}} \bibinfo{volume}{18}, \bibinfo{number}{1} (\bibinfo{year}{1994}), \bibinfo{pages}{87--122}.
\newblock


\bibitem[Zhang et~al\mbox{.}(2025d)]%
        {zhang2025modeling}
\bibfield{author}{\bibinfo{person}{Michael~JQ Zhang}, \bibinfo{person}{W.~Bradley Knox}, {and} \bibinfo{person}{Eunsol Choi}.} \bibinfo{year}{2025}\natexlab{d}.
\newblock \showarticletitle{Modeling Future Conversation Turns to Teach {LLM}s to Ask Clarifying Questions}. In \bibinfo{booktitle}{\emph{The Thirteenth International Conference on Learning Representations}}.
\newblock
\urldef\tempurl%
\url{https://openreview.net/forum?id=cwuSAR7EKd}
\showURL{%
\tempurl}


\bibitem[Zhang et~al\mbox{.}(2025g)]%
        {zhang2024collaboration}
\bibfield{author}{\bibinfo{person}{Shuning Zhang}, \bibinfo{person}{Hui Wang}, {and} \bibinfo{person}{Xin Yi}.} \bibinfo{year}{2025}\natexlab{g}.
\newblock \showarticletitle{Exploring Collaboration Patterns and Strategies in Human-AI Co-creation through the Lens of Agency: A Scoping Review of the Top-tier HCI Literature}.
\newblock \bibinfo{journal}{\emph{Proc. ACM Hum.-Comput. Interact.}} \bibinfo{volume}{9}, \bibinfo{number}{7}, Article \bibinfo{articleno}{CSCW413} (\bibinfo{date}{Oct.} \bibinfo{year}{2025}), \bibinfo{numpages}{43}~pages.
\newblock
\href{https://doi.org/10.1145/3757594}{doi:\nolinkurl{10.1145/3757594}}


\bibitem[Zhang et~al\mbox{.}(2025b)]%
        {10.5555/3737916.3739859}
\bibfield{author}{\bibinfo{person}{Zhenyu Zhang}, \bibinfo{person}{Runjin Chen}, \bibinfo{person}{Shiwei Liu}, \bibinfo{person}{Zhewei Yao}, \bibinfo{person}{Olatunji Ruwase}, \bibinfo{person}{Beidi Chen}, \bibinfo{person}{Xiaoxia Wu}, {and} \bibinfo{person}{Zhangyang Wang}.} \bibinfo{year}{2025}\natexlab{b}.
\newblock \showarticletitle{Found in the middle: how language models use long contexts better via plug-and-play positional encoding}. In \bibinfo{booktitle}{\emph{Proceedings of the 38th International Conference on Neural Information Processing Systems}} (Vancouver, BC, Canada) \emph{(\bibinfo{series}{NIPS '24})}. \bibinfo{publisher}{Curran Associates Inc.}, \bibinfo{address}{Red Hook, NY, USA}, Article \bibinfo{articleno}{1943}, \bibinfo{numpages}{21}~pages.
\newblock
\showISBNx{9798331314385}


\bibitem[Zhang et~al\mbox{.}(2025f)]%
        {zhang2025personalizationlargelanguagemodels}
\bibfield{author}{\bibinfo{person}{Zhehao Zhang}, \bibinfo{person}{Ryan~A. Rossi}, \bibinfo{person}{Branislav Kveton}, \bibinfo{person}{Yijia Shao}, \bibinfo{person}{Diyi Yang}, \bibinfo{person}{Hamed Zamani}, \bibinfo{person}{Franck Dernoncourt}, \bibinfo{person}{Joe Barrow}, \bibinfo{person}{Tong Yu}, \bibinfo{person}{Sungchul Kim}, \bibinfo{person}{Ruiyi Zhang}, \bibinfo{person}{Jiuxiang Gu}, \bibinfo{person}{Tyler Derr}, \bibinfo{person}{Hongjie Chen}, \bibinfo{person}{Junda Wu}, \bibinfo{person}{Xiang Chen}, \bibinfo{person}{Zichao Wang}, \bibinfo{person}{Subrata Mitra}, \bibinfo{person}{Nedim Lipka}, \bibinfo{person}{Nesreen~K. Ahmed}, {and} \bibinfo{person}{Yu Wang}.} \bibinfo{year}{2025}\natexlab{f}.
\newblock \showarticletitle{Personalization of Large Language Models: A Survey}.
\newblock \bibinfo{journal}{\emph{Transactions on Machine Learning Research}} (\bibinfo{year}{2025}).
\newblock
\showISSN{2835-8856}
\urldef\tempurl%
\url{https://openreview.net/forum?id=tf6A9EYMo6}
\showURL{%
\tempurl}
\newblock
\shownote{Survey Certification}.


\bibitem[Zhang et~al\mbox{.}(2020)]%
        {zhang2020recent}
\bibfield{author}{\bibinfo{person}{Zheng Zhang}, \bibinfo{person}{Ryuichi Takanobu}, \bibinfo{person}{Qi Zhu}, \bibinfo{person}{MinLie Huang}, {and} \bibinfo{person}{XiaoYan Zhu}.} \bibinfo{year}{2020}\natexlab{}.
\newblock \showarticletitle{Recent advances and challenges in task-oriented dialog systems}.
\newblock \bibinfo{journal}{\emph{Science China Technological Sciences}} \bibinfo{volume}{63}, \bibinfo{number}{10} (\bibinfo{year}{2020}), \bibinfo{pages}{2011--2027}.
\newblock


\bibitem[Zhao et~al\mbox{.}(2024)]%
        {zhao2024risk}
\bibfield{author}{\bibinfo{person}{Yukun Zhao}, \bibinfo{person}{Zhen Huang}, \bibinfo{person}{Martin Seligman}, {and} \bibinfo{person}{Kaiping Peng}.} \bibinfo{year}{2024}\natexlab{}.
\newblock \showarticletitle{Risk and prosocial behavioural cues elicit human-like response patterns from AI chatbots}.
\newblock \bibinfo{journal}{\emph{Scientific Reports}}  \bibinfo{volume}{14} (\bibinfo{year}{2024}), \bibinfo{pages}{7095}.
\newblock
\href{https://doi.org/10.1038/s41598-024-55949-y}{doi:\nolinkurl{10.1038/s41598-024-55949-y}}


\bibitem[Zhao et~al\mbox{.}(2025)]%
        {zhao-etal-2025-personalens}
\bibfield{author}{\bibinfo{person}{Zheng Zhao}, \bibinfo{person}{Clara Vania}, \bibinfo{person}{Subhradeep Kayal}, \bibinfo{person}{Naila Khan}, \bibinfo{person}{Shay~B Cohen}, {and} \bibinfo{person}{Emine Yilmaz}.} \bibinfo{year}{2025}\natexlab{}.
\newblock \showarticletitle{{P}ersona{L}ens: A Benchmark for Personalization Evaluation in Conversational {AI} Assistants}. In \bibinfo{booktitle}{\emph{Findings of the Association for Computational Linguistics: ACL 2025}}, \bibfield{editor}{\bibinfo{person}{Wanxiang Che}, \bibinfo{person}{Joyce Nabende}, \bibinfo{person}{Ekaterina Shutova}, {and} \bibinfo{person}{Mohammad~Taher Pilehvar}} (Eds.). \bibinfo{publisher}{Association for Computational Linguistics}, \bibinfo{address}{Vienna, Austria}, \bibinfo{pages}{18023--18055}.
\newblock
\showISBNx{979-8-89176-256-5}
\href{https://doi.org/10.18653/v1/2025.findings-acl.927}{doi:\nolinkurl{10.18653/v1/2025.findings-acl.927}}


\bibitem[Zou et~al\mbox{.}(2025)]%
        {zou2025collaborativeintelligencehumanagentsystems}
\bibfield{author}{\bibinfo{person}{Henry~Peng Zou}, \bibinfo{person}{Wei-Chieh Huang}, \bibinfo{person}{Yaozu Wu}, \bibinfo{person}{Chunyu Miao}, \bibinfo{person}{Dongyuan Li}, \bibinfo{person}{Aiwei Liu}, \bibinfo{person}{Yue Zhou}, \bibinfo{person}{Yankai Chen}, \bibinfo{person}{Weizhi Zhang}, \bibinfo{person}{Yangning Li}, \bibinfo{person}{Liancheng Fang}, \bibinfo{person}{Renhe Jiang}, {and} \bibinfo{person}{Philip~S. Yu}.} \bibinfo{year}{2025}\natexlab{}.
\newblock \bibinfo{title}{A Call for Collaborative Intelligence: Why Human-Agent Systems Should Precede AI Autonomy}.
\newblock
\showeprint[arxiv]{2506.09420}~[cs.AI]
\urldef\tempurl%
\url{https://arxiv.org/abs/2506.09420}
\showURL{%
\tempurl}


\end{thebibliography}
